\documentclass[aps,prd,preprint,superscriptaddress,amssymb,amsmath,nofootinbib]{revtex4}

\usepackage{color}
\usepackage{epsfig}
\usepackage{graphicx}
\usepackage{float}
\usepackage{epstopdf}
\usepackage{hyperref} 
\usepackage{comment}
\usepackage[usenames,dvipsnames]{xcolor}
\usepackage{multirow}
\usepackage[normalem]{ulem}
\usepackage{slashed}
\usepackage{feynmf}

\begin{document}

\preprint{}

\title{Predictions for muon electric and magnetic dipole moments from $h \rightarrow \mu^+ \mu^-$ in two-Higgs-doubletmModels with new leptons}

\author{Radovan Dermisek}
\email[]{dermisek@indiana.edu}
\affiliation{Physics Department, Indiana University, Bloomington, Indiana 47405, USA}

\author{Keith Hermanek}
\email[]{khermane@iu.edu}
\affiliation{Physics Department, Indiana University, Bloomington, Indiana 47405, USA}

\author{Navin McGinnis}
\email[]{nmcginnis@triumf.ca}
\affiliation{TRIUMF, 4004 Westbrook Mall, Vancouver, British Columbia V6T 2A3, Canada}

\author{Sangsik Yoon}
\email[]{yoon12@iu.edu}
\affiliation{Physics Department, Indiana University, Bloomington, Indiana 47405, USA}

\date{\today}

\begin{abstract}

We calculate chirally enhanced corrections to the muon's electric and magnetic dipole moments in two-Higgs-doublet models extended by vector-like leptons, and we explore a sharp correlation between $h \rightarrow \mu^+ \mu^-$ and the muon's dipole moments in these models. Among many detailed predictions,
for a model with new leptons with the same quantum numbers as standard model leptons, we find that $0.38 \lesssim \tan \beta \lesssim 21$ necessarily requires a muon electric dipole moment to be observed at near-future experiments, assuming $h \rightarrow \mu^+ \mu^-$ is measured within $1\%$ of the standard model prediction for the current central value of the measured muon magnetic moment. In all studied models, the predicted values of the electric dipole moment can reach up to current experimental limits. Moreover, we show that in some models there can be two sources of chiral enhancement, parametrizing the correlation between $h \rightarrow \mu^+ \mu^-$ and the dipole moments by a complex number. 
This leads to sign-preferred predictions for the electric dipole moment.

\end{abstract}


\maketitle

\section{Introduction}

Recently, measurements of fundamental properties of the muon have gained renewed interest in serving as harbingers of new physics. Measurements of the standard model (SM) Higgs boson decay into muons allow for a deviation by more than a factor of 2 compared to the SM prediction~\cite{ATLAS:2020fzp}, while the most recent measurement \cite{Muong-2:2023cdq} of the muon's anomalous magnetic moment, $\Delta a_{\mu}$, suggests that the experimental world average is discrepant by more than five standard deviations from the SM prediction of 2020 \cite{Aoyama:2020ynm}. Current theoretical and experimental efforts aim to understand this result \cite{Borsanyi:2020mff,CMD-3:2023alj}. Additionally, the muon electric dipole moment has been a popular tool in studying the possible effects of large $CP$ violation from new physics~\cite{Fukuyama:2012np}.

In this paper, we calculate contributions to the electric and magnetic dipole moments of the muon in a two-Higgs-doublet model (2HDM), assuming a $Z_2$ symmetry which 
enforces type-II couplings to SM leptons, extended with vector-like leptons. We explore representations of new leptons up to and including $SU(2)$ triplets that were previously studied in SM extensions \cite{Kannike:2011ng,Dermisek:2013gta}. Such models (and other possibilities, see for example Refs. \cite{Crivellin:2021rbq,Capdevilla:2021rwo,Capdevilla:2020qel,Khaw:2022qxh}) can generate simultaneous corrections to the muon mass and dipole moments as $\delta m_{\mu} \simeq \lambda^3 v^3 / M^2$, $\Delta a_{\mu} \simeq m_{\mu} v \textrm{Re}[\lambda^3] / 16 \pi^2 M^2$, and $d_{\mu} \simeq e v \textrm{Im}[\lambda^3] / 32 \pi^2 M^2$, where $\lambda^3$ and $M^2$ are the products of the couplings and mass scales of new particles. The corrections to dipole moments enjoy a \textit{chiral enhancement} of $\lambda v / m_{\mu}$ compared to the naive estimate, $\Delta a_{\mu} \simeq m_{\mu}^2 \textrm{Re}[\lambda^2] / 16 \pi^2 M^2$ and $d_{\mu} \simeq e m_{\mu} \textrm{Im}[\lambda^2] / 32 \pi^2 M^2$, and allow for the largest scales of new physics to explain the anomaly. Chirally enhanced corrections to the muon dipole moments from vector-like leptons may also be connected to new physics relevant for models of dark matter~\cite{Angelescu:2016mhl,Jana:2020joi,Arcadi:2021cwg,Frank:2021nkq,Athron:2021iuf,Wojcik:2022woa} and the Cabibbo anomaly~\cite{Endo:2020tkb,Crivellin:2020ebi}.\footnote{Additional explanations for $\Delta a_{\mu}$ consistent with searches for new Higgses can be accomplished exclusively in 2HDMs, requiring light scalar and pseudoscalar masses below the electroweak scale and large couplings to leptons, such as in type-X and flavor-aligned 2HDMs~\cite{Broggio:2014mna,Cherchiglia:2017uwv}, and even in the general 2HDM where flavor-changing currents are permitted~\cite{Athron:2021auq}. A muon-specific 2HDM can also accomplish this~\cite{Nakai:2022vgp}.}

In a 2HDM with type-II couplings (2HDM-II), these contributions can be even further enhanced by factors of $\tan^2 \beta$ \cite{Dermisek:2021ajd,Dermisek:2020cod} [where $\tan \beta$ is the ratio of the vacuum expectation values (VEV) of the two-Higgs doublets]. For the models not explored in \cite{Dermisek:2021ajd,Dermisek:2020cod}, we find yet another source of chiral enhancement. For all of the models that we consider, we present complete expressions of both dipole moments in the mass eigenstate basis. These formulas are completely generic and can be applied to any model with new leptons and an extended Higgs sector. These could be applied, for instance, to other types of 2HDMs~\cite{Frank:2020smf,Chun:2020uzw,DelleRose:2020oaa,Ferreira:2021gke,Ghosh:2021jeg,Han:2021gfu,Dey:2021pyn,Raju:2022zlv} or more generic extensions of the SM with vector-like leptons and new scalar particles~\cite{Chakrabarty:2020jro,Chakrabarty:2021ztf,Liu:2021kqp,Mondal:2021vou,Dey:2021pyn,Bharadwaj:2021tgp}. 

In addition to calculations in the mass eigenstate basis, we also present results using a SM effective field theory (SMEFT)-like 2HDM effective field theory working at dimension six, where the $SU(2)_L \times U(1)_Y$ symmetry is linearly realized and both Higgs doublets are kept light while vector-like leptons are integrated out.
These results reproduce the leading-order $v^2/M^2$ corrections in the mass eigenstate basis. We find that working in the so-called Higgs basis \cite{Lavoura:1994fv,Lavoura:1994yu,Botella:1994cs,Davidson:2005cw} greatly simplifies the matching calculation, leading to strikingly simple parametrizations for the muon dipole moments. 

Although chirally enhanced dipole moments can allow for extremely heavy scales of new physics currently out of reach at the Large Hadron Collider (LHC) and possible future colliders, it was recently shown that these corrections are highly correlated to the $h \rightarrow \mu^+ \mu^-$ decay rate \cite{Crivellin:2020tsz,Fajfer:2021cxa,Dermisek:2021mhi,Dermisek:2022aec,Dermisek:2023nhe,Crivellin:2021rbq,Crivellin:2022wzw,Hamaguchi:2022byw}, although this was explored earlier for lighter new leptons \cite{Kannike:2011ng,Dermisek:2013gta}. In particular, the dimension-six operator generating the muon dipole operators is directly proportional to the one that modifies the muon mass by a real factor, $k$. This allows for a simultaneous parametrization of three observables$-\Delta a_{\mu}$, $d_{\mu}$, and $h \rightarrow \mu^+ \mu^- -$through a single model-dependent factor, providing a novel way to test high scales in the theory that are directly inaccessible to current and future colliders. In a 2HDM-II extended with vector-like leptons, the proportionality factor can now be complex due to the interference of multiple sources of chiral enhancement. This was first pointed out in a model-independent fashion in \cite{Dermisek:2022aec,Dermisek:2023nhe}.
The complex factor leads to asymmetric and sign-preferred predictions of $d_{\mu}$, previously not investigated in the literature. We find a similar behavior can also occur when including subleading loop corrections to the muon mass. The correlation of muon observables can be used to obtain novel bounds on the masses and parameter space of 2HDMs with vector-like leptons inferred from current and future limits of $h \rightarrow \mu^+ \mu^-$ and projected measurements of $d_{\mu}$.

The paper is organized as follows. In Sec.~\ref{sec:params} we introduce models of vector-like leptons in the context of 2HDM-II allowing for complex Yukawa couplings and discuss our conventions for the scalar potential. In Sec.~\ref{sec:dipoles} we present general formulas for contributions to $\Delta a_{\mu}$ and $d_{\mu}$ in the mass eigenstate basis. In Sec.~\ref{sec:2hdm_eft} we present calculations of the dipole moments in the 2HDM-II effective field theory. In Sec.~\ref{sec:results} we explore the connection between $\Delta a_{\mu}$, $d_{\mu}$, and $h \rightarrow \mu^+ \mu^-$, present detailed results, and extend discussions from previous works. We conclude in Sec.~\ref{sec:conclusion}. We also include lengthy appendices providing details of the 2HDM scalar potential, representations for new leptons and our conventions for calculations in the mass eigenstate basis, approximate formulas for couplings of leptons, and further details of the effective field theory calculations.

\section{2HDM-II with Vectorlike Leptons}
\label{sec:params}
\subsection{Model setup and parameters}

We consider a 2HDM extended with charged vector-like lepton doublets $L_{L,R}$ and singlets $E_{L,R}$, as well as neutral vector-like singlets $N_{L,R}$. 
We assume a $Z_2$ symmetry, enforcing couplings of the SM leptons to the Higgs doublets as in a type-II 2HDM. This model was first considered in \cite{Dermisek:2015oja} and studied extensively in \cite{Dermisek:2020cod, Dermisek:2021ajd} in order to explain the anomalous magnetic moment of the muon. The most general Lagrangian assuming only mixing of second-generation leptons with new leptons is described by
\begin{equation}
\begin{split}
    \mathcal{L} \supset & - y_{\mu} \bar{l}_L \mu_R H_d - \lambda_E \bar{l}_L E_R H_d - \lambda_L \bar{L}_L \mu_R H_d - \lambda \bar{L}_L E_R H_d - \bar{\lambda} H_d^{\dagger} \bar{E}_L L_R \\
    & - \kappa_N \bar{l}_L N_R H_u -\kappa \bar{L}_L N_R H_u - \bar{\kappa} H_u^{\dagger} \bar{N}_L L_R \\
    & - M_L \bar{L}_L L_R - M_E \bar{E}_L E_R - M_N \bar{N}_L N_R + h.c.
\label{eq:vll_lag}
\end{split}
\end{equation} 
Quantum numbers of the fields are given in Table~\ref{tab:Q_numbers}. 
Couplings of $H_d$ and $H_u$ to other SM fermions are like those in the typical type-II 2HDM. Additional models with different representations of vectorlike leptons are described in Appendix \ref{sec:approx_form}. An alternate version of these models exists motivated by the minimal supersymmetric SM extended with vector-like leptons \cite{Dermisek:2021ajd}. Because the superpotential is holomorphic, terms involving $H_d^{\dagger}$ and $H_u^{\dagger}$ are forbidden (but similar terms appear through $H_u$ and $H_d$, respectively). In principle, each of the eight couplings and three vector-like masses can be complex. Field redefinitions can be chosen such that several phases of the parameters are unphysical. The Lagrangian in Eq.~(\ref{eq:vll_lag}) admits four additional physical phases (two in the charged and two in the neutral lepton sectors) that cannot be rotated away. For example, we could take the mass parameters $M_{L,E,N}$ and all Yukawa couplings to be real except for $\overline{\lambda}, \lambda, \overline{\kappa}$, and $\kappa$ (although we do not make this assumption). We explore the impact of these complex couplings later in Sec.~\ref{sec:dipoles}.

The doublet components are defined as 
\begin{equation}
    \begin{split}
        l_L = \begin{pmatrix}
        \nu_{\mu} \\ \mu_L 
        \end{pmatrix}, \ \ \ \ \  L_{L,R} = \begin{pmatrix}
        L_{L,R}^0 \\ L_{L,R}^{-}
        \end{pmatrix}, \ \ \ \ \ H_d = \begin{pmatrix}
        H_d^+ \\ H_d^0
        \end{pmatrix}, \ \ \ \ \ 
        H_u = \begin{pmatrix}
        H_u^0 \\ H_u^-
        \end{pmatrix}.
    \end{split}
\end{equation}
After electroweak symmetry breaking (EWSB), the neutral components of the Higgs doublets develop VEVs $\langle H_d^0 \rangle = v_d$ and $\langle H_u^0 \rangle = v_u$, whereby $v = \sqrt{v_d^2 + v_u^2} = 174$ GeV and their ratio is parametrized by $v_u / v_d = \tan \beta$. The diagonalization procedure for rotating these fields to the physical basis is outlined in Appendix \ref{sec:approx_form} and also provided for other representations.
\begin{table}[t]
\centering
 \begin{tabular}{||c | c c c c c c c||} 
     \hline
     & $l_L$ \ \ & $\mu_R$ \ \ & $H_u$ \ \ & $H_d$ \ \ & $L_{L,R}$ \ \ & $N_{L,R}$ \ \ & $E_{L,R}$ \\
    \hline
    $SU(2)_L$ & $\textbf{2}$ \ \ & $\textbf{1}$ \ \ & $\textbf{2}$ \ \ & $\textbf{2}$ \ \ & $\textbf{2}$ \ \ & $\textbf{1}$ \ \ & $\textbf{1}$ \\
    $U(1)_Y $ & $ -\frac{1}{2}$ \ \ & $-1$ \ \ & $ -\frac{1}{2}$ \ \ & $\frac{1}{2}$ \ \ & $ -\frac{1}{2}$ \ \ & 0 \ \ & $-1$ \\
    $Z_2 $ & $+$ \ \ & $-$ \ \ & $ + $ \ \ & $-$ \ \ & $ + $ \ \ & + \ \ & $-$ \\ [1ex] 
 \hline    
\end{tabular}
    \caption{$SU(2)_L \times U(1)_Y \times Z_2$ quantum numbers of standard model leptons, Higgs doublets, and vector-like leptons. The electric charge generated after EWSB is $Q = T^3 + Y$. }
    \label{tab:Q_numbers}
\end{table}
Integrating out the heavy lepton fields reduces the above Lagrangian to
\begin{equation}
    \mathcal{L} \supset - y_{\mu} \bar{l}_L \mu_R H_d - C_{\mu H_d} \bar{l}_L \mu_R H_d  \left( H_d^{\dagger} H_d \right) + h.c., 
\end{equation}
where
\begin{equation}
    C_{\mu H_d} = \left( \frac{\lambda_L \lambda_E \bar{\lambda}}{M_L M_E }\right) \equiv \left( \frac{m_{\mu}^{LE}}{v_d^3} \right).
    \label{eq:cmuH}
\end{equation}
The parameter $m_{\mu}^{LE}$ would be the muon mass if $y_{\mu}$ were zero. The effective field theory Lagrangian additionally contains dimension-six operators such as $C_{\mu H_u} \bar{l}_L \mu_R H_d  \left( H_u^{\dagger} H_u \right)$; however, after EWSB, they vanish at tree level and do not contribute to the muon mass. In Sec.~\ref{sec:2hdm_eft} we explore non-zero contributions of these types of operators to the muon mass at loop level via quartic interactions in the scalar sector.

After EWSB, the physical muon mass $m_{\mu}$ is given by
\begin{equation}
    m_{\mu}e^{i \phi_{m_{\mu}}} \simeq y_{\mu} v_d + m_{\mu}^{LE},
\label{eq:mumass}
\end{equation}
and the muon Yukawa coupling is
\begin{equation}
    \lambda^h_{\mu \mu} \simeq y_{\mu} \cos \beta + 3 m_{\mu}^{LE} /v \simeq (m_{\mu}e^{i\phi_{m_{\mu}}} + 2 m_{\mu}^{LE}) / v. 
\end{equation}
Since $(\lambda_{\mu \mu}^h)_{SM} = m_{\mu} / v$, the decay rate of $h \rightarrow \mu^+ \mu^-$ compared to its SM rate is
\begin{equation}
\begin{split}
    R_{h \rightarrow \mu^+ \mu^-} & \equiv \frac{\textrm{BR}\left(h \rightarrow \mu^+ \mu^- \right)}{\textrm{BR}\left(h \rightarrow \mu^+ \mu^- \right)_{SM}} = \frac{|\lambda_{\mu \mu}^h|^2}{m_{\mu}^2/v^2} \\
    & = 1 + 4 \left( \frac{\textrm{Re} \left[m_{\mu}^{LE} e^{-i \phi_{m_{\mu}}} \right]}{m_{\mu} }\right) + 4 \left( \frac{\textrm{Re} \left[m_{\mu}^{LE} e^{-i \phi_{m_{\mu}}} \right]}{m_{\mu} }\right)^2 + 4 \left( \frac{\textrm{Im} \left[m_{\mu}^{LE} e^{-i \phi_{m_{\mu}}} \right]}{m_{\mu} }\right)^2, 
\end{split}
\label{eq:hmumu}
\end{equation}
which is currently limited by $R_{h \rightarrow \mu^+ \mu^-} \leq 2.2$ \cite{ATLAS:2020fzp}.

\subsection{Softly broken 2HDM potential}

The most general scalar potential of the two-Higgs doublets, $H_d$ and $H_u$, consistent with our conventions\footnote{In 2HDM models with a $Z_2$ symmetry, the doublets are usually defined as $\Phi_i = (\phi^{+}_i, v_i + \phi_i^0 / \sqrt{2})^T$, which transform as $\Phi_1 \rightarrow - \Phi_1$ and $\Phi_2 \rightarrow + \Phi_2$ \cite{Gunion:2002zf}. Matching to our definitions of $H_{d,u}$ requires rotating the fields via $\Phi_1 \rightarrow H_d$ and $\Phi_2 \rightarrow -i \sigma^2 H_u^{\dagger}$, obtaining the potential through $V(\Phi_1, \Phi_2) \rightarrow V(H_d, H_u)$.} while allowing soft $Z_2$-breaking terms is given by
\begin{equation}
\begin{split}
        V(H_d, H_u) & = m_1^2 \left( H_d^{\dagger} H_d \right) + m_2^2 \left( H_u^{\dagger} H_u \right) + m_{12}^2 \left(H_d^{\dagger} \cdot H_u^{\dagger} + H_d \cdot H_u \right) \\
        & + \frac{1}{2} \lambda_1 \left(H_d^{\dagger} H_d\right)^2 + \frac{1}{2} \lambda_2 \left(H_u^{\dagger} H_u \right)^2 + \lambda_3 \left(H_d^{\dagger} H_d \right) \left(H_u^{\dagger} H_u \right) \\ 
        & + \lambda_4 \left(H_d^{\dagger} \cdot H_u^{\dagger} \right)  \left(H_d \cdot H_u \right) + \frac{1}{2} \lambda_5 \left[ \left(H_d^{\dagger} \cdot H_u^{\dagger} \right)^2 + \left(H_d \cdot H_u \right)^2 \right],
\end{split}
\label{eq:potential}
\end{equation}
where the explicit $``\cdot"$ contracts $SU(2)$ doublets through the antisymmetric $\epsilon_{ij}$, e.g., $H_d \cdot H_u = \epsilon_{12} (H_d)_1 (H_u)_2 + \epsilon_{21} (H_d)_2 (H_u)_1 = H_d^+ H_u^- - H_d^0 H_u^0 $. The exact $Z_2$-symmetric potential is modified by the additional free parameter $m_{12}^2$ which is required to softly break the symmetry. For further details, see Appendix~\ref{sec:higgs_basis}. 

After diagonalization of the scalar fields to the mass eigenstate basis, the quartic couplings $\lambda_1, \lambda_2, \lambda_3, \lambda_4,$ and $\lambda_5$ can be written as 
\begin{equation}
    \begin{split}
        & \lambda_1 = \left( \frac{1}{2v_d^2} \right) \left( m_H^2 \cos^2 \alpha + m_h^2 \sin^2 \alpha - m_{12}^2 \tan \beta \right), \\
        & \lambda_2 = \left( \frac{1}{2v_u^2} \right) \left( m_H^2 \sin^2 \alpha + m_h^2 \cos^2 \alpha - \frac{m_{12}^2}{\tan \beta} \right), \\
        & \lambda_3 = \left( \frac{1}{2v^2} \right) \left(2m_{H^{\pm}}^2 + \left(\frac{\sin 2 \alpha}{\sin 2\beta}\right) (m_H^2 - m_h^2) - \frac{m_{12}^2}{\sin \beta \cos \beta} \right),\\
        & \lambda_4 = \left( \frac{1}{2v^2} \right) \left(m_A^2 - 2 m_{H^{\pm}}^2 + \frac{m_{12}^2}{\sin \beta \cos \beta} \right), \\
        & \lambda_5 = -\left(\frac{1}{2v^2} \right) \left(m_A^2 - \frac{m_{12}^2}{\sin \beta \cos \beta} \right). \\
    \end{split}
\label{eq:exact_couplings}
\end{equation}
If we consider $m_H^2 \simeq m_A^2 \simeq m_{H^{\pm}}^2 \gg m_h^2$, this regime also enforces that $\alpha \rightarrow \beta - \pi/2$ (i.e. the alignment limit), in which the light eigenstate $h$ behaves as the SM Higgs boson \cite{Gunion:2002zf}. In this limit, the above couplings highly simplify to
\begin{equation}
    \begin{split}
        & \lambda_1 \simeq  \left( \frac{1}{2v_d^2} \right) \left(m_{H,A,H^{\pm}}^2 \sin^2 \beta - m_{12}^2 \tan \beta \right), \\
        & \lambda_2 \simeq \left( \frac{1}{2v_u^2} \right) \left(m_{H,A,H^{\pm}}^2 \cos^2 \beta - \frac{m_{12}^2}{\tan \beta} \right), \\
        & \lambda_3 = - \lambda_4 = - \lambda_5 \simeq \left( \frac{1}{2v^2} \right) \left(m_{H,A,H^{\pm}}^2 - \frac{m_{12}^2}{\sin \beta \cos \beta} \right).
    \end{split}
\label{eq:q_couplings}
\end{equation}
Notice that these couplings are not independent from one another in the decoupling limit, but rather $\lambda_2 = \lambda_1 / \tan^4 \beta$ and $\lambda_3 = \lambda_1 / \tan^2 \beta$.
In addition, the stability conditions of Eq.~(\ref{eq:stab}) all reduce to the same lower bound:
\begin{equation}
    m_{H,A,H^{\pm}}^2 > \frac{m_{12}^2}{ \sin \beta \cos \beta}.
\end{equation}

Now, this range is further restricted by imposing perturbative unitarity constraints on the above couplings. The condition for unitarity is expressed in the form of partial-wave scattering from $W^{\pm}$ and $Z$ bosons in the high-energy limit, which are realized by the Goldstone boson equivalence theorem through their longitudinal (scalar) modes in $2 \rightarrow 2$ scattering \cite{Lee:1977eg,Lee:1977yc,Cornwall:1974km}. Applying this analysis to the softly broken 2HDM-II (see \cite{Ginzburg:2005dt,Kanemura:2015ska,Jurciukonis:2018skr} for details while using the condition for unitarity in \cite{Marciano:1989ns}), partial-wave unitarity and perturbativity require $|\lambda_{1, \cdots, 5}| \leq 4 \pi$.

Applying these unitarity constraints on each coupling, we find that the heavy Higgs mass is bounded above:
\begin{equation}
    \begin{split}
        & |\lambda_1|: \ m_{H,A,H^{\pm}}^2 \leq \frac{8 \pi v^2}{\tan^2 \beta} + \frac{m_{12}^2}{\sin \beta \cos \beta}, \\
        & |\lambda_2|: \ m_{H,A,H^{\pm}}^2 \leq 8 \pi v^2 \tan^2 \beta + \frac{m_{12}^2}{\sin \beta \cos \beta}, \\
        & |\lambda_{3,4,5}|: \ m_{H,A,H^{\pm}}^2 \leq 8 \pi v^2 + \frac{m_{12}^2}{\sin \beta \cos \beta}.
    \end{split}
\label{eq:ubound}
\end{equation}
For an exact $Z_2$ symmetry, these unitarity conditions lead to upper limits on each individual Higgs mass discussed in \cite{Maalampi:1991fb,Kanemura:1993hm}, similar to the Lee-Quigg-Thacker upper mass limit obtained in a single-Higgs-doublet model \cite{Lee:1977eg,Lee:1977yc}. However, with a softly broken $Z_2$ symmetry, $m_H^2 \simeq m_A^2 \simeq m_{H^{\pm}}^2$ can be arbitrarily large when appropriately choosing $m_{12}^2$ for a given $\tan \beta$. In the limit $m_{12}^2 \gg 8 \pi v^2$, the bounds in Eq.~(\ref{eq:ubound}) all become equal. For sufficiently heavy $m_H^2 \simeq m_A^2 \simeq m_{H^{\pm}}^2$, the entire range of $\tan \beta$ is allowed by direct searches \cite{ATLAS:2020zms,ATLAS:2021upq}.

\section{Contributions to the Muon Dipole Moments}
\label{sec:dipoles}
\begin{figure}[t]
\centering
\includegraphics[scale=1.1]{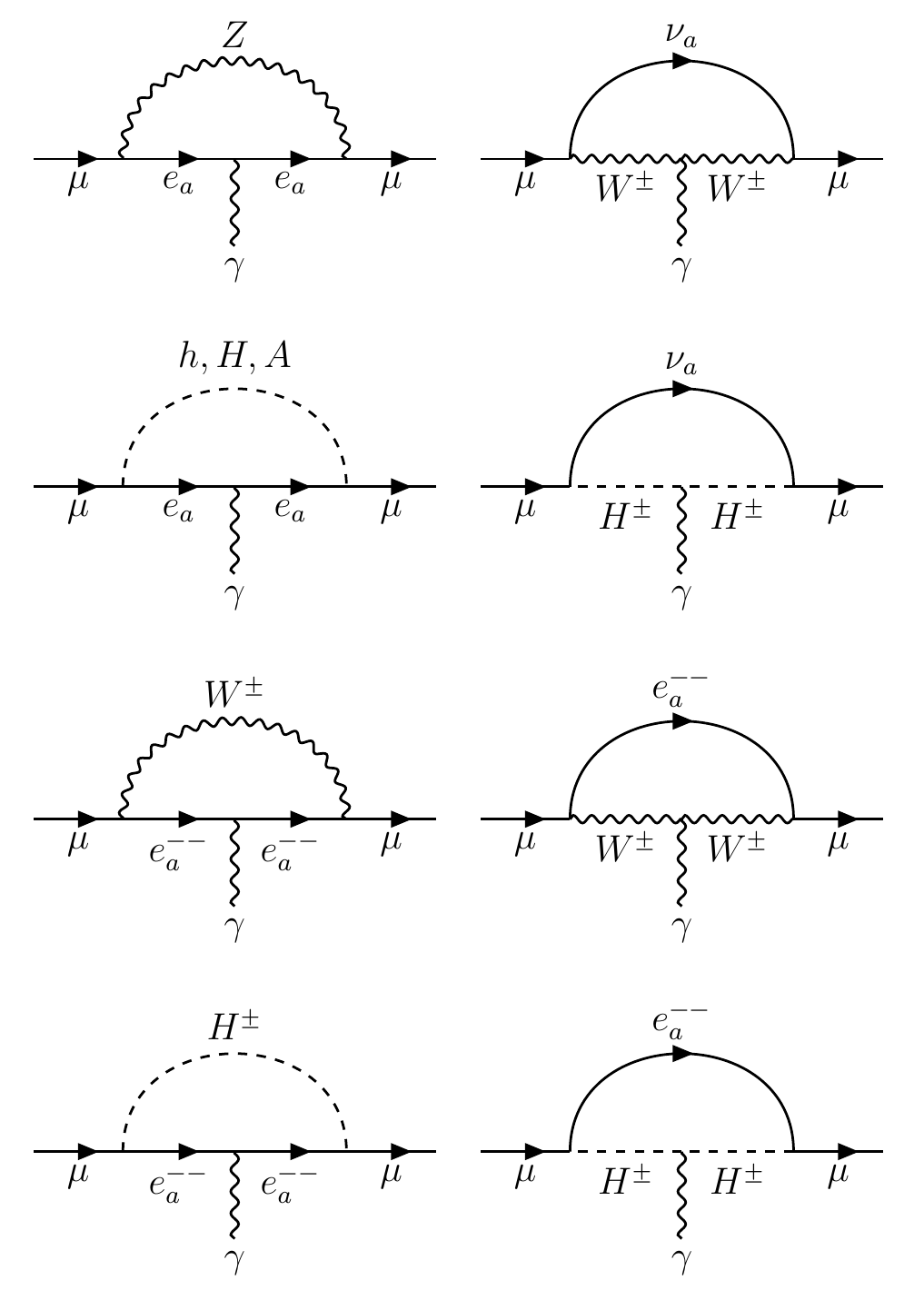}
\caption{Contributions to $\Delta a_{\mu}$ and $d_{\mu}$ from diagrams involving $Z$, $W^{\pm}$, and Higgs bosons with new fermion eigenstates. The bottom four diagrams are for representations involving doubly charged fermions.}
\label{fig:diags}
\end{figure}

We turn our attention to calculating the anomalous magnetic $(\Delta a_{\mu})$ and electric $(d_{\mu})$ dipole moments of the muon. They can be extracted via the effective Lagrangian (where we use $e > 0$ and the time-like metric)

\begin{equation}
    \mathcal{L} \supset \frac{1}{2} \left(\frac{e}{2 m_{\mu}} \right) \Delta a_{\mu} \bar{\mu} \sigma^{\mu \nu} \mu F_{\mu \nu} - \frac{i}{2} d_{\mu} \bar{\mu} \sigma^{\mu \nu} \gamma_5 \mu F_{\mu \nu}, 
\label{eq:dipole_lag}
\end{equation}
which are both calculated via Fig.~\ref{fig:diags}. The contributions to $(g-2)_{\mu}$ discussed in \cite{Dermisek:2021ajd} are not specific to any type of 2HDM and, in a similar manner, the contributions to $d_{\mu}$ presented here are general for any model with an extended Higgs sector. For completeness, we list contributions to $\Delta a_{\mu}$ in addition to $d_{\mu}$. Couplings to mass eigenstates are defined identically as therein, but for complex Lagrangian parameters studied here and are collected in Appendix~\ref{sec:meb_couplings}. Approximate formulas for couplings relevant to $(g-2)_{\mu}$ and $d_{\mu}$ are given in Appendix~\ref{sec:approx_form}.

In the mass eigenstate basis, charged or neutral lepton eigenstates $f_a$ couple to the $Z$ boson through
\begin{equation}
    \mathcal{L} \supset \left(\bar{f}_{L a} \gamma^{\mu} g_L^{Z f_a f_b} f_{L b} + \bar{f}_{R a} \gamma^{\mu} g_R^{Z f_a f_b} f_{R b} \right) Z_{\mu},
\end{equation}
and generate contributions to $\Delta a_{\mu}$ and $d_{\mu}$ as
\begin{equation}
    \begin{split}
        \Delta a_{\mu}^Z = \left( \frac{-m_{\mu}}{8 \pi^2 m_Z^2} \right) \sum_{a = 4, 5} & \left[ m_{\mu} \left( \left\lvert g_{R}^{Z \mu e_a} \right\rvert ^2 +  \left\lvert g_{L}^{Z \mu e_a} \right\rvert ^2 \right) F_Z(x_{Z}^{a}) \right. \\
        & \left. - m_{e_a} \ \textrm{Re} \left[ g_{R}^{Z \mu e_{a}} (g_{L}^{Z \mu e_{a}})^* \right] G_Z(x_{Z}^{a}) \right],
    \end{split}
\end{equation}
\begin{equation}
    d_{\mu}^Z = \left( \frac{e}{16 \pi^2 m_Z^2} \right) \sum_{a = 4,5} m_{e_a} \ \textrm{Im} \left[ g_{R}^{Z \mu e_{a}} (g_{L}^{Z \mu e_{a}})^* \right] G_{Z}(x_Z^a),
    \end{equation}
where $x_Z^a = m_{e_a}^2 / M_Z^2$ parametrizes the following loop functions $F_Z(x)$ and $G_Z(x)$:
\begin{equation}
    F_Z(x) = \frac{5x^4 - 14x^3 + 39x^2 - 38x + 8 - 18 x^2 \ \textrm{ln}(x)}{12 (1 - x)^4},
\end{equation}
 \begin{equation}
      G_Z(x) = - \ \frac{x^3 + 3x - 4 - 6x \ \textrm{ln}(x)}{2 (1 - x)^3}.
 \end{equation}
In a similar way, couplings of charged and neutral leptons to the $W^{\pm}$ boson are given by
\begin{equation}
    \mathcal{L} \supset \left( \bar{\hat{\nu}}_{L a} \gamma^{\mu} g_{L}^{W \nu_a e_b }
    \hat{e}_{L b} + \bar{\hat{\nu}}_{R a} \gamma^{\mu} g_{R}^{W \nu_a e_b } \hat{e}_{R b} \right) W_{\mu}^{+} + h.c. 
\end{equation}
The corresponding contributions to $\Delta a_{\mu}$ and $d_{\mu}$ are
\begin{equation}
    \begin{split}
        \Delta a_{\mu}^W = \left( \frac{m_{\mu}}{16 \pi^2 m_W^2} \right) \sum_{a = 4, 5} & \left[ m_{\mu} \left( \left\lvert g_R^{W \nu_a \mu} \right\rvert^2 + \left\lvert g_L^{W \nu_a \mu} \right\rvert^2 \right) F_W(x_{W}^{a}) \right. \\ 
        & \left. - m_{\nu_a} \ \textrm{Re} \ \left[ g_R^{W \nu_a \mu } (g_L^{W \nu_a \mu} )^* \right] G_W(x_{W}^{a}) \right],
    \end{split}
\end{equation}
\begin{equation}
    d_{\mu}^W = \left(\frac{e}{32 \pi^2 m_W^2} \right) \sum_{a = 4,5} m_{\nu_a} \ \textrm{Im} \left[  g_R^{W \nu_a \mu } (g_L^{W \nu_a \mu })^* \right] G_W(x_W^a),
\end{equation}
where $x_W^a = m_{\nu_a}^2 / M_W^2$ parametrizes the loop functions $F_W(x)$ and $G_W(x)$,
\begin{equation}
    F_W(x) = \frac{4x^4 - 49x^3 + 78x^2 - 43x + 10 + 18 x^3 \ \textrm{ln}(x)}{6(1 - x)^4},
 \end{equation}
 \begin{equation}
    G_W(x) = \frac{-x^3 + 12x^2 - 15x + 4 - 6x^2 \ \textrm{ln}(x)}{(1 - x)^3}.
 \end{equation}
For neutral Higgs scalars $\phi = h, H, A$, their couplings to charged leptons are defined as
\begin{equation}
    \mathcal{L} \supset - \frac{1}{\sqrt{2}}  \bar{\hat{e}}_{L a} \lambda_{e_a e_b}^{\phi} \hat{e}_{R b} \phi + h.c., 
\end{equation}
where their contributions to $\Delta a_{\mu}$ and $d_{\mu}$ are given by 
\begin{equation}
    \Delta a_{\mu}^{\phi} = \left( \frac{m_{\mu}}{32 \pi^2 m_{\phi}^2} \right) \sum_{a = 4, 5}  \left[ m_{\mu} \left( \left\lvert \lambda_{\mu e_a}^{\phi} \right\rvert^2 + \left\lvert \lambda_{e_a \mu}^{\phi} \right\rvert^2 \right) F_{\phi}(x_{\phi}^{a}) + m_{e_a}\ \textrm{Re} \left[ \lambda_{\mu e_a}^{\phi} \lambda_{e_a \mu}^{\phi} \right] G_{\phi}(x_{\phi}^{a}) \right],
\end{equation}
\begin{equation}
        d_{\mu}^{\phi} = - \left(\frac{e}{64 \pi^2 m_{\phi}^2} \right) \sum_{a = 4,5} m_{e_a} \ \textrm{Im} \left[ \lambda_{\mu e_a}^{\phi} \lambda_{e_a \mu}^{\phi} \right] G_{\phi}(x_{\phi}^{a}), 
\end{equation}
with $x_{\phi}^a = m_{e_a}^2 / m_{\phi}^2$ and 
\begin{equation}
     F_{\phi}(x) = \frac{x^3 - 6x^2 + 3x + 2 + 6x\ \textrm{ln} (x)}{6 (1 - x)^4},
\end{equation}
\begin{equation}
        G_{\phi}(x) = \frac{-x^2 + 4x - 3 - 2\ \textrm{ln}(x)}{(1 - x)^3}.
\end{equation}
The couplings that describe interactions between charged and neutral leptons to the charged Higgs boson $H^{\pm}$ are
\begin{equation}
    \mathcal{L} \supset -\bar{\hat{\nu}}_{L a} \lambda_{\nu_a e_b}^{H^{\pm}} \hat{e}_{R b} H^+ - \bar{\hat{e}}_{L a} \lambda_{e_a \nu_b}^{H^{\pm}} \hat{\nu}_{R b} H^- + h.c. \ .
\end{equation}
The contribution from the charged Higgs to $\Delta a_{\mu}$ and $d_{\mu}$ are
\begin{equation}
    \begin{split}
        \Delta a_{\mu}^{H^{\pm}} =  \left( \frac{-m_{\mu}}{16 \pi^2 m_{H^{\pm}}^2} \right)\sum_{a = 4, 5} & \left[ m_{\mu} \left( \left\lvert \lambda_{\nu_a \mu}^{H^{\pm}}\right\rvert^2 + \left\lvert \lambda_{\mu \nu_a}^{H^{\pm}}\right\rvert^2 \right) F_{H^{\pm}}(x_{H^{\pm}}^{a}) \right. \\
        & \left. + m_{\nu_a} \ \textrm{Re} \left[  \lambda_{\nu_a \mu}^{H^{\pm}} \lambda_{\mu \nu_a}^{H^{\pm}} \right] G_{H^{\pm}}(x_{H^{\pm}}^{a}) \right],
    \end{split}
\end{equation}
\begin{equation}
        d_{\mu}^{H^{\pm}} = \left(\frac{e}{32 \pi^2 m_{H^{\pm}}^2} \right) \sum_{a = 4,5} m_{\nu_a} \ \textrm{Im} \left[ \lambda_{\nu_a \mu}^{H^{\pm}} \lambda_{\mu \nu_a}^{H^{\pm}} \right] G_{H^{\pm}}(x_{H^{\pm}}^{a}),
\end{equation}
whereby $x_{H^{\pm}}^{a} = m_{\nu_a}^2 / m_{H^{\pm}}^2$ and 
\begin{equation}
    F_{H^{\pm}}(x) = \frac{2x^3 + 3x^2 - 6x + 1 - 6x^2 \ \textrm{ln}(x)}{6 (1 - x)^4},
\end{equation}
\begin{equation}
    G_{H^{\pm}}(x) = \frac{-x^2 + 1 + 2x \ \textrm{ln}(x)}{(1 - x)^3}.
\end{equation}

Models involving doubly charged fermions coupling to the $W^{\pm}$ boson and a singly charged fermion are described by
\begin{equation}
    \mathcal{L} \supset \left( \bar{\hat{e}}_{L a}^{--} \gamma^{\mu} g_{L}^{W e_b e_a^{--} }
    \hat{e}_{L b} + \bar{\hat{e}}_{R a}^{--} \gamma^{\mu} g_{R}^{W e_b e_a^{--} } \hat{e}_{R b}\right) W_{\mu}^{-} + h.c. 
\end{equation}
Contributions from the $W^{\pm}$ boson to $\Delta a_{\mu}$ and $d_{\mu}$ are given by
\begin{equation}
\begin{split}
    \Delta a_{\mu}^W = \left( \frac{m_{\mu}}{16 \pi^2 m_W^2} \right) \sum_{a = 4,5} & \left[ m_{\mu} \left( \left\lvert g_R^{W \mu e_a^{--}} \right\rvert^2 + \left\lvert g_L^{W \mu e_a^{--}} \right\rvert^2 \right) F'_W(x_{W}^{a--}) \right. \\
    & \left. - m_{e_a^{--}} \ \textrm{Re} \ \left[  g_R^{W \mu e_a^{--}} (g_L^{W \mu e_a^{--}})^* \right] G'_W(x_{W}^{a--}) \right],
\end{split}
\end{equation}
\begin{equation}
    d_{\mu}^W = \left(\frac{e}{32 \pi^2 m_W^2} \right) \sum_{a = 4,5} m_{e_a^{--}} \ \textrm{Im} \left[  g_R^{W \mu e_a^{--}} (g_L^{W \mu e_a^{--}})^* \right] G'_W(x_W^{a --}),
\end{equation}
where $x_W^{a --} = m_{e_a^{--}}^2 / M_W^2$ parametrizes the loop functions $F'_W(x)$ and $G'_W(x)$,
\begin{equation}
    F'_W(x) = - \left( F_W(x) + 4 F_Z(x) \right),
 \end{equation}
 \begin{equation}
    G'_W(x) = - \left( G_W(x) + 4 G_Z(x) \right),
 \end{equation}
as defined above. Finally, the couplings that describe interactions between singly and doubly charged fermions to the charged Higgs boson $H^{\pm}$ are
\begin{equation}
    \mathcal{L} \supset -\bar{\hat{e}}_{L a} \lambda_{e_a e_b^{--}}^{H^{\pm}} \hat{e}_{R b}^{--} H^+ - \bar{\hat{e}}_{L a}^{--} \lambda_{e_a^{--} e_b}^{H^{\pm}} \hat{e}_{R b} H^- + h.c. 
\end{equation}
The charged Higgs contributes to $\Delta a_{\mu}$ and $d_{\mu}$ by 
\begin{equation}
\begin{split}
    \Delta a_{\mu}^{H^{\pm}} =  \left( \frac{-m_{\mu}}{16 \pi^2 m_{H^{\pm}}^2} \right) \sum_{a = 4,5} & \left[ m_{\mu} \left( \left\lvert \lambda_{e_a^{--} \mu}^{H^{\pm}}\right\rvert^2 + \left\lvert \lambda_{\mu e_a^{--}}^{H^{\pm}}\right\rvert^2 \right) F'_{H^{\pm}}(x_{H^{\pm}}^{a--}) \right. \\
    & \left. + m_{e_a^{--}} \ \textrm{Re} \left[ \lambda_{e_a^{--} \mu}^{H^{\pm}} \lambda_{\mu e_a^{--}}^{H^{\pm}} \right] G'_{H^{\pm}}(x_{H^{\pm}}^{a--}) \right],
\end{split}
\end{equation}
\begin{equation}
        d_{\mu}^{H^{\pm}} = \left(\frac{e}{32 \pi^2 m_{H^{\pm}}^2} \right) \sum_{a = 4,5} m_{e_a ^{--}} \ \textrm{Im} \left[ \lambda_{e_a^{--} \mu}^{H^{\pm}} \lambda_{\mu e_a^{--}}^{H^{\pm}} \right] G'_{H^{\pm}}(x_{H^{\pm}}^{a --}),
\end{equation}
whereby $x_{H^{\pm}}^{a--} = m_{e_a^{--}}^2 / m_{H^{\pm}}^2$ and with the functions defined above,
\begin{equation}
    F'_{H^{\pm}}(x) = - \left(F_{H^{\pm}}(x) + 2 F_{\phi}(x) \right),
\end{equation}
\begin{equation}
    G'_{H^{\pm}}(x) =  - \left(G_{H^{\pm}}(x) + 2 G_{\phi}(x) \right).
\end{equation}

\section{2HDM-II Effective Field Theory}
\label{sec:2hdm_eft}
From Eqs.~(\ref{eq:hd0})-(\ref{eq:huc}), one can see that the SM degrees of freedom $h,G,G^{\pm}$ are mixed with the new scalars $H,A,H^{\pm}$ whose masses we consider as being anywhere between the EW scale and the masses of new leptons. It is more convenient to work in the Higgs basis \cite{Lavoura:1994fv,Lavoura:1994yu,Botella:1994cs,Davidson:2005cw}, where the SM and additional Higgs fields are separated into two doublets $H_1$ and $H_2$, respectively. Details of the Higgs basis are listed in Appendix~\ref{sec:higgs_basis} and will be used exclusively in the effective field theory calculations from now on. 

\begin{figure}[t]
\includegraphics[scale=0.8]{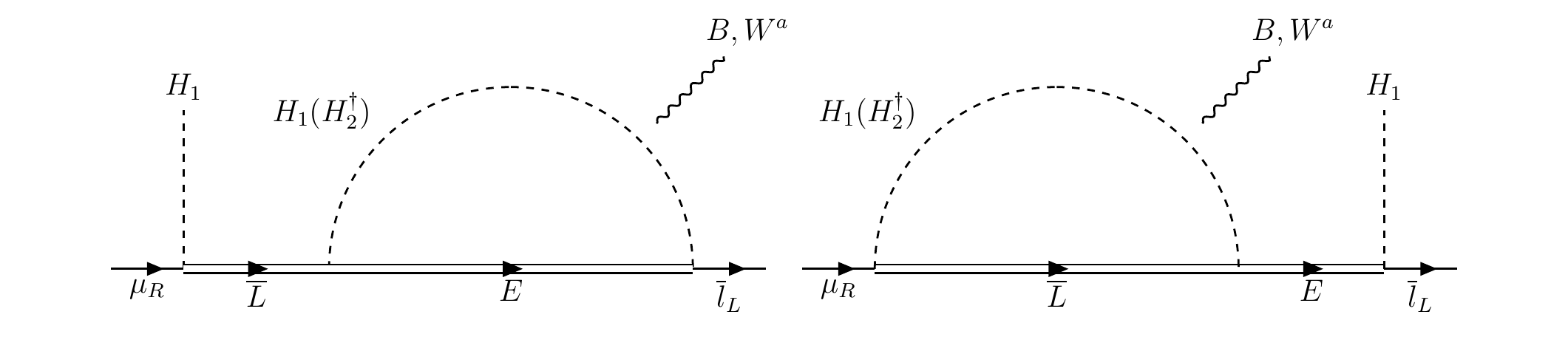}
\includegraphics[scale=0.8]{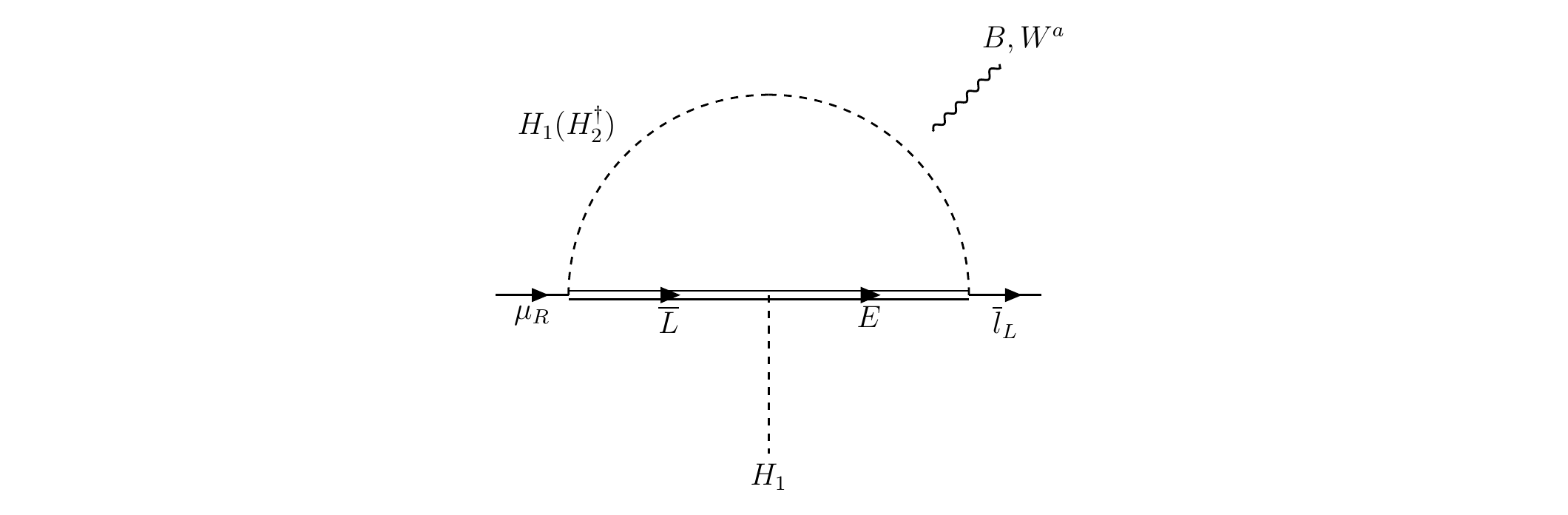}
\caption{Generic diagrams in the Higgs basis contributing to $C_{\mu B}$ and $C_{\mu W}$ containing heavy leptons coupling to the muon. Different diagrams are generated for different representations of vector-like leptons. The $B$ and $W^a$ fields are understood to attach to all possible locations in the loop.}
\label{fig:eft_diags}
\end{figure}

Dimension-six operators that modify the Higgs coupling to the muon as well as generate contributions to the muon's dipole moments via a chiral enhancement were studied extensively in \cite{Crivellin:2021rbq, Guedes:2022cfy, Dermisek:2022aec, Dermisek:2023nhe}. Models involving UV completions where two new fermions mix with left- and right-handed muon fields through the SM Higgs, which generate $C_{\mu H}$ at tree level at the matching scale, are called \textit{tree models}. Completions involving either a single scalar and two fermions (fermion-fermion-scalar [FFS]) or two scalars and one fermion (scalar-scalar-fermion [SSF]), where $C_{\mu H}$ is generated at one loop, are referred to as \textit{loop models}. Other kinds of models include \textit{bridge models}, where UV completions generate $C_{\mu H}$ at one loop purely from new leptons; however, a tree-level contribution is also present, proportional to the muon Yukawa coupling. In this paper, the 2HDM-II extended with vector-like leptons is a type of tree model that exhibits this same behavior, whose relevant Lagrangian is
\begin{equation}
    \begin{split}
    \mathcal{L} \supset & - y_{\mu} \overline{l}_L \mu_R H_d - C_{\mu H_d} \overline{l}_L \mu_R H_d (H_d^{\dagger} H_d) \\
    & - C_{\mu H_u}^{(1)} \overline{l}_L \mu_R H_d (H_u^{\dagger} H_u) - C_{\mu H_u}^{(2)} \overline{l}_L \mu_R \cdot H_u^{\dagger} (H_d \cdot H_u) - C_{\mu H_u}^{(3)} \overline{l}_L \mu_R \cdot H_u^{\dagger} (H_d^{\dagger} \cdot H_u^{\dagger}) \\
    & - C_{\mu B} \overline{l}_L\sigma^{\mu \nu} \mu_R H_d B_{\mu \nu} - C_{\mu W} \overline{l}_L \sigma^{\mu \nu} \mu_R \tau^{a} H_d W^a_{\mu \nu} + h.c..
    \end{split}
\end{equation}
The mass operators with $C^{(1,2,3)}_{\mu H_u}$ are generated at one loop for all models we consider. In the Higgs basis, the relevant part of the Lagrangian above containing the light doublet $H_1$ is
\begin{equation}
    \begin{split}
    \mathcal{L} \supset & - y_{\mu} \overline{l}_L \mu_R H_1 \cos \beta - C_{\mu H_1} \overline{l}_L \mu_R H_1 (H_1^{\dagger} H_1) \cos^3 \beta \\
    & - C_{\mu B} \overline{l}_L\sigma^{\mu \nu} \mu_R H_1 B_{\mu \nu} \cos \beta - C_{\mu W} \overline{l}_L \sigma^{\mu \nu} \mu_R \tau^{a} H_1 W^a_{\mu \nu} \cos \beta + h.c.,
    \end{split}
\end{equation}
where $C_{\mu H_1} = C_{\mu H_d} + (C_{\mu H_u}^{(1)} + C_{\mu H_u}^{(2)} + C_{\mu H_u}^{(3)}) \tan^2 \beta$, including loop corrections. Note that there are additional operators involving $H_2$; however, they do not contribute to the mass or dipole moments. After EWSB, the Wilson coefficients $C_{\mu B}$ and $C_{\mu W}$ combine to generate the muon's dipole moments in Eq.~(\ref{eq:dipole_lag}) and we find
\begin{subequations}
    \begin{equation}
        \Delta a_{\mu} = - \left(\frac{4 m_{\mu} v_d}{e} \right) \textrm{Re} \left[C_{\mu \gamma} e^{-i \phi_{m_{\mu}}}\right],
    \end{equation}
    \begin{equation}
        d_{\mu} = 2 v_d \textrm{Im} \left[C_{\mu \gamma} e^{-i \phi_{m_{\mu}}} \right],
    \end{equation}
\label{eq:dipoles_op}
\end{subequations}
where $C_{\mu \gamma} = \cos \theta_W C_{\mu B} - \sin \theta_W C_{\mu W}$ and $\theta_W$ is the weak mixing angle. We now present the contribution to $\Delta a_{\mu}$ and $d_{\mu}$ for the five representations of vector-like leptons listed in \cite{Kannike:2011ng}, calculated in the unbroken $SU(2)_L \times U(1)_Y$ theory from diagrams in Fig.~\ref{fig:eft_diags}. Allowing for only down-type couplings in the limit $M_{L,E} \gg v_d \times (\lambda_L, \lambda_E, \lambda, \overline{\lambda})$, $\Delta a_{\mu}$ and $d_{\mu}$ become
\begin{subequations}
    \begin{equation}
    \Delta a_{\mu} \simeq - \frac{1}{16 \pi^2} \left(\frac{m_{\mu}}{v^2} \right) \textrm{Re} \left[m_{\mu}^{LE} e^{-i \phi_{m_{\mu}}} \right] \left(\mathcal{Q}_1 +  \mathcal{Q}_2 \tan^2 \beta \right),
    \end{equation}
    \begin{equation}
    d_{\mu} \simeq \frac{1}{32 \pi^2} \left(\frac{e}{v^2} \right) \textrm{Im} \left[m_{\mu}^{LE} e^{-i \phi_{m_{\mu}}} \right] \left(\mathcal{Q}_1 +  \mathcal{Q}_2 \tan^2 \beta \right).
\end{equation} 
\label{eq:dipoles}
\end{subequations}
\begin{table}[t]
\centering
 \begin{tabular}{||c c c||} 
 \hline
 $L \oplus E$ & \ \ \  $\lambda^* = 0$ & $\lambda^* = \bar{\lambda}$ \\ [1.0ex] 
 \hline
 $\mathbf{2}_{-1/2} \oplus \mathbf{1}_{-1}$ & \ \ \ \ $ 1 + \tan^2 \beta$ & \ \ \ \ $ 1 + \tan^2 \beta$ \\
 $\mathbf{2}_{-1/2} \oplus \mathbf{3}_{-1}$ & \ \ \ \ $9 + 5 \tan^2 \beta$ & \ \ \ \ $9 + 5 \tan^2 \beta$ \\
 $\mathbf{2}_{-3/2} \oplus \mathbf{1}_{-1}$ & \ \ \ \ $5 + \frac{17}{6} \tan^2 \beta $ & \ \ \ \ $5 + 3 \tan^2 \beta$ \\
 $\mathbf{2}_{-3/2} \oplus \mathbf{3}_{-1}$ & \ \ \ \ $ 5 + \frac{11}{6} \tan^2 \beta$ & \ \ \ \ $5 + 3 \tan^2 \beta$ \\
 $\mathbf{2}_{-1/2} \oplus \mathbf{3}_{0}$ & \ \ \ \ $1 + \frac{11}{12} \tan^2 \beta$ & \ \ \ \ $1 + \tan^2 \beta$ \\ [1ex] 
 \hline
 \end{tabular}
 \caption{Summary of $\mathcal{Q}_1 + \mathcal{Q}_2 \tan^2 \beta$ factors for representations of $L \oplus E$ fields entering $\Delta a_{\mu}$ and $d_{\mu}$, evaluated in the decoupling limit and assuming one scale of new physics, $x_M^{(1)} \rightarrow \infty$ and $x_M^{(2)} \rightarrow 1$, with $\lambda^* = 0$ (second column) and $\lambda^* = \bar{\lambda}$ (third column). The $Z_2$ charge assignments for $L$ and $E$ are as they are in Table~\ref{tab:Q_numbers}.}
\label{tab:2hdmQ}
\end{table}
Notice how each representation experiences a $\tan^2 \beta$ enhancement in addition to the standard chiral enhancement. The factors $\mathcal{Q}_1 = \mathcal{Q}_1(x_{L,E}^{(1)})$ and $\mathcal{Q}_2 = \mathcal{Q}_2(x_{L,E}^{(2)})$ represent contributions from light (SM) and new scalar fields, respectively, parametrized by $x_{L,E}^{(1,2)} = M_{L,E}^2 / M_{1,2}^2$ for arbitrary masses of new leptons in each representation. Depending on the model and how the diagram is closed in Fig.~\ref{fig:eft_diags}, contributions may also involve the $\lambda$ coupling. To obtain Eq.~(\ref{eq:dipoles}) in a simplified way, we distinguish situations where $\lambda^* = 0$ and $\lambda^* = \bar{\lambda}$ in the following text, while expressions containing the full mass dependence on $\mathcal{Q}_{1,2}$ with arbitrary couplings are presented in Appendix~\ref{sec:tree_contributions}. 

If we consider the decoupling limit where $m_H^2 \simeq m_A^2 \simeq m_{H^{\pm}}^2 \gg m_h^2$, then the mass parameter $M_2^2$ in the Higgs basis becomes $M_2^2 = m_{H,A,H^{\pm}}^2 - 2 \lambda_1 v^2 / \tan^2 \beta$. We find that $M_2 \simeq m_{H,A,H^{\pm}} \gg m_h$ can be easily achieved for masses $m_{H,A,H^{\pm}}$ and $M_{L,E} \gtrsim 3$ TeV and couplings up to the perturbativity limit.\footnote{This regime is a valid range for explaining $\Delta a_{\mu}$ for couplings $\gtrsim 0.5$. See \cite{Dermisek:2021ajd} for details.} If there is one scale of new physics such that $M \equiv M_{L,E} \simeq M_2$, then we may take $x_M^{(2)} \rightarrow 1$ in $\mathcal{Q}_2$. In this limit, the SM fields are significantly lighter compared to the vector-like leptons and $\mathcal{Q}_1$ is well approximated when $x_M^{(1)} \rightarrow \infty$. The obtained expressions for $(\mathcal{Q}_1 + \mathcal{Q}_2 \tan^2 \beta)$ agree with the leading-order contributions of $\Delta a_{\mu}$ and $d_{\mu}$ calculated in the mass eigenstate basis in each representation\footnote{Refer to Sec.~\ref{sec:dipoles} using approximate diagonalization matrices and couplings in the mass eigenstate basis given in Appendix~\ref{sec:approx_form}, expanding to $\sim \mathcal{O}(v^2 / M_{L,E}^2)$.} and are collected in Table~\ref{tab:2hdmQ}.

Contributions involving the $\bar{\lambda}$ coupling arise from diagrams where each heavy fermion propagator contains its mass insertion. However, diagrams where each propagator contains its momentum term are responsible for generating contributions from $\lambda$. In the latter situation, diagrams involving $\lambda$ in the top row of Fig.~\ref{fig:eft_diags} vanish due to on-shell equations of motion for the spinors, whereas the bottom diagram remains nonzero when integrating over loop momentum entering the numerator. This happens for $\textbf{2}_{-3/2} \oplus \textbf{1}_{-1}, \textbf{2}_{-3/2} \oplus \textbf{3}_{-1},$ and $\textbf{2}_{-1/2} \oplus \textbf{3}_{0} $ representations. Additionally, note that only the $\tan^2 \beta$ term is affected by $\lambda$; contributions from the light Higgs doublet precisely cancel in the limit where the masses of new leptons are much heavier than the EW scale. However, contributions from the now-heavy doublet $H_2$ remain nonzero when the masses of leptons are comparable to $M_2$.

When we consider the other limit, $m_h \lesssim M_2 \ll M$, all terms with $\lambda$ vanish and $\mathcal{Q}_2 \rightarrow \mathcal{Q}_1 \equiv \mathcal{Q}$. For all representations given in Table~\ref{tab:2hdmQ}, $\mathcal{Q}$ reduces to 1, 5, or 9. Note that the heavy Higgses are not required to be near the EW scale; this is already a good approximation for $M_2 \simeq M / 4$. 

Scenarios involving the mixing between new heavy leptons and the muon via only the SM Higgs were first studied in the mass eigenstate basis in \cite{Dermisek:2013gta,Kannike:2011ng,Freitas:2014pua} and revisited in \cite{Guedes:2022cfy, Dermisek:2023nhe} in the SMEFT landscape. To provide insight into these scenarios, although unphysical, one can take the limit $\tan \beta \rightarrow 0$ in Eq.~(\ref{eq:dipoles}) which isolates loop contributions only from the SM doublet $H_1$. We find agreement with results in \cite{Guedes:2022cfy} in every representation. However, we disagree with \cite{Kannike:2011ng}, notably, in representations including doubly charged vector-like fields.
\begin{figure}[t]
\includegraphics[scale=0.9]{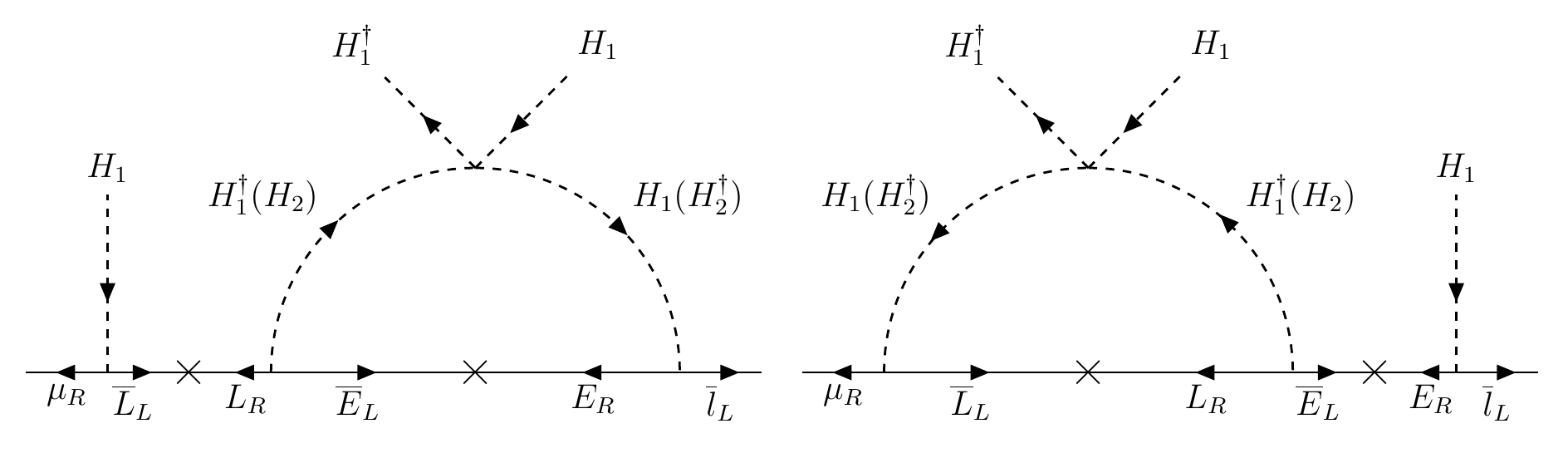}
\includegraphics[scale=0.8]{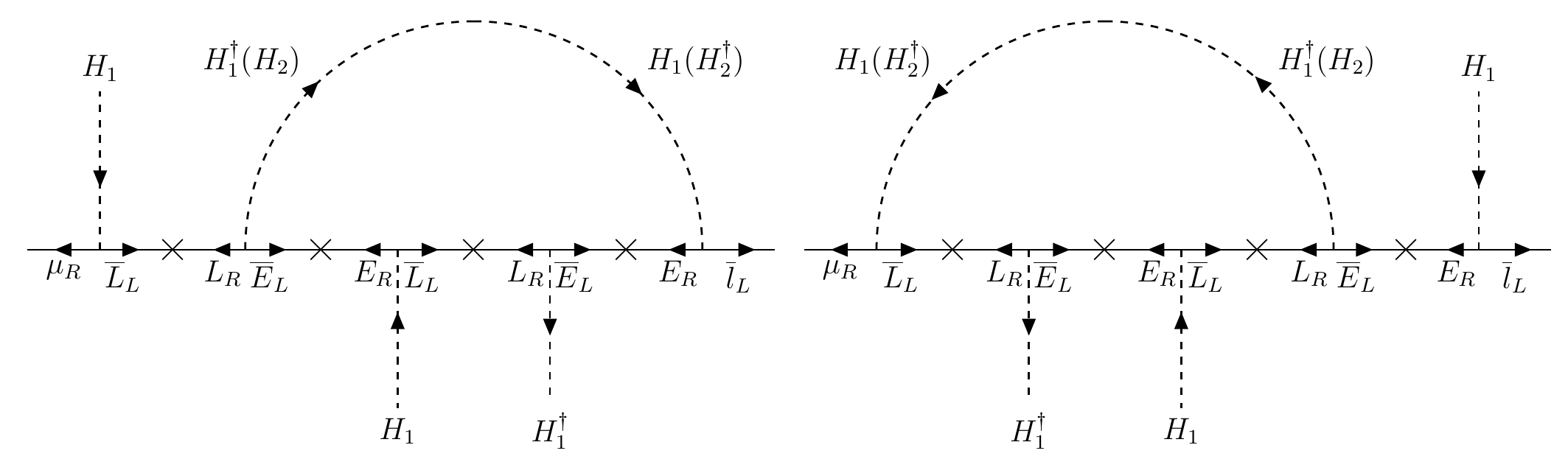}
\includegraphics[scale=0.8]{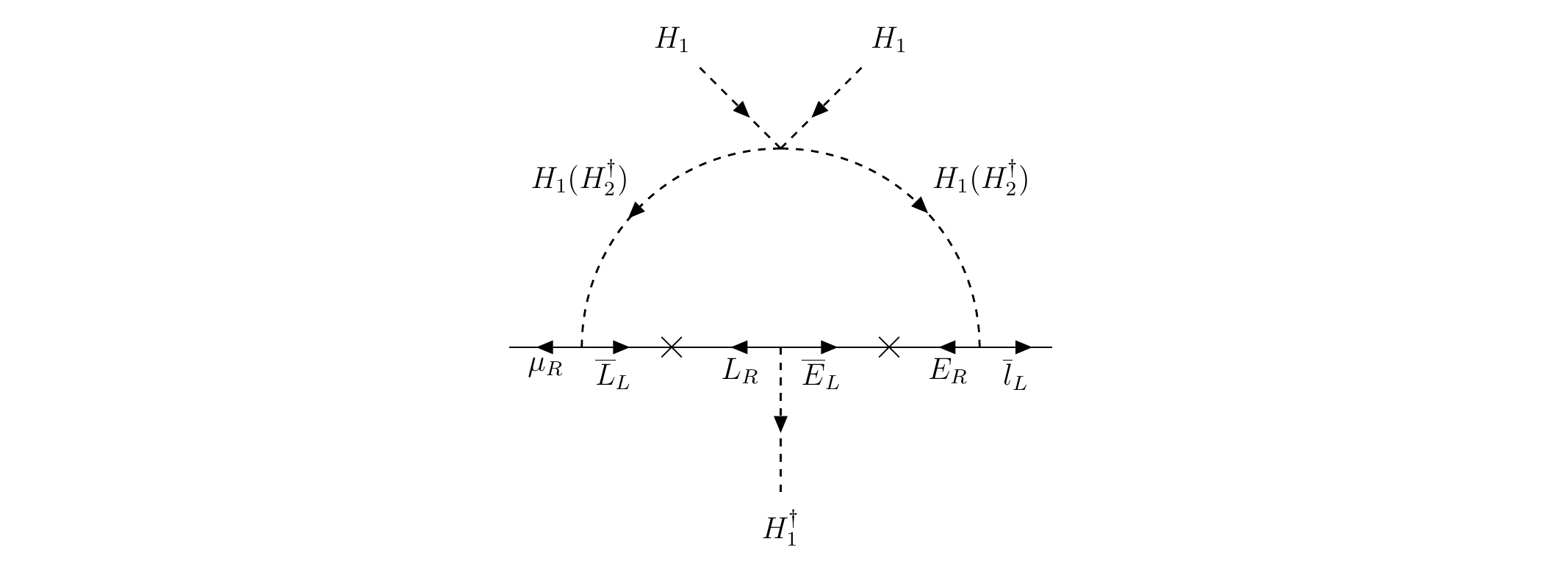}\caption{Diagrams contributing to the Wilson coefficient $C_{\mu H_1}$ at 1-loop in the $\mathbf{2}_{-1/2} \oplus \mathbf{1}_{-1}$ representation. Arrows denote chiral flow. Note that there are additional diagrams similar to the second and third row where two mass insertions are replaced with momentum lines in the loop.}
\label{fig:scalar_diags}
\end{figure}

The quartic couplings $\lambda_1, \lambda_2, \lambda_3, \lambda_4,$ and $\lambda_5$ in the scalar potential are relevant contributions at the one-loop level to the Wilson coefficient $C_{\mu H_1}$. In principle, the combination of $\tan \beta$, $M_2$, and large couplings
can generate sufficiently large contributions to overcome loop-suppressed effects, competing with the tree-level contribution $C_{\mu H_1}^{(tree)} = m_{\mu}^{LE} / v_d^3$. Turning our attention to the main representation of the paper, the contribution to $C_{\mu H_1}$ assuming only down-type couplings and the same common mass $M_{L,E} = M$ for new leptons up to one loop in Fig.~\ref{fig:scalar_diags} is
\begin{equation} 
    \begin{split}
        C_{\mu H_1} & = \left( \frac{m_{\mu}^{LE}}{v_d^3}\right) \left[1 - \left(\frac{3 \Lambda_3}{16 \pi^2} \right) K(x_M^{(2)}) \tan^2 \beta + \left(\frac{\lambda \overline{\lambda}}{16 \pi^2} \right) \left(L(x_M^{(1)}) \cos^2 \beta + L(x_M^{(2)}) \sin^2 \beta \right) \right. \\
        & \left. \ \ \ \ \ \ \ \ \ \ \ \ \ \ \  - \left(\frac{1}{8 \pi^2} \right) (|\lambda|^2 + (\lambda \overline{\lambda})^* + |\overline{\lambda}|^2) \left(N(x_M^{(1)}) \cos^2 \beta + N(x_M^{(2)}) \sin^2 \beta \right) \right],
    \end{split}
\label{eq:cmuH_1loop}
\end{equation}
where $x_M^{(1,2)}$ parametrizes the loop functions
\begin{equation}
    K(x) = \frac{-x^2 + x + x^2 \ \textrm{ln} (x)}{(1-x)^2},
\end{equation}
\begin{equation}
    L(x) = \frac{-x^3 + x + 2 x^2 \ \textrm{ln}(x)}{(1-x)^3},
\end{equation}
\begin{equation}
    N(x) = - \frac{x^3 - 4 x^2 + 3x + 2x \ \textrm{ln}(x)}{2(1-x)^3}.
\end{equation}
If we assume once more that $x_M^{(1)} \rightarrow \infty$ while using $\Lambda_3 = 2 \lambda_1 / \tan^2 \beta$ in Appendix~\ref{sec:higgs_basis}, the muon mass is additionally modified by the above Wilson coefficient:
\begin{equation}
    \begin{split}
        m_{\mu} e^{i \phi_{m_{\mu}}} \simeq y_{\mu} v_d & + m_{\mu}^{LE} \left[1 - \left(\frac{3\lambda_1}{8 \pi^2} \right) K(x_M^{(2)}) + \left(\frac{\lambda \overline{\lambda}}{16 \pi^2} \right) \left( \cos^2 \beta + L(x_M^{(2)}) \sin^2 \beta \right) \right. \\
        & \left. - \left(\frac{1}{16 \pi^2} \right) (|\lambda|^2 + (\lambda \overline{\lambda})^* + |\overline{\lambda}|^2) \left(\cos^2 \beta + 2 N(x_M^{(2)}) \sin^2 \beta \right) \right].
    \end{split}
\end{equation}

In principle, for the representations listed in \cite{Kannike:2011ng}, one can construct similar diagrams to Fig.~\ref{fig:scalar_diags} and calculate its full one-loop correction to the muon mass from the same dimension-six operators. However, we expect their contributions to be of similar order in all cases and it suffices to present results in the $\textbf{2}_{-1/2} \oplus \textbf{1}_{-1}$ representation. We will discuss the impact of these loop effects in the following section.

\section{Results}
\label{sec:results}
The current experimental world average of the muon's anomalous magnetic moment deviates from the SM prediction by 5.1 $\sigma$ \cite{Muong-2:2023cdq,Aoyama:2020ynm}, 
\begin{equation}
    \Delta a_{\mu} \equiv (2.49 \pm 0.48) \times 10^{-9}, 
\end{equation} 
and the current upper bound on the muon's electric dipole moment is \cite{Muong-2:2008ebm}
\begin{equation}
    |d_{\mu}| \leq 1.9 \times 10^{-20} e \cdot \textrm{cm}. 
\end{equation}
The Fermilab Muon $g-2$ Collaboration estimates an improvement in $d_{\mu}$ up to a level of \cite{Lukicov:2019ibv} 
\begin{equation}
    |d_{\mu}| < 10^{-21} e \cdot \textrm{cm}, 
\end{equation}
whereas the Paul Scherrer Institute (PSI) intends to host a new experiment involving the frozen-spin technique \cite{Adelmann:2021udj}, and projects a measurement of
\begin{equation}
        |d_{\mu}| < 6 \times 10^{-23} e \cdot \textrm{cm}. 
\end{equation}

Precision electroweak measurements of $Z$ and $W^{\pm}$ bosons constrain possible modifications of couplings to the muon at the $0.1 \%$ level which, in the limit of small mixing, translate to the following bounds on $\lambda_L$ and $\lambda_E$ \cite{Dermisek:2021ajd} specifically in the $\textbf{2}_{-1/2} \oplus \textbf{1}_{-1}$ model:
\begin{equation}
    \Big| \frac{\lambda_L v_d}{M_L} \Big| \lesssim 0.04, \ \ \ \ \ \Big| \frac{\lambda_E v_d}{M_E} \Big| \lesssim 0.03,
\label{eq:ew_cons}
\end{equation}
respectively, at $95 \%$ C.L. Additionally, we impose a bound on the mass parameters $M_L$ and $M_E$, such that $M_L > 800$ GeV and $M_E > 200$ GeV to generically satisfy constraints placed by searches for new leptons \cite{CMS:2019hsm,ATLAS:2022yhd,ATLAS:2015qoy}. 

\subsection{2HDM-II ellipse of dipole moments}

It was first pointed out in \cite{Dermisek:2022aec,Dermisek:2023nhe} that the connection between the three observables $h\rightarrow \mu^+ \mu^-, \ \Delta a_{\mu}, $ and $d_{\mu}$ is a byproduct of a correlation between the Wilson coefficients that modify the muon mass $(C_{\mu H_1})$ and generate the muon dipole moments $(C_{\mu \gamma})$ through a real model-dependent factor $k$. In the 2HDM-II, this relation is
\begin{equation}
    C_{\mu H_1} = \frac{k}{e \cos^2 \beta} \ C_{\mu \gamma}.
\end{equation}
This connection parametrizes the new physics contributions to the dipole moments and modification of $h \rightarrow \mu^+ \mu^-$ in a given model on an ellipse defined by
\begin{equation}
    R_{h \rightarrow \mu^+ \mu^-} = \left(\frac{k v^2 \Delta a_{\mu}}{2m_{\mu}^2} - 1 \right)^2 + \left(\frac{k v^2 d_{\mu}}{e m_{\mu}} \right)^2.
\label{eq:ellipse}
\end{equation}
This is derived from rewriting the definition of $R_{h \rightarrow \mu^+ \mu^-}$ in Eq.~(\ref{eq:hmumu}) in terms of the moments in Eq.~(\ref{eq:dipoles_op}). The $k$ factor is defined for a class of models in which $C_{\mu H_1}$ is generated at either tree or loop level~\cite{Dermisek:2022aec,Dermisek:2023nhe}. Although $k$ may include radiative corrections, we will not consider their effects until Sec.~\ref{sec:kloop}.

From Eq.~(\ref{eq:dipoles}), the $k$ factor for the 2HDM-II extended with vector-like leptons is
\begin{equation}
    k = \frac{64 \pi^2}{\left(\mathcal{Q}_1 + \mathcal{Q}_2 \tan^2 \beta \right)},
\label{eq:k_tree}
\end{equation}
which is generalized for different representations as summarized in Table~\ref{tab:2hdmQ}. As follows from Eq.~(\ref{eq:ellipse}) as well as mentioned in \cite{Dermisek:2022aec}, only models with $k \lesssim 19$ or $483 \lesssim k \lesssim 750$ are consistent with $\Delta a_{\mu}$ within $1 \sigma$ and $|d_{\mu}| = 0$ assuming $R_{h \rightarrow \mu^+ \mu^-} = 1 \pm 10\%$. In other words, models with $ 1.31 \lesssim (\mathcal{Q}_1 + \mathcal{Q}_2 \tan^2 \beta) \lesssim 33$ or $ (\mathcal{Q}_1 + \mathcal{Q}_2 \tan^2 \beta) \lesssim 0.84$ \textit{necessarily} require a nonzero value for $|d_{\mu}|$ with the expected precision level of $R_{h \rightarrow \mu^+ \mu^-}$ at the LHC. Further improvement of the precision of $R_{h \rightarrow \mu^+ \mu^-}$ to $1\%$ would extend these ranges to $1.28 \lesssim (\mathcal{Q}_1 + \mathcal{Q}_2 \tan^2 \beta) \lesssim 427$ or $(\mathcal{Q}_1 + \mathcal{Q}_2 \tan^2 \beta) \lesssim 0.86$. 
\begin{figure}[t]
\centering
\includegraphics[scale=0.5]{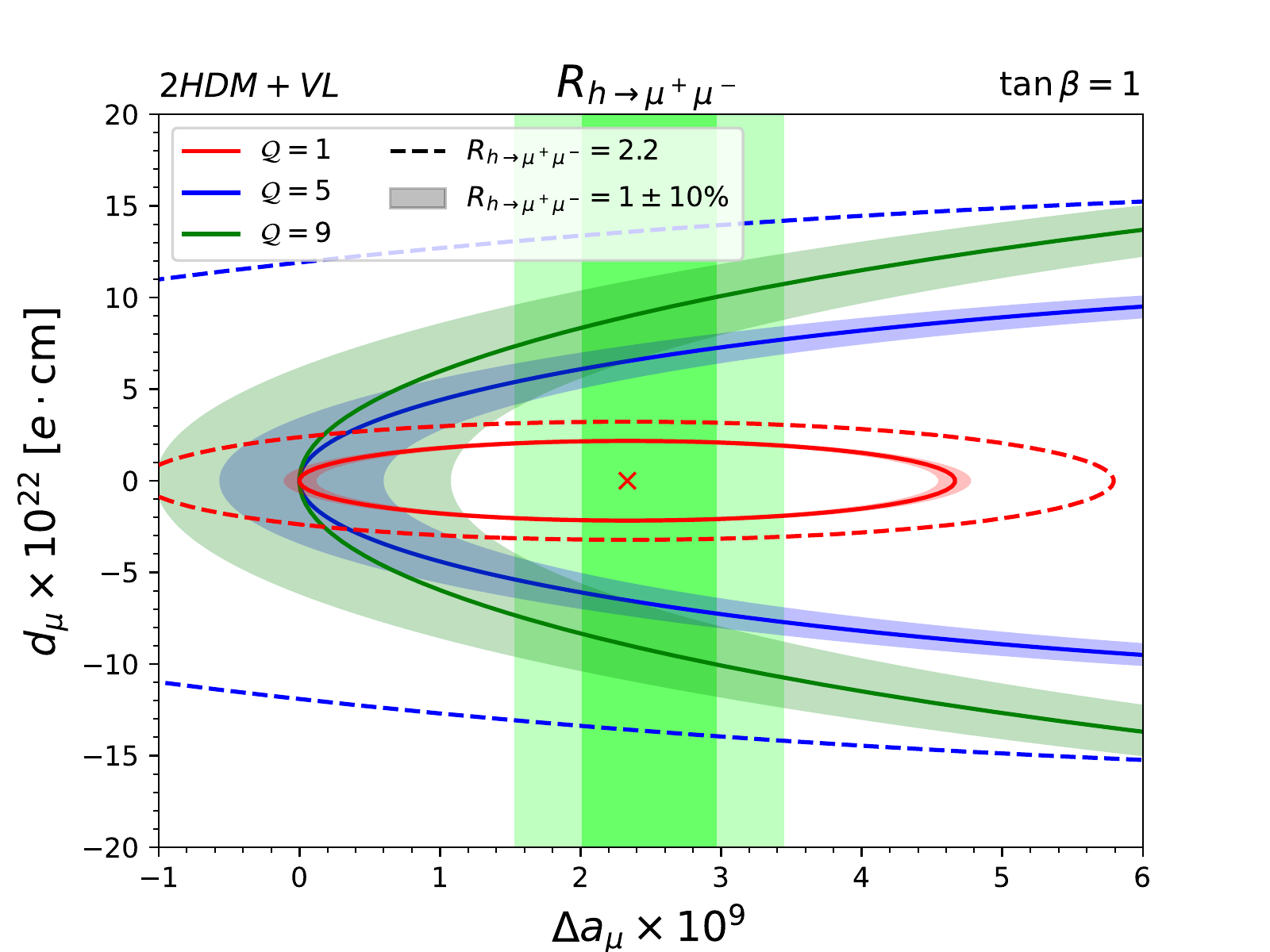}
\includegraphics[scale=0.5]{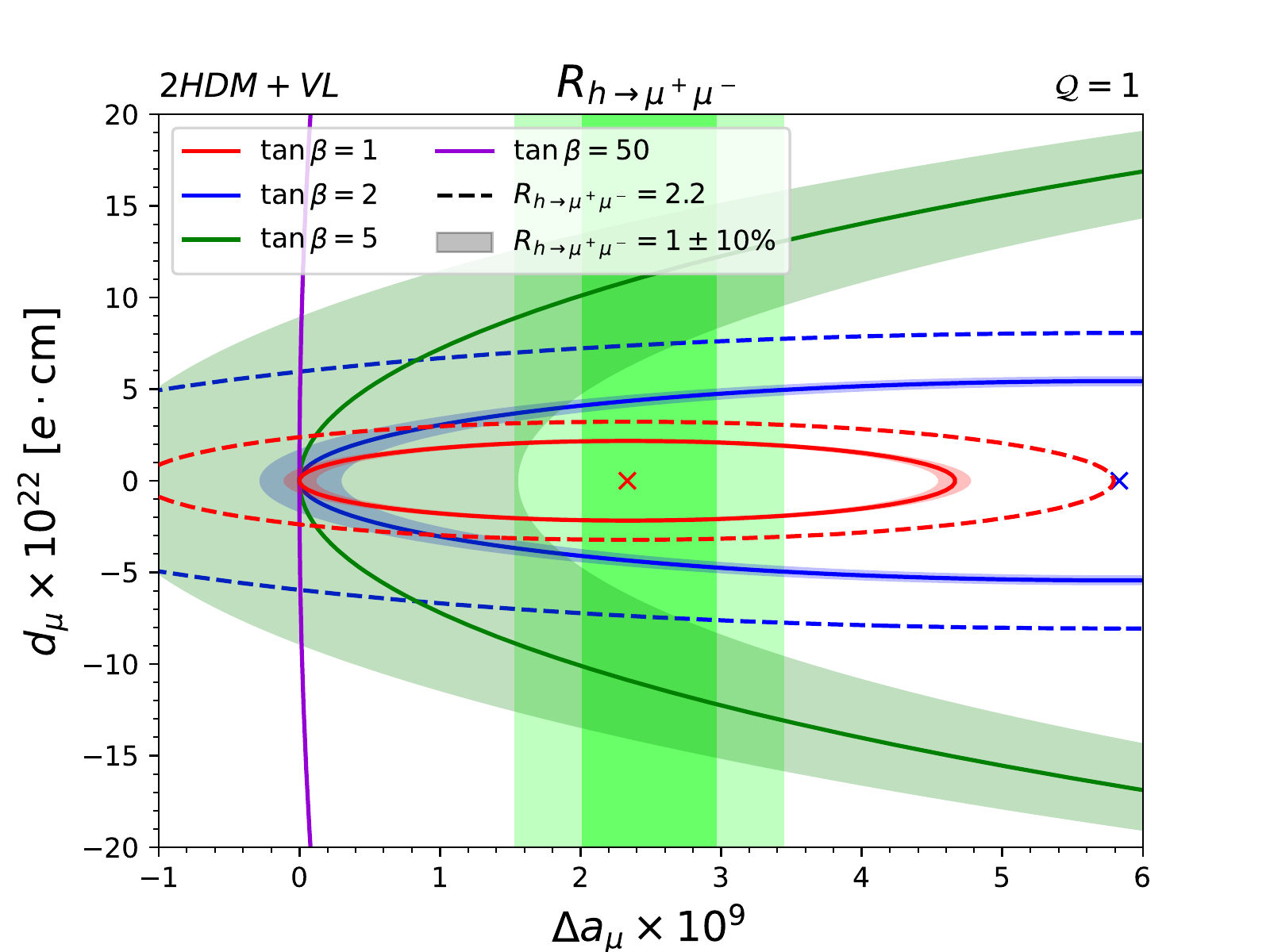}
\caption{Contours of $R_{h \rightarrow \mu^+ \mu^-}$ in the $\Delta a_{\mu}-d_{\mu}$ plane for $\tan \beta = 1$ when $\mathcal{Q} = 1,5,$ and $9$ (left) and $\mathcal{Q} = 1$ when $\tan \beta = 1, 2, 5$ and 50 (right). Solid lines correspond to the SM prediction, $R_{h \rightarrow \mu^+ \mu^-} = 1$, while the shaded region extends to $\pm 10\%$. Dashed lines correspond to the current upper limit on $R_{h \rightarrow \mu^+ \mu^-} = 2.2$. A colored $\times$ corresponds to the center of its ellipse. Dark and light green regions correspond to the $\pm 1 \sigma$ and $\pm 2 \sigma$ regions of $\Delta a_{\mu}$, respectively.}
\label{fig:contour_R_tb_c}
\end{figure}

When heavy Higgses are significantly lighter than vector-like leptons, we saw in the previous section that $\mathcal{Q}_{1} = \mathcal{Q}_{2} \equiv \mathcal{Q} = 1, 5,$ or 9 for all five representations. In this limit, the $k$ factors become $k = 64 \pi^2 / \mathcal{Q}(1+\tan^2 \beta)$. In Fig.~\ref{fig:contour_R_tb_c}, we see that as $\mathcal{Q}$ increases for fixed $\tan \beta$, the center of the ellipse shifts to larger vales of $\Delta a_{\mu}$ and $d_{\mu}$. The experimental value of $\Delta a_{\mu}$ restricts the parameter space of models to specific ranges of $d_{\mu}$ and $R_{h \rightarrow \mu^+ \mu^-}$. For example, for $\tan \beta = 1$, models with $\mathcal{Q} = 9, 5,$ and 1 predict $|d_{\mu}|$ between $\sim (6 - 12)\times 10^{-22} e \cdot \textrm{cm}$, $\sim (5 - 8)\times 10^{-22} e \cdot \textrm{cm}$, and $\sim 2 \times 10^{-22} e \cdot \textrm{cm}$, respectively, while explaining $\Delta a_{\mu}$ and assuming $R_{h \rightarrow \mu^+ \mu^-}$ lies within $10\%$ of the SM prediction (left plot). Predictions from an extensive range of $\tan \beta$ for $\mathcal{Q} = 1$ are shown in the right plot. Note that in different ranges of $\tan \beta$, there are constraints on $m_{H, A, H^{\pm}}$. The results do not depend significantly on these masses and we assume they are sufficiently large to evade the limits. 
\begin{figure}[t]
\includegraphics[scale=0.5]{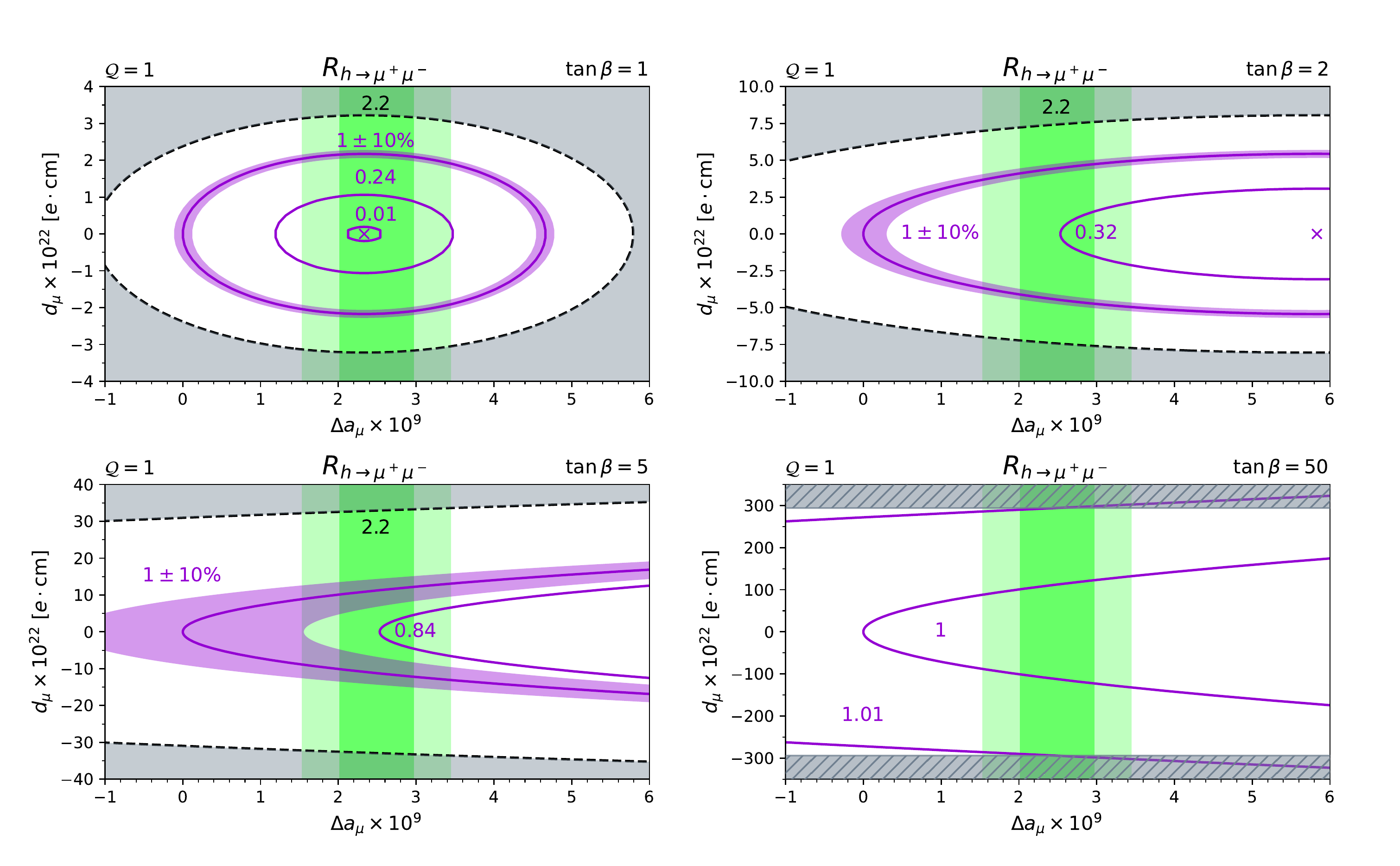}
\caption{Contours of $R_{h \rightarrow \mu^+ \mu^-}$ in the $\Delta a_{\mu} - d_{\mu}$ plane for $\mathcal{Q} = 1$ when $\tan \beta = 1$ (top left), $\tan \beta = 2$ (top right), $\tan \beta = 5$ (bottom left), and $\tan \beta = 50$ (bottom right). Solid purple lines correspond to $R_{h \rightarrow \mu^+ \mu^-} = 1$ (SM prediction), while the purple shaded region extends to $\pm 10\%$, if possible. An $\times$ corresponds to the center of the ellipses. Dashed black lines represent the boundary excluded by $R_{h \rightarrow \mu^+ \mu^-} > 2.2$ in shaded grey (this region is not relevant for $\tan \beta = 50$). The hatched grey region shown for $\tan \beta = 50 $ is restricted by both perturbativity and EW precision constraints on couplings.}
\label{fig:contour_tanb}
\end{figure}

In Fig.~\ref{fig:contour_tanb}, we show the full range of $d_{\mu}$ allowed for $R_{h \rightarrow \mu^+ \mu^-} \leq 2.2$ with $ \mathcal{Q} = 1$. For $\tan \beta = 1$ and 5, the region is limited up to $|d_{\mu}| \sim 3 \times 10^{-22} e \cdot \textrm{cm}$ and up to $|d_{\mu}| \sim 33 \times 10^{-22} e \cdot \textrm{cm}$, respectively. As $\tan \beta$ increases, larger couplings are needed to explain $\Delta a_{\mu}$ for fixed masses of new leptons. Limiting couplings only by perturbativity [while satisfying precision measurements as in Eq.~(\ref{eq:ew_cons})], $|d_{\mu}|$ can reach up to $\sim 300 \times 10^{-22} e \cdot \textrm{cm}$ for $\tan \beta = 50$. This is just above the current experimental limit.
\begin{figure}[t]
\includegraphics[scale=0.5]{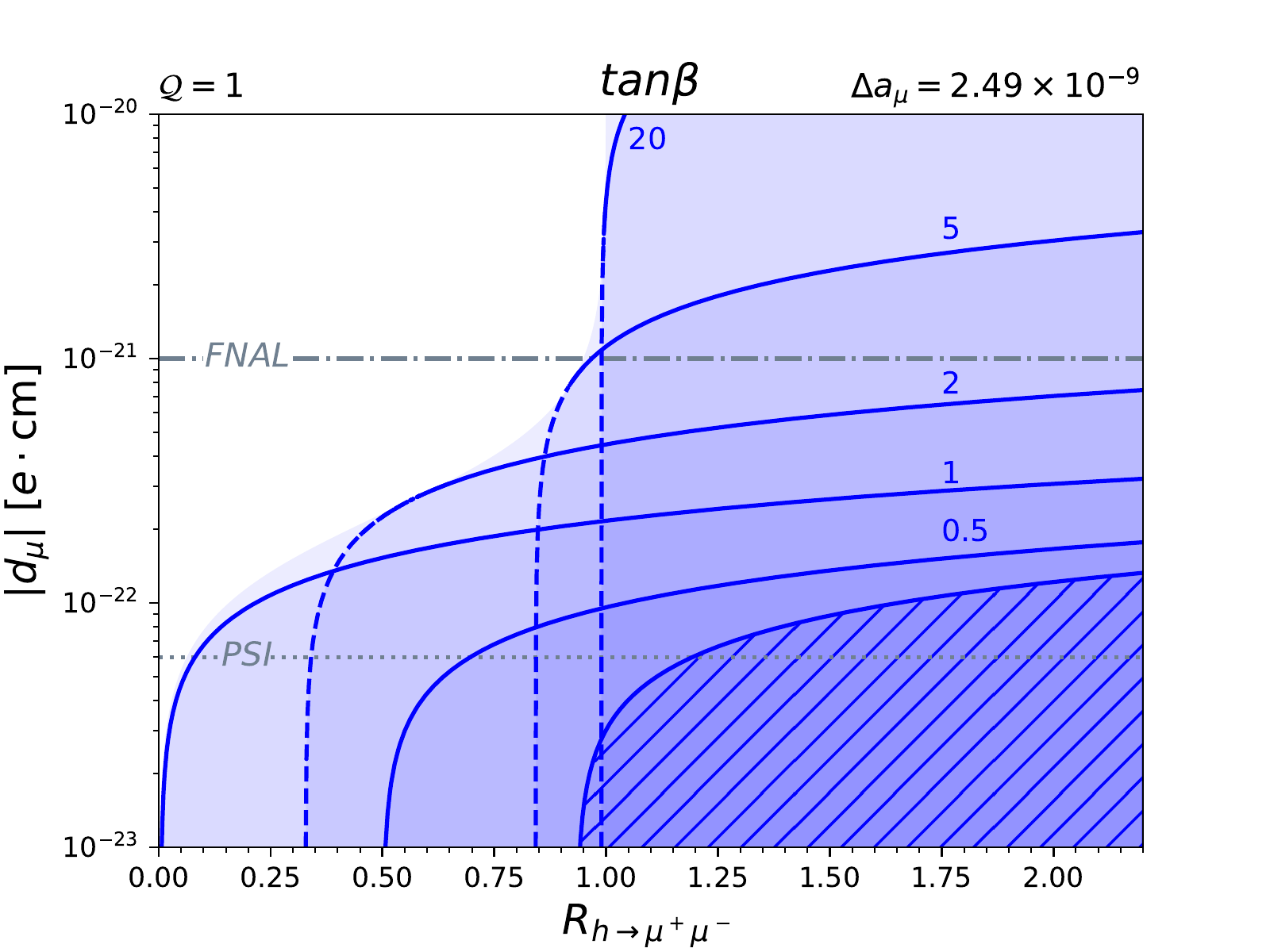}
\includegraphics[scale=0.5]{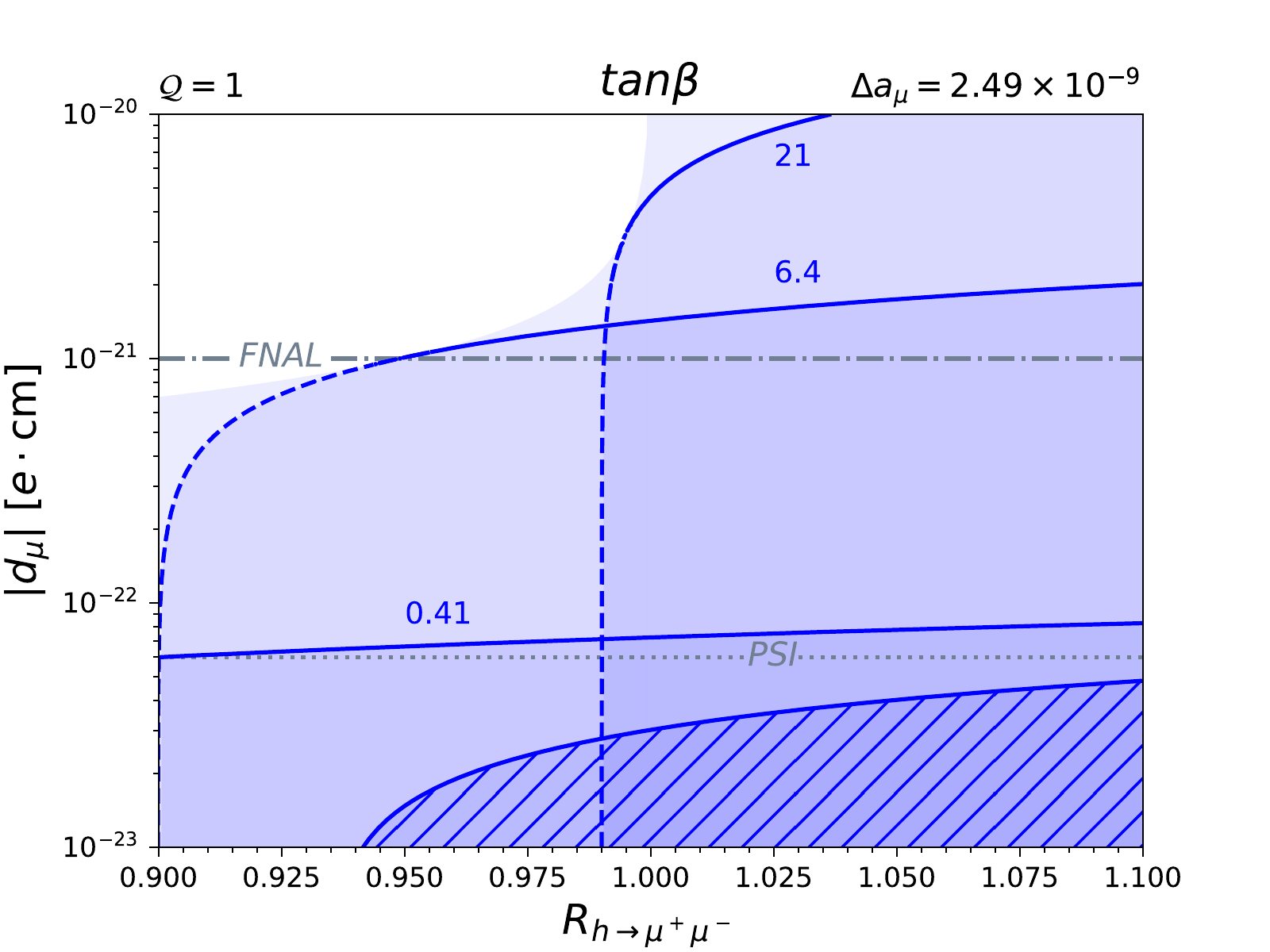}
\caption{Left: contours of $\tan \beta$ for predictions of $|d_{\mu}|$ with respect to $R_{h \rightarrow \mu^+ \mu^-}$ assuming the central value of $\Delta a_{\mu}$ for $\mathcal{Q} = 1$. Projected sensitivities from Fermilab and PSI experiments are shown respectively by dash-dotted and dotted lines. The blue hatched region is where the top Yukawa coupling becomes nonperturbative. Right: contours for $\tan \beta$ when $R_{h \rightarrow \mu^+ \mu^-} = 1 \pm 0.1$.}
\label{fig:dvsr_contours}
\end{figure}

We show the full range of $R_{h \rightarrow \mu^+ \mu^-}$ consistent with experiment and predicted values of $|d_{\mu}|$ for the central value of $\Delta a_{\mu}$ as $\tan \beta$ varies in Fig.~\ref{fig:dvsr_contours} for models with $\mathcal{Q} = 1$. The solid and dashed lines represent the smaller and larger of the two solutions for $\tan \beta$ [or $k$ through Eq.~(\ref{eq:k_tree})] that predict the same value of $|d_{\mu}|$ and $R_{h \rightarrow \mu^+ \mu^-}$. When $0.41 \lesssim \tan \beta \lesssim 6.4$, $R_{h \rightarrow \mu^+ \mu^-} = 1 \pm 10\%$ necessarily requires values of $d_{\mu}$ that can be seen at PSI. As $\tan \beta$ increases, the predicted range of $|d_{\mu}|$ extends from zero to current experimental limits. Further assuming $R_{h \rightarrow \mu^+ \mu^-} = 1 \pm 1\%$ extends the range of $\tan \beta$ that can be fully tested at PSI to $0.38 \lesssim \tan \beta \lesssim 21$. Note that the blue hatched region is where the top Yukawa coupling becomes nonperturbative.

\begin{figure}[t]
\includegraphics[scale=0.5]{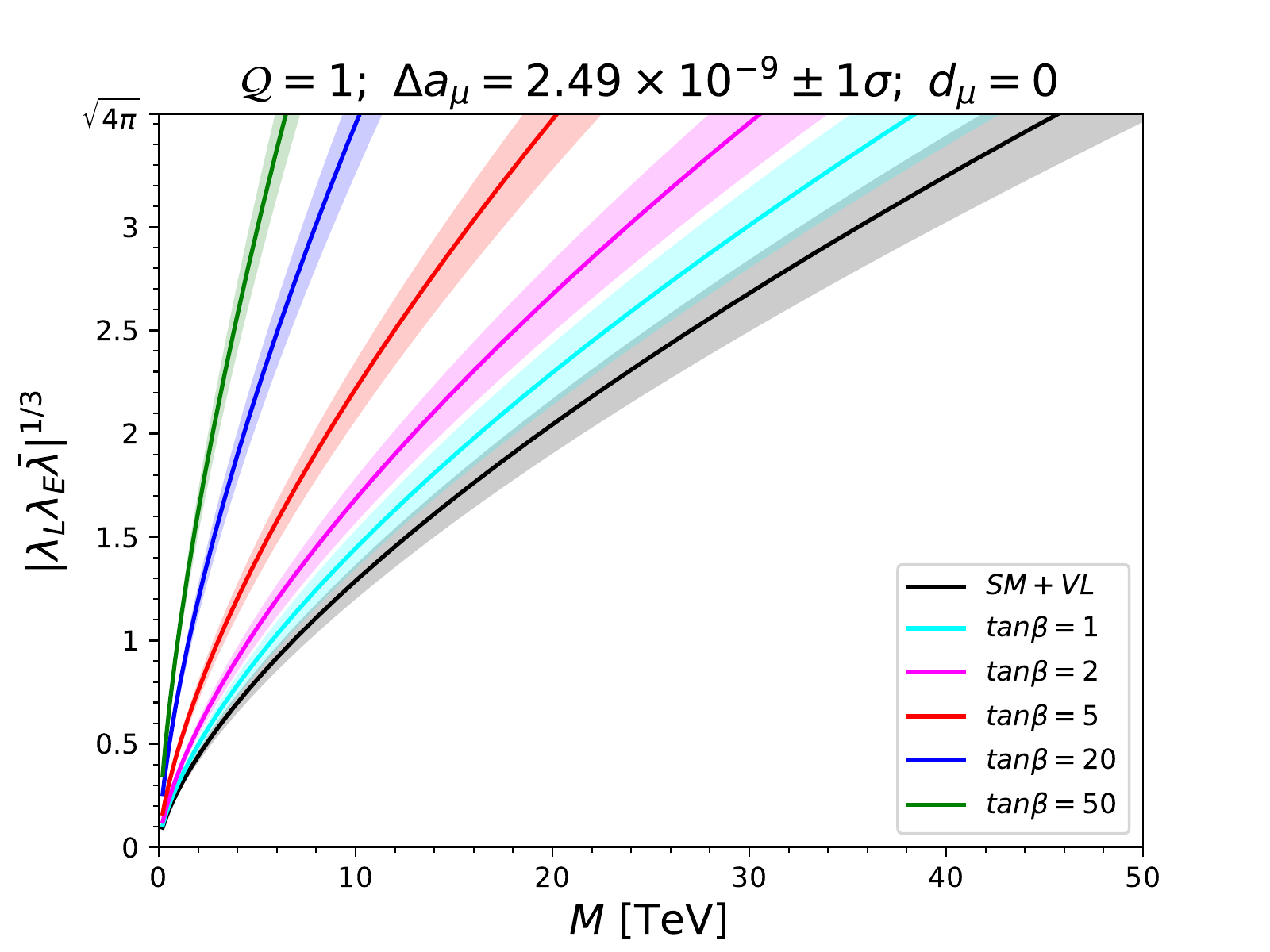}
\includegraphics[scale=0.5]{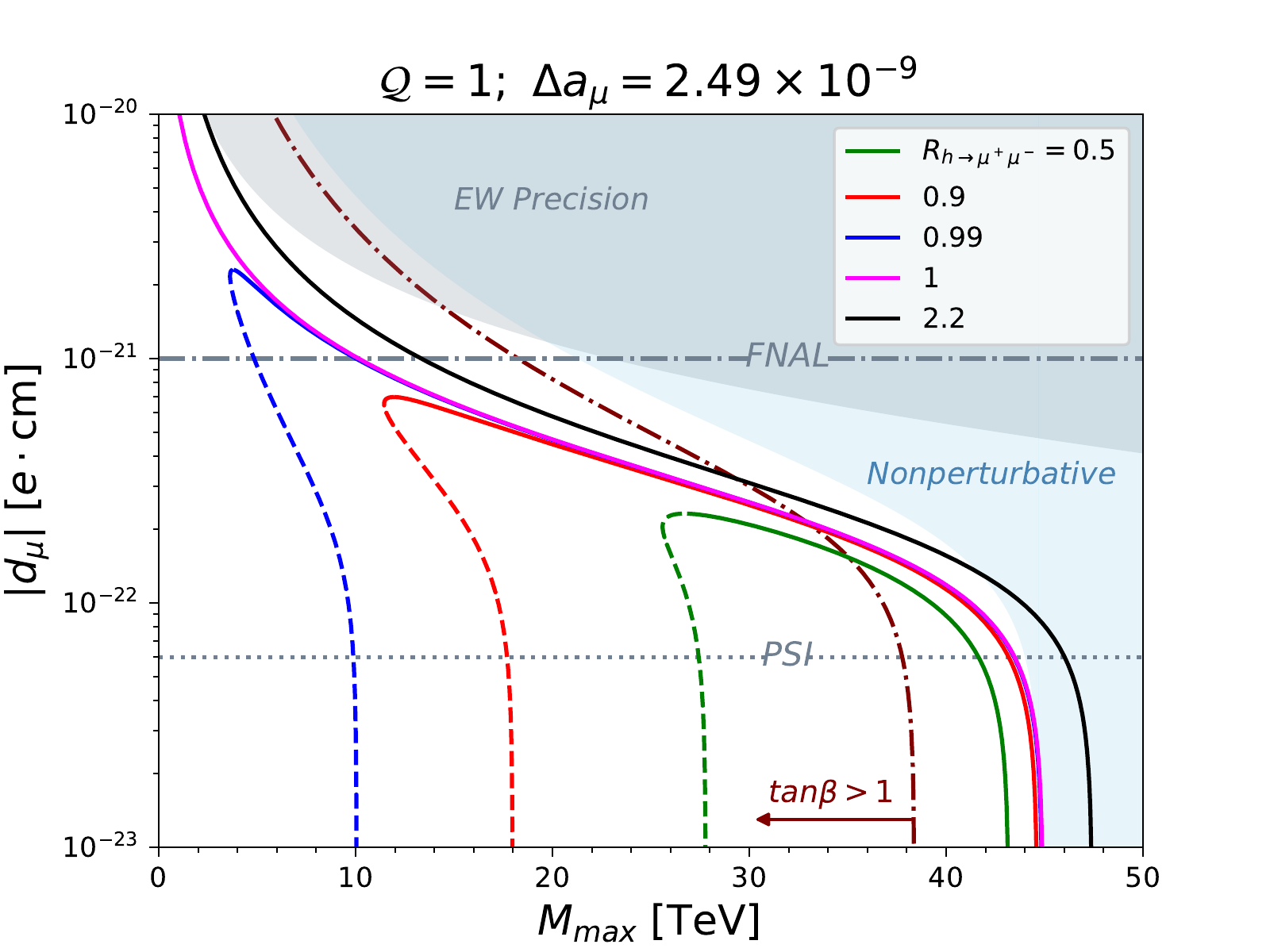}
\caption{Left: common scale of new physics for models with $\mathcal{Q} = 1$ for the SM + vector-like leptons (VL) case, $ \tan \beta = 1, 2, 5, 20$, and $50$ required to explain the central value of $\Delta a_{\mu}$ (solid) and the $1 \sigma$ region (shaded) for a given overall size of couplings with $d_{\mu}= 0$. Right: range of $|d_{\mu}|$ with respect to the maximum mass needed to explain the central value of $\Delta a_{\mu}$ for several values of $R_{h \rightarrow \mu^+ \mu^-} = 0.5, 0.9, 0.99, 1$, and $2.2$. Dashed and solid lines represent the smaller and larger of the two solutions of Eq.~(\ref{eq:ellipse}). The dash-dotted maroon line corresponds to $\tan \beta = 1$. The shaded light blue region is where the top Yukawa coupling becomes nonperturbative. Projected sensitivities from Fermilab and PSI experiments are given by grey dash-dotted and dotted lines, respectively. The grey shaded region is where the model violates electroweak precision constraints.}
\label{fig:tanb_limits}
\end{figure}

In the left plot of Fig.~\ref{fig:tanb_limits}, we plot curves explaining $\Delta a_{\mu}$ within $1 \sigma$ for a common scale of new physics versus the size of couplings for different values of $\tan \beta = 1, 2, 5, 20$, and $50$ as well as the SM+VL scenario for $\mathcal{Q} = 1$. As $\tan \beta$ increases, the slope of the contours increases and further limits the allowed range of the predicted mass spectrum. In the right panel, the red and blue curves highlight the behavior of, respectively, $R_{h \rightarrow \mu^+ \mu^-} = 0.9 $ and 0.99. With increasing levels of precision on $R_{h \rightarrow \mu^+ \mu^-}$, we see that the maximum scale of new physics becomes more limited, decreasing as $R_{h \rightarrow \mu^+ \mu^-} \rightarrow 1$. According to Eq.~(\ref{eq:ellipse}), a nonzero $d_{\mu}$ may increase $R_{h \rightarrow \mu^+ \mu^-}$, meaning that measuring $d_{\mu} \neq 0$ will infer a scale of new physics that is always lower than the corresponding maximum scale allowed in the left plot of Fig.~\ref{fig:tanb_limits}. Given the current upper limit on $R_{h \rightarrow \mu^+ \mu^-} \leq 2.2$, the absolute maximum scale of new physics occurs at $\sim 47$ TeV, agreeing with numerical results for randomized scenarios with couplings up to the perturbativity limit in \cite{Dermisek:2020cod,Dermisek:2021ajd}. Even when the precision increases closer to the SM-like scenario $R_{h \rightarrow \mu^+ \mu^-} = 1$, the scale will only slightly decrease to $\sim 45$ TeV from above. Due to the quadratic form of Eq.~(\ref{eq:ellipse}), when a given value of $R_{h \rightarrow \mu^+ \mu^-} < 1$, there are two positive, distinct solutions for $k$ [or for $\tan \beta$ through Eq.~(\ref{eq:k_tree})] corresponding to the same value of $R_{h \rightarrow \mu^+ \mu^-}$. In the right panel, the dashed (solid) curves represent the smaller (larger) of the two $k$ solutions. Conversely, when $R_{h \rightarrow \mu^+ \mu^-} > 1$, there is only one solution that is positive and one that is negative. From Eq.~(\ref{eq:k_tree}) and Table~\ref{tab:2hdmQ}, we see that $k$ is always positive [even when considering the full mass dependence of $\mathcal{Q}_2(x_M^{(2)})$], meaning that solutions where $k < 0$ for $R_{h \rightarrow \mu^+ \mu^-} > 1$ are never possible (as opposed to some specific FFS- and SSF-type scenarios considered in \cite{Dermisek:2023nhe}). For example, when $d_{\mu} \rightarrow 0$ for the dashed red (blue) curves, future LHC precision measurements of $\lambda_{\mu \mu}^h$ at the $10 \%  \ (1\%)$ level will rule out models with $\tan \beta \lesssim 6.4 \ (21)$, restricting the mass spectrum to $M \lesssim 18 \ (10)$ TeV for Yukawa couplings up to the perturbativity limit, $\sqrt{4 \pi}$. However, if we consider the other solid red (blue) branches, the mass limit approaches $M \sim 45$ TeV for both $R_{h \rightarrow \mu^+ \mu^-} = 0.9 \ (0.99)$ , restricting $\tan \beta$ to $\lesssim 0.31 \ (0.26)$, respectively, which is close to (in) the region where the top Yukawa coupling is nonperturbative, as indicated by the light blue region. These branches can be extrapolated to the left figure and lie between the black and cyan curves.

When comparing with the FFS-type models with doublets and singlets discussed in \cite{Dermisek:2023nhe}, both model types have $k \propto M^4$ and thus have identical profiles and shapes up to some scaling. For the same factor of $\mathcal{Q}$ between models, one can obtain the FFS-type curves through scaling the contours in the right panel of Fig.~\ref{fig:tanb_limits} by $(\pi/9)^{1/4}$.

\subsection{Complex $k$ factor}

In the previous section, contributions involving $\lambda$ vanished in the limit of leptons being the heaviest particles in the spectrum. Because of this, the correlation was direct between $C_{\mu H_1}$ and $C_{\mu \gamma}$, enforcing $k \in \mathbb{R}$. However, for a general spectrum of leptons and heavy Higgses, contributions from $\lambda$ will not vanish and attribute to the moments, as in Eqs.~(\ref{eq:cmugamma_lam1}),~(\ref{eq:cmugamma_lam2}), and~(\ref{eq:cmugamma_lam3}) in Appendix~\ref{sec:tree_contributions}. Hence, if $\arg(\overline{\lambda}) \neq \arg(\lambda)$, $k$ will necessarily be complex. When $k = |k|e^{i \phi_k}$ becomes complex, the ellipse equation in Eq.~(\ref{eq:ellipse}) is modified to
\begin{equation}
    R_{h \rightarrow \mu^+ \mu^-} = \left(\frac{|k|v^2 \Delta a_{\mu}}{2 m_{\mu}^2} - \cos \phi_k \right)^2 + \left(\frac{|k|v^2 d_{\mu}}{e m_{\mu}} - \sin \phi_k \right)^2,
\label{eq:rot_ellipse}
\end{equation}
where the axes of the ellipse in the $\Delta a_{\mu} - d_{\mu}$ plane are shifted by the phase angle $\phi_k$. The center of the ellipse is now located at $(2 m_{\mu}^2 \cos \phi_k / |k| v^2, e m_{\mu} \sin \phi_k / |k| v^2)$, which is no longer symmetric along $|d_{\mu}| = 0$. The asymmetry along with a large complex phase suggests that certain models may shift the entire ellipse to a region where a nonzero $|d_{\mu}|$ is always predicted. This occurs when the edge of semiminor axis intersects with $|d_{\mu}| = 0$ and is satisfied whenever $R_{h \rightarrow \mu^+ \mu^-} \leq \sin^2 \phi_k$. 
\begin{figure}[t]
\includegraphics[scale=0.6]{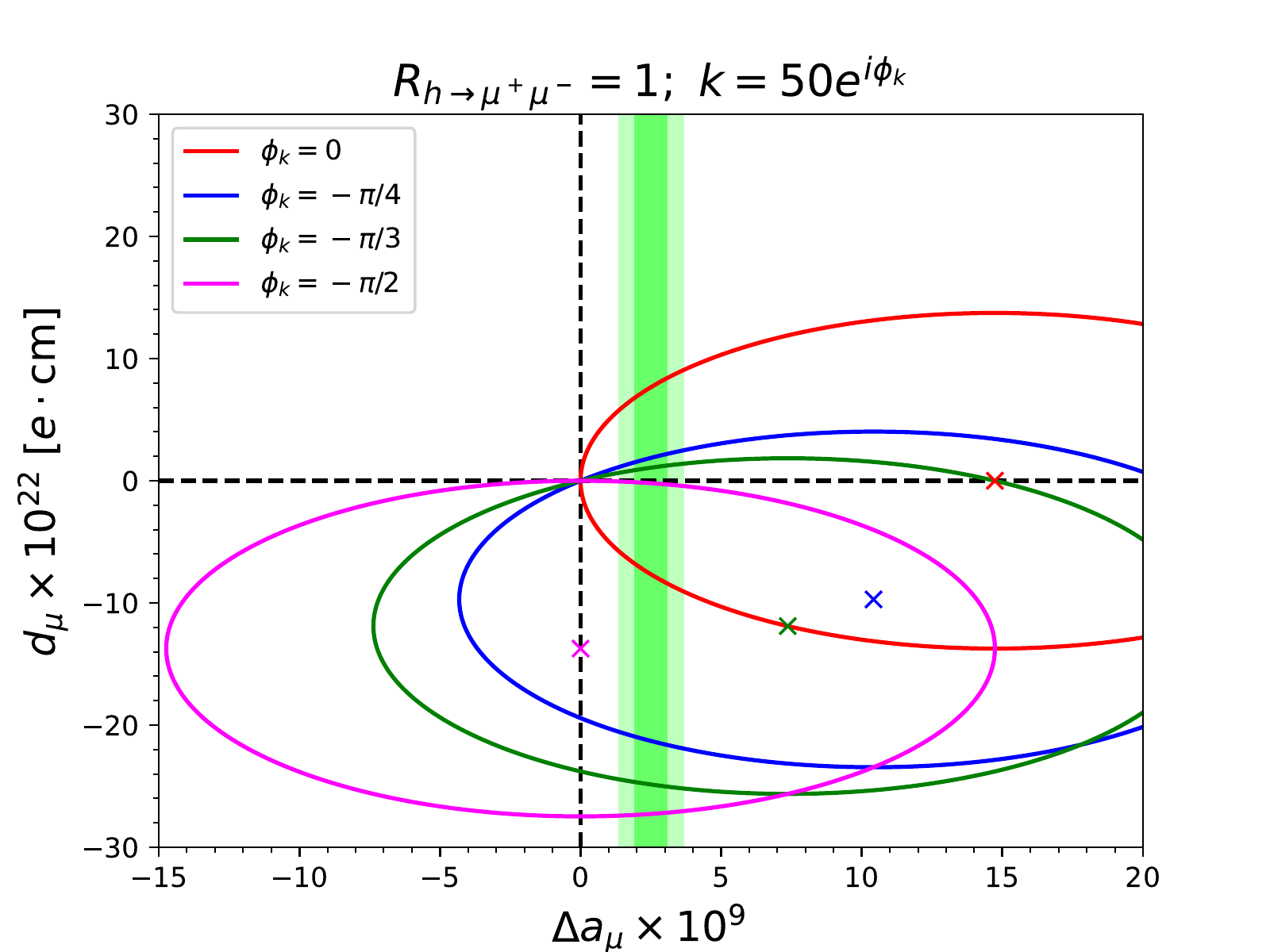}
\caption{Ellipses of $R_{h \rightarrow \mu^+ \mu^-} = 1$ when $k = 50 e^{i \phi_k}$ for phases $\phi_k = 0, - \pi/4, - \pi/3,$ and $-\pi/2$. Each colored $\times$ marks the center of the ellipse. The dark and light green bands represent the region of $\Delta a_{\mu}$ within $1 \sigma$ and $2 \sigma$, respectively.}
\label{fig:complex_k_tree}
\end{figure}

In Fig.~\ref{fig:complex_k_tree}, we demonstrate this behavior for a variety of phases of $k = 50 e^{i \phi_k}$ for $\phi_k = 0, -\pi /4, -\pi/3, - \pi/2$. Models generating a large complex phase can be extended to ones where the $\lambda$ coupling is present alongside $\overline{\lambda}$, such as the $\textbf{2}_{-3/2} \oplus \textbf{1}_{-1}$ model, and similar arguments can apply to $\textbf{2}_{-3/2} \oplus \textbf{3}_{-1}, \textbf{2}_{-1/2} \oplus \textbf{3}_{0}$ models as well. In the limit that $M \simeq M_2$ for arbitrary $\lambda$, $k = 64 \pi^2 / (5 + \tan^2 \beta (17 + \lambda^* / \overline{\lambda} ) / 6)$. In order to generate $|k| = 50$ with phases between $\phi_k = - \pi/4$ and $- \pi / 2$ for $\tan \beta = 2$, the magnitude of $|\lambda^* / \overline{\lambda}| \simeq 17.4 - 31.0$ and its phase should lie between $\phi_{\lambda^*/\overline{\lambda}} \simeq -0.88 + \pi$ and $-0.66 + \pi$ rad, respectively. Notice that the $\pi$ phase shift is required for $\textrm{Re}[\lambda^* / \overline{\lambda}]$ to cancel with the other real terms in order to generate a large magnitude of $k$, which prefers a larger $\tan \beta$ for a more comparable size of couplings (or vice versa as presented here). $\textrm{Im}[\lambda^* / \overline{\lambda}]$ is responsible for generating the phase of $k$. We note that a wide variety of scenarios can accommodate these large ratios of couplings while respecting perturbativity and EW precision constraints needed to explain $\Delta a_{\mu}$. However, for $\Delta a_{\mu}$, because of the partial cancellation from the real part of $\lambda^*/\overline{\lambda}$, a large complex phase in the combination $\lambda_L \lambda_E \overline{\lambda}$ is required.\footnote{On a side note, models may also generate a complex $k$ factor induced by loop corrections (see, for example, the $\textbf{2}_{-1/2} \oplus \textbf{1}_{-1}$ model in the next section).}

\subsection{One-loop corrections to $k$}
\label{sec:kloop}

In the Higgs basis, additional contributions from quartic couplings can appear from one-loop corrections to the mass operator $C_{\mu H_1}$ [Eq.~(\ref{eq:cmuH_1loop})]. The full one-loop corrected $k$ factor for our main model of interest, new leptons with quantum numbers identical to SM leptons, is
\begin{equation}
\begin{split}
    k = \frac{64 \pi^2}{\left(1 + \mathcal{Q}_2(x_M^{(2)}) \tan^2 \beta \right)} & \left[1 - \left(\frac{3\lambda_1}{8 \pi^2} \right) K(x_M^{(2)}) + \left(\frac{\lambda \overline{\lambda}}{16 \pi^2} \right) \left( \cos^2 \beta + L(x_M^{(2)}) \sin^2 \beta \right) \right. \\ 
    & \left. - \left(\frac{1}{16 \pi^2} \right) (|\lambda|^2 + (\lambda \overline{\lambda})^* + |\overline{\lambda}|^2) \left( \cos^2 \beta + 2 N(x_M^{(2)}) \sin^2 \beta \right) \right].
\end{split}
\label{eq:k_loop}
\end{equation}
Notice that $C_{\mu H_1}$ now also involves $\lambda$ at loop level, which in general  can generate a nonzero phase for complex Lagrangian parameters. 

\begin{figure}[t]
\includegraphics[scale=0.5]{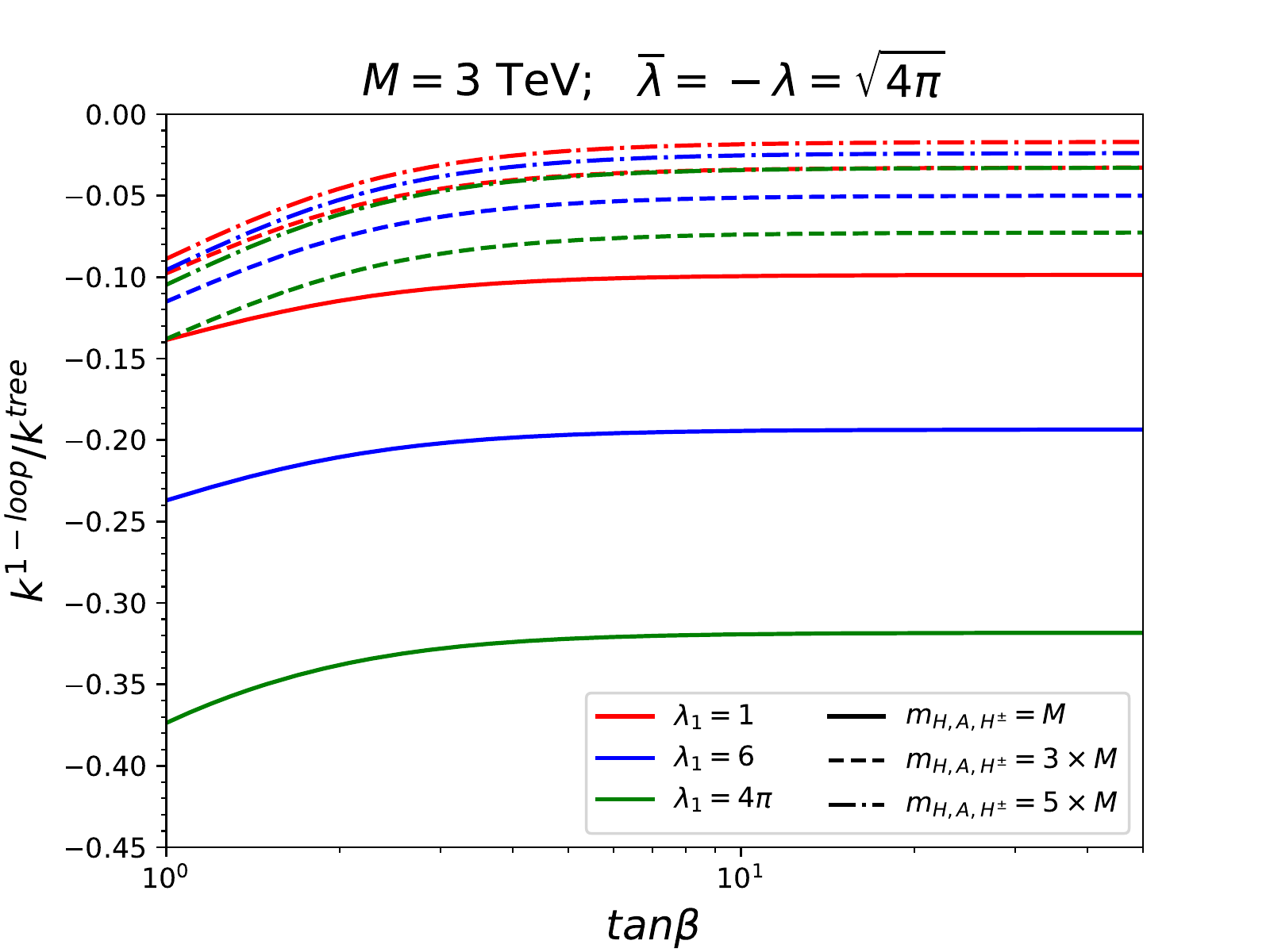}
\includegraphics[scale=0.5]{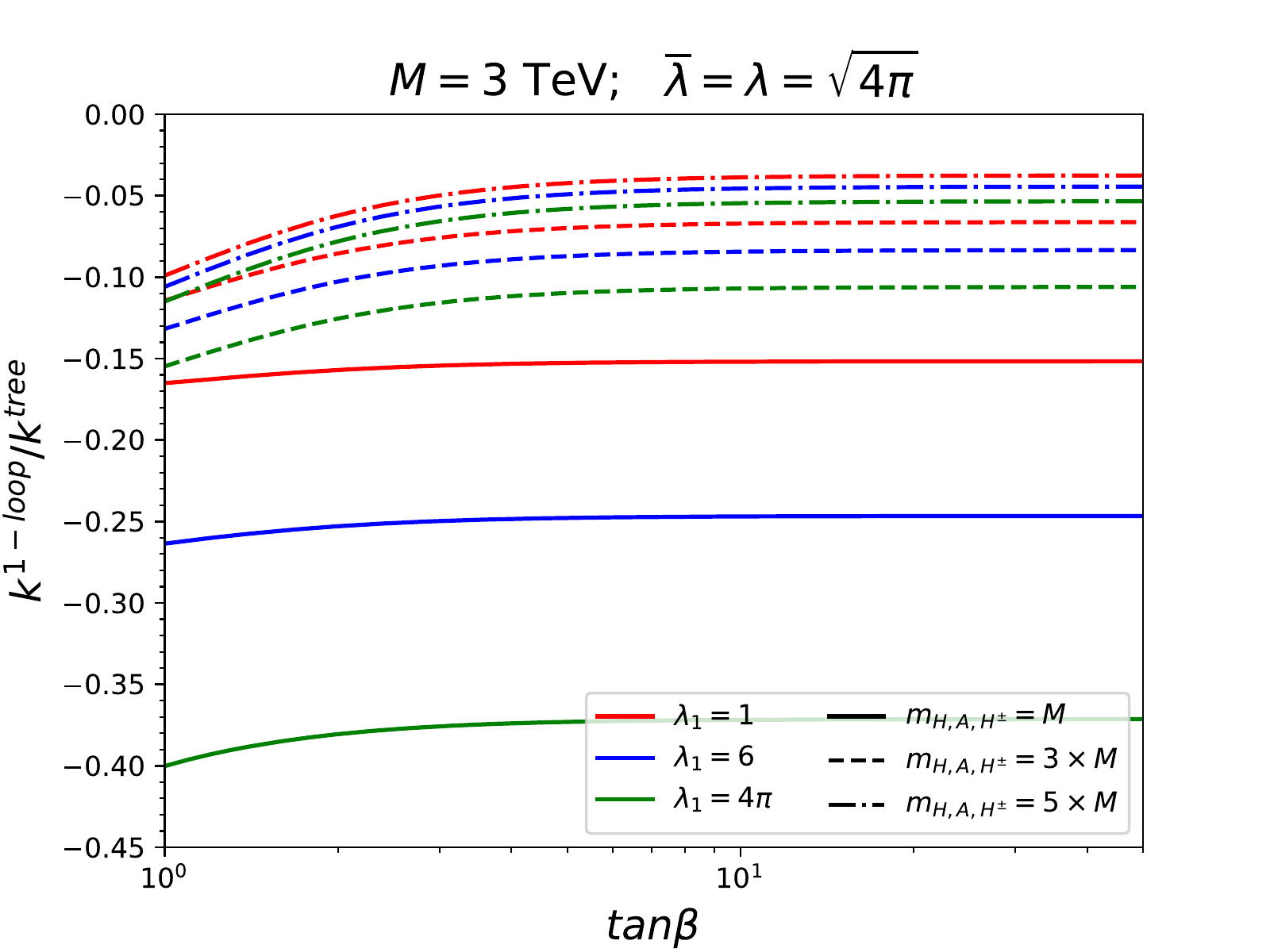}
\caption{Curves representing the ratio of the one-loop contributions to $k$ normalized to its tree-level value $k^{one-loop} / k^{tree}$ when $\lambda = - \overline{\lambda} = \sqrt{4 \pi}$ (left) and $\lambda = \overline{\lambda} = \sqrt{4 \pi}$ (right) are real, assuming $M = 3$ TeV. The dashed and dash-dotted lines correspond to $m_{H,A,H^{\pm}} = 3 \times M$ and $5 \times M$, respectively, in both plots.}
\label{fig:k_ratio}
\end{figure}

The one-loop contribution can be relatively large due to the fact that $K(x),L(x)$, and $2N(x) > 0$ when $x > 0$ and of similar order for $x \leq 1$, while the range of the quartic coupling $\lambda_1$ is limited to $0 < \lambda_1 \leq 4 \pi$, required by the stability of the scalar potential and perturbativity. The Yukawa couplings are only limited by perturbativity in both directions: $|\lambda|, |\overline{\lambda}| \leq \sqrt{4 \pi}$. Additionally, when the Higgs masses are split from the leptons, in the limit when $M_2^2 \simeq m_{H,A,H^{\pm}}^2$ becomes large, $x_M^{(2)} \rightarrow 0$ and the functions $K(x), L(x),$ and $N(x) \rightarrow 0$ when the heavy Higgses decouple, leaving only contributions from the light $H_1$ doublet. Hence, in Fig.~\ref{fig:k_ratio} the dashed and dash-dotted lines when $m_{H,A,H^{\pm}} = 3 \times M$ and $5 \times M$, respectively, reduce the relative loop contribution compared to $m_{H,A,H^{\pm}} = M$ (solid lines) in both plots. Notice that the behaviors in both plots plateau for $\tan \beta \gtrsim 4$ as both the quartic piece and heavy scalar contribution remain when $\cos \beta \rightarrow 0 $ and $\sin \beta \rightarrow 1$. 

According to Fig.~\ref{fig:k_ratio}, when $m_{H,A,H^{\pm}} = M$, we see that the one-loop corrections can affect its tree-level contribution at most when all couplings are near their perturbativity limits, such as $\sim 37 \% - 32 \%$ for $\tan \beta = 1 - 50$ when $\lambda_1 = 4 \pi$ (green solid line) and $\overline{\lambda} = -\lambda = \sqrt{4 \pi}$ (left plot). For an even smaller quartic coupling $\lambda_1 = 1$ (red solid line), the contribution is at most $\sim 14 \%$. In the other direction, $\overline{\lambda} = \lambda = \sqrt{4 \pi}$ (right plot) when $\lambda_1 = 4 \pi$ (green solid line), the relative contribution is almost $\sim 40 \%$ throughout the entire domain of $\tan \beta$, while the contribution is nearly $\sim 16 \%$ everywhere for $\lambda_1 = 1$. We note that the behavior is similar in both plots because, irrespective of sign, there is a partial cancellation between $\lambda \overline{\lambda}$ and $(\lambda \overline{\lambda})^*$ terms, leaving terms such as $|\lambda|^2, |\overline{\lambda}|^2$ unaffected by their relative sign. In fact, the difference $L(x) - 2N(x) < 0$, meaning that when $\lambda \overline{\lambda} = - 4 \pi$ the contribution in each scenario is slightly reduced when comparing to $\lambda \overline{\lambda} = 4 \pi.$ In any case, the magnitude of $k$ will be reduced by the loop corrections.

\section{Conclusions}
\label{sec:conclusion}

Building upon previous works, we demonstrated that in the 2HDM-II extended with vector-like leptons, there can be multiple sources of chiral enhancement to the muon dipole moments, generating a correlation to $h \rightarrow \mu^+ \mu^-$ through a single, model-dependent complex factor. This correlation parametrizes deviations of these observables from their SM values through an ellipse. We explored representations of new leptons that were previously considered in SM extensions and, by expanding on calculations of previous works, we found new radiative corrections relevant for this correlation. 

The ellipse of muon dipole moments allows for novel constraints on the scale of new physics and the 2HDM-II parameter space that are not accessible by colliders. For example, assuming that the central value of $\Delta a_{\mu}$ stays the same, future measurements of $h \rightarrow \mu^+ \mu^-$ imply predicted values of $d_{\mu}$ that could be probed at future muon electric dipole experiments. We found that for models with new leptons that have analogous quantum numbers to SM leptons, $\mathcal{Q} = 1$, measuring $R_{h \rightarrow \mu^+ \mu^-}$ within $10\%$ leads to values of $d_{\mu}$ up to the projected sensitivity at PSI for $0.41 \lesssim \tan \beta \lesssim 6.4$. Increasing the precision to $\pm 1\%$ leads to a range of $0.38 \lesssim \tan \beta \lesssim 21$. These bounds imply an upper bound on the scale of new leptons that is $\sim 18$ and $10$ TeV, respectively. On the other hand, if $d_{\mu}$ is not seen at PSI, the precision of $h \rightarrow \mu^+ \mu^-$ at $10\%$ and $1\%$ requires $\tan \beta \gtrsim 6.4$ and $21$, respectively, and with the same upper limit on the scale of new leptons. Furthermore, other solutions allowed by the ellipse would lead to an upper bound on the scale of new leptons, $\sim 45$ TeV, but would require pushing the values of $\tan \beta$ to the limit of perturbativity related to the top Yukawa coupling. 

We obtained similar bounds for all other models we explored. For example, if the central value of $\Delta a_{\mu}$ is assumed with $R_{h \rightarrow \mu^+ \mu^-} = 1 \pm 10\%$, models with $1.31 \lesssim (\mathcal{Q}_1 + \mathcal{Q}_2 \tan^2 \beta) \lesssim 33$ necessarily require nonzero $d_{\mu}$. More specifically, in the cases with $\mathcal{Q}_1 = \mathcal{Q}_2 = 1,5,9$ and $\tan \beta = 1$, these models predict $|d_{\mu}| = 2 \times 10^{-22} e \cdot \textrm{cm}, (5-8) \times 10^{-22} e \cdot \textrm{cm}$, and $(6-12) \times 10^{-22} e \cdot \textrm{cm}$, respectively. For $\tan \beta = 50$ and limiting couplings only by perturbativity, the maximum predicted value of $|d_{\mu}|$ is $ \sim 300 \times 10^{-22} e \cdot \textrm{cm}$, for $\mathcal{Q} = 1$. This limit is already at the range of current sensitivity and other models would only give larger $|d_{\mu}|$. 

In addition to our main results, we explored new radiative corrections to the muon mass and dipole moments, appearing only in models where new leptons transform as $\mathbf{2}_{-3/2} \oplus \textbf{1}_{-1}, \mathbf{2}_{-3/2} \oplus \textbf{3}_{-1},$ and $\mathbf{2}_{-1/2} \oplus \textbf{3}_{0}$. We found that these corrections become clear in the Higgs basis, where contributions to the dipole moments scale as $(\mathcal{Q}_1 + \mathcal{Q}_2 \tan^2 \beta)$, where each factor behaves differently when leptons are much heavier than the SM doublet, $\mathcal{Q}_1(x_M^{(1)} \rightarrow \infty)$, or comparable to heavy Higgses, $\mathcal{Q}_2(x_M^{(2)} \rightarrow 1)$. By happenstance, if we compare the SM contributions of each representation with that in the 2HDM-II as in Table~\ref{tab:2hdmQ}, the same pattern of contributions occurs for \textit{both} when $\lambda^* = \overline{\lambda}$ and produce integer values. 

The additional corrections modify the ellipse equation through an additional phase in the $k$ factor. In fact, for $k \in \mathbb{C}$, a striking feature occurs: situations exist where the prediction for $d_{\mu}$ is asymmetric and sign-preferred, shifting the ellipse to a region where a portion of the parameter space can be ruled out while being consistent with $\Delta a_{\mu}$. This behavior appears for models whose contribution to $C_{\mu \gamma}$ also comes from an additional source for generally complex $\lambda^*$, whenever $\phi_{\lambda^*} \neq \phi_{\overline{\lambda}} + \pi n$ for $n = 0,1, \cdots$ and $M_2 \simeq M$. Other sources of complex $k$ can occur via subleading corrections to the muon mass, generating up to $40\%$ of the contribution compared to the tree-level piece when all couplings are near their perturbativity limits in nearly the entire range of $\tan \beta$. However, this will reduce the magnitude of $k$ by the same amount, causing the center of the ellipse to shift to larger $\Delta a_{\mu}$ while predicting  a large nonzero $d_{\mu}$ within the range of $R_{h \rightarrow \mu^+ \mu^-} = 1 \pm 10\%$ \cite{Dermisek:2022aec}, increasing the discovery potential of these effects in near-future experiments. 

The mass and couplings of the muon remain among the last vestiges of indirect hints of new physics beyond the SM. The need to precisely measure these quantities is currently driving new experimental efforts~\cite{Lukicov:2019ibv,Adelmann:2021udj} and inspiring the next generation of particle colliders~\cite{Buttazzo:2020ibd,Han:2020pif,Han:2021lnp,deBlas:2022aow,Forslund:2022xjq,Arakawa:2022mkr}. The complementarity of future measurements of the muon dipole moments and $h \rightarrow \mu^+ \mu^-$ will provide an important road map to understanding high scales of new physics whose presence may already be leaving clues in experiments.

\acknowledgments
The work of R.D. was supported in part by the U.S. Department of Energy under Award No. {DE}-SC0010120. TRIUMF receives federal funding via a contribution agreement with the National Research Council of Canada.

\appendix
\section{Details about the 2HDM scalar potential}
\label{sec:higgs_basis}

The parameters $m_{1}^2, m_{2}^2, \lambda_1, \lambda_2, \lambda_3,$ and $\lambda_4$ of the scalar potential in Eq.~(\ref{eq:potential}) are real by Hermiticity, while $m_{12}^2$ and $\lambda_5$ are taken to be real to preserve $CP$ symmetry in the Higgs sector. The stability of the potential further requires \cite{Gunion:2002zf,Nie:1998yn,Deshpande:1977rw,Barroso:2013awa}
\begin{equation}
    \lambda_1, \lambda_2 > 0, \ \ \  \sqrt{\lambda_1 \lambda_2} + \lambda_3 > 0, \ \ \textrm{and} \ \ \sqrt{\lambda_1 \lambda_2} + \lambda_3 + \lambda_4 -|\lambda_5| > 0.
\label{eq:stab}
\end{equation}

After EWSB when the neutral components of the doublets acquire a VEV, $\langle H_d^0 \rangle = v_d$ and  $\langle H_u^0 \rangle = v_u$, the rotation angle $\alpha$ diagonalizes the $CP$-even scalar fields $h$ and $H$ to the physical basis and yields
\begin{equation}
    H_d^0 = v_d + \frac{1}{\sqrt{2}} \left(-h \sin \alpha + H \cos \alpha \right) + \frac{i}{\sqrt{2}} \left(G \cos \beta - A \sin \beta \right),
\label{eq:hd0}
\end{equation}
\begin{equation}
    H_u^0 = v_u + \frac{1}{\sqrt{2}} \left(h \cos \alpha + H \sin \alpha \right) - \frac{i}{\sqrt{2}} \left(G \sin \beta + A \cos \beta \right),
\label{eq:hu0}
\end{equation}
\begin{equation}
    H_d^{\pm} = G^{\pm}  \cos \beta - H^{\pm} \sin \beta,
\label{eq:hdc}
\end{equation}
\begin{equation}
    H_u^{\pm} = - G^{\pm} \sin \beta - H^{\pm} \cos \beta.
\label{eq:huc}
\end{equation}

From the original basis, we can rotate via the angle $\beta$ to the Higgs basis where the SM and additional scalar fields become separated:
\begin{equation}
    \begin{pmatrix}
        H_1 \\ 
        -i \sigma^2 H_2^{\dagger}
    \end{pmatrix} = \begin{pmatrix}
        \cos \beta & \sin \beta \\ 
        - \sin \beta & \cos \beta
    \end{pmatrix} \begin{pmatrix}
        H_d \\ 
        -i \sigma^2 H_u^{\dagger}
    \end{pmatrix}, 
\label{eq:hb_trans}
\end{equation}
We see that in the alignment limit when $\beta - \alpha = \pi/2$, the two doublets simplify to
\begin{equation}
    H_1 = \begin{pmatrix}
    G^+ \\ 
    v + \frac{1}{\sqrt{2}} (h \sin (\beta - \alpha) + H \cos (\beta - \alpha) + i G)
    \end{pmatrix} \rightarrow \begin{pmatrix}
    G^+ \\ 
    v + \frac{1}{\sqrt{2}} (h + i G)
    \end{pmatrix}, 
\label{eq:transdoub1}
\end{equation}
\begin{equation}
    H_2 = \begin{pmatrix}
    \frac{1}{\sqrt{2}} (h \cos (\beta - \alpha) - H \sin (\beta - \alpha) - i A) \\ 
    -H^-
    \end{pmatrix} \rightarrow \begin{pmatrix}
     -\frac{1}{\sqrt{2}} (H + i A) \\ 
    -H^-
    \end{pmatrix},
\label{eq:transdoub2}
\end{equation}
where $H_1$ contains only the SM degrees of freedom and $H_2$ contains the additional Higgs fields in this new basis.
From Eqs.~(\ref{eq:transdoub1}) and~(\ref{eq:transdoub2}), one notices that $\langle H_1^0 \rangle = v$ and $\langle H_2^0 \rangle = 0$, such that $H_1$ becomes the SM Higgs doublet in the alignment limit. Inverting Eq.~(\ref{eq:hb_trans}), the transformations $(H_d,H_u) \rightarrow (H_1,H_2)$ are
\begin{equation}
    H_d \rightarrow H_1 \cos \beta + \epsilon H_2^{\dagger} \sin \beta,
\end{equation}
\begin{equation}
    H_u \rightarrow - H_1^{\dagger} \epsilon \sin \beta + H_2 \cos \beta,
\end{equation}
where we used the fact that $i \sigma^2 = \epsilon$ and $\epsilon_{12} = +1$. Inserting these definitions into the 2HDM-II scalar potential~(\ref{eq:potential}), we find in this basis
\begin{equation}
\begin{split}
        V(H_1, H_2) & = M_1^2 \left( H_1^{\dagger} H_1 \right) + M_2^2 \left( H_2^{\dagger} H_2 \right) + M_{12}^2 \left(H_1^{\dagger} \cdot H_2^{\dagger} + H_1 \cdot H_2 \right) \\
        & + \frac{1}{2} \Lambda_1 \left(H_1^{\dagger} H_1\right)^2 + \frac{1}{2} \Lambda_2 \left(H_2^{\dagger} H_2 \right)^2 + \Lambda_3 \left(H_1^{\dagger} H_1 \right) \left(H_2^{\dagger} H_2 \right) \\ 
        & + \Lambda_4 \left(H_1^{\dagger} \cdot H_2^{\dagger} \right)  \left(H_1 \cdot H_2 \right) + \frac{1}{2} \Lambda_5 \left[ \left(H_1^{\dagger} \cdot H_2^{\dagger} \right)^2 + \left(H_1 \cdot H_2 \right)^2 \right] \\
        & + \Lambda_6 \left(H_1^{\dagger} H_1 \right) \left[H_1^{\dagger} \cdot H_2^{\dagger} + H_1 \cdot H_2 \right] + \Lambda_7 \left(H_2^{\dagger} H_2 \right) \left[H_1^{\dagger} \cdot H_2^{\dagger} + H_1 \cdot H_2 \right].
\end{split}
\end{equation}
If we utilize the extrema conditions of the potential in Eq.~(\ref{eq:potential}), $ \frac{1}{v_d} \frac{\partial V}{\partial H_d} \Big|_{v} = m_1^2 - m_{12}^2 \tan \beta + \lambda_1 v_d^2 + \lambda_{345} v_u^2 = 0$ and $\frac{1}{v_u} \frac{\partial V}{\partial H_u} \Big|_{v} = m_2^2 - \frac{m_{12}^2}{\tan \beta} + \lambda_2 v_u^2 + \lambda_{345} v_d^2 = 0$ using the shorthand notation $\lambda_{345} \equiv \lambda_3 + \lambda_4 + \lambda_5$, the dimensionful parameters listed below can be simplified as
\begin{equation}
    \begin{split}
        & M_1^2 = m_1^2 \cos^2 \beta + m_2^2 \sin^2 \beta - 2 m_{12}^2 \sin \beta \cos \beta \\
        & \ \ \ \ \ = - v^2 \left(\lambda_1 \cos^4 \beta + \lambda_2 \sin^4 \beta + 2 \lambda_{345} \sin^2 \beta \cos^2 \beta \right), \\
        & M_2^2 = m_1^2 \sin^2 \beta + m_2^2 \cos^2 \beta + 2 m_{12}^2 \sin \beta \cos \beta \\
        & \ \ \ \ \ = \frac{m_{12}^2}{\sin \beta \cos \beta} - v^2 \left((\lambda_1 + \lambda_2 - 2 \lambda_{345}) \sin^2 \beta \cos^2 \beta + \lambda_{345} \right), \\
        & M_{12}^2 = (m_1^2 - m_2^2) \sin \beta \cos \beta + m_{12}^2 (\cos^2 \beta - \sin^2 \beta) \\
        & \ \ \ \ \ = v^2 \left((-\lambda_1 + \lambda_{345}) \sin \beta \cos^3 \beta + (\lambda_2 - \lambda_{345}) \sin^3 \beta \cos \beta \right),
    \end{split}
\label{eq:hbasis_mass}
\end{equation}
as well as the quartic couplings,
\begin{equation}
    \begin{split}
        & \Lambda_1 = \lambda_1 \cos^4 \beta + \lambda_2 \sin^4 \beta + 2 \lambda_{345} \sin^2 \beta \cos^2 \beta, \\
        & \Lambda_2 = \lambda_1 \sin^4 \beta + \lambda_2 \cos^4 \beta + 2 \lambda_{345} \sin^2 \beta \cos^2 \beta, \\
        & \Lambda_3 = \lambda_3 + (\lambda_1 + \lambda_2 - 2 \lambda_{345}) \sin^2 \beta \cos^2 \beta, \\
        & \Lambda_4 = \lambda_4 + (\lambda_1 + \lambda_2 - 2 \lambda_{345}) \sin^2 \beta \cos^2 \beta, \\
        & \Lambda_5 = \lambda_5 + (\lambda_1 + \lambda_2 - 2 \lambda_{345}) \sin^2 \beta \cos^2 \beta, \\
        & \Lambda_6 = (\lambda_1 - \lambda_{345}) \sin \beta \cos^3 \beta + (-\lambda_2 + \lambda_{345}) \sin^3 \beta \cos \beta, \\ 
        & \Lambda_7 = (\lambda_1 - \lambda_{345}) \sin^3 \beta \cos \beta + (-\lambda_2 + \lambda_{345}) \sin \beta \cos^3 \beta,
    \end{split}
\label{eq:hbasis_couplings}
\end{equation}
which agree with results in \cite{Davidson:2005cw}. Using the exact definitions of couplings in Eq.~(\ref{eq:exact_couplings}), we may also express these parameters in terms of the physical Higgs masses and mixing angles:
\begin{equation}
    \begin{split}
        & M_1^2 = - \frac{1}{2} \left( m_H^2 \cos^2(\beta - \alpha) + m_h^2 \sin^2 (\beta - \alpha) \right), \\
        & M_2^2 = \frac{m_{12}^2}{\sin \beta \cos \beta} - \frac{1}{2} \left[m_H^2\left(\sin^2 (\beta - \alpha) + 
        \frac{\sin 2 \alpha}{\sin 2 \beta}\right) +  m_h^2\left(\cos^2 (\beta - \alpha) - 
        \frac{\sin 2 \alpha}{\sin 2 \beta}\right) \right], \\
        & M_{12}^2 = - \frac{1}{2} \left(m_H^2 - m_h^2 \right) \sin(\beta - \alpha) \cos(\beta - \alpha), 
    \end{split}
\end{equation}
and
\begin{equation}
    \begin{split}
        & \Lambda_1 = -M_1^2/v^2 , \\
        & \Lambda_2 = \left(\frac{1}{8v^2 \sin^2 \beta \cos^2 \beta} \right) \left[ -\frac{4 m_{12}^2 \cos^2 2 \beta}{\sin \beta \cos \beta} + m_H^2 \left(\sin(\beta + \alpha) - \sin(\beta - \alpha) \cos 2 \beta \right)^2 \right. \\ 
        & \left. ~~~~~~~~~~~~~~~~~~~~~~~~~~~~~~~~ + m_h^2 \left(\cos(\beta + \alpha) + \cos(\beta - \alpha) \cos 2 \beta \right)^2 \right], \\
        & \Lambda_3 = \left( \frac{1}{2v^2}\right) \left[2m_{H^{\pm}}^2 - \frac{2 m_{12}^2}{\sin \beta \cos \beta} + m_H^2\left(\sin^2 (\beta - \alpha) + 
        \frac{\sin 2 \alpha}{\sin 2 \beta}\right) +  m_h^2\left(\cos^2 (\beta - \alpha) - 
        \frac{\sin 2 \alpha}{\sin 2 \beta}\right)\right], \\
        & \Lambda_4 = \left(\frac{1}{2v^2} \right) \left[m_A^2 - 2 m_{H^{\pm}}^2 + m_H^2 \sin^2 (\beta - \alpha) + m_h^2 \cos^2 (\beta - \alpha) \right], \\
        & \Lambda_5 = \left(\frac{1}{2v^2} \right) \left[-m_A^2 + m_H^2 \sin^2 (\beta - \alpha) + m_h^2 \cos^2 (\beta - \alpha) \right], \\
        & \Lambda_6 = -M_{12}^2/v^2, \\
        & \Lambda_7 = \left(\frac{1}{2v^2 \sin \beta \cos \beta} \right) \left[ \frac{m_{12}^2 \cos 2 \beta}{\sin \beta \cos \beta} - m_H^2 \sin(\beta - \alpha) \left( \cos(\beta - \alpha) \sin \beta \cos \beta - \sin(\beta + \alpha) \right) \right. \\
        & \left. ~~~~~~~~~~~~~~~~~~~~~~~~~~~~~ + m_h^2 \cos(\beta - \alpha) \left( \sin(\beta - \alpha) \sin \beta \cos \beta - \cos(\beta + \alpha) \right) \right].
    \end{split}
\end{equation}
Finally, working in the decoupling limit ($m_H^2 \simeq m_A^2 \simeq m_{H^{\pm}}^2 \gg m_h^2$, which also enforces $\beta - \alpha \rightarrow \pi/2$) while writing the parameters in terms of a single quartic coupling $\lambda_1$ from Eq.~(\ref{eq:q_couplings}), we find
\begin{equation}
    \begin{split}
        & M_1^2 = 0, \\
        & M_2^2 = m_{H,A,H^{\pm}}^2 - 2 \lambda_1 v^2 / \tan^2 \beta, \\
        & M_{12}^2 = 0, \\
        & \Lambda_1 = 0, \\
        & \Lambda_2 = \lambda_1 (2 - 1 / \sin^2 \beta)^2, \\
        & \Lambda_3 = 2 \lambda_1 / \tan^2 \beta, \\
        & \Lambda_4 = \Lambda_5 = \Lambda_6 = 0, \\
        & \Lambda_7 = \lambda_1 (2 - 1 / \sin^2 \beta) / \tan \beta. \\
    \end{split}
\end{equation}
These are the relevant formulas used for the discussion in the main text.

\section{Models and mass eigenstate basis}
\label{sec:meb_couplings}
First, we introduce models and define their mass eigenstate basis in each of the five representations. We additionally present couplings of lepton eigenstates to SM bosons $Z,W^{\pm}, h$ and new bosons $H,A,H^{\pm}$. These couplings are defined with a common notation and will be further characterized by diagonalization and Yukawa matrices in each individual representation. 

\subsection{$\textbf{2}_{-1} \oplus \textbf{1}_{-1}$}

After EWSB, the Lagrangian of Eq.~(\ref{eq:vll_lag}) gives the charged lepton mass matrix:
\begin{equation}
    \left( \bar{\mu}_L,  \bar{L}_L^-, \bar{E}_L \right) M_e \begin{pmatrix}
    \mu_R \\ 
    L_R^- \\
    E_R
    \end{pmatrix} = \left( \bar{\mu}_L, \bar{L}_L^-, \bar{E}_L \right) \begin{pmatrix} 
    y_{\mu} v_d & 0 & \lambda_E v_d \\
    \lambda_L v_d & M_L & \lambda v_d \\
    0 & \bar{\lambda} v_d & M_E 
    \end{pmatrix} \begin{pmatrix}
    \mu_R \\ 
    L_R^- \\
    E_R \end{pmatrix}, 
\end{equation}
and the neutral lepton mass matrix,
\begin{equation}
    \left( \bar{\nu}_L, \bar{L}_L^0, \bar{N}_L \right) M_{\nu} \begin{pmatrix}
    \nu_R = 0 \\ 
    L_R^0 \\
    N_R
    \end{pmatrix} = \left( \bar{\nu}_L, \bar{L}_L^0, \bar{N}_L \right) \begin{pmatrix} 
    0 & 0 & \kappa_N v_u \\
    0 & M_L & \kappa v_u \\
    0 & \bar{\kappa} v_u & M_N 
    \end{pmatrix} \begin{pmatrix}
    \nu_R = 0 \\ 
    L_R^0 \\
    N_R \end{pmatrix}.
\end{equation}
We construct the mass eigenstate basis in the following steps. First, we find $U_{L,R}$ such that $U_L^{\dagger} M M^{\dagger} U_L$ and $U_R^{\dagger} M^{\dagger} M U_R$ are diagonal. In this procedure, $U_L^{\dagger} M U_R$ becomes diagonal where the diagonal entries are the physical masses up to possible phases. Applying this procedure to the mass matrices $M_e$ and $M_{\nu}$ above, we have

\begin{equation}
    U_L^{e \dagger}
    \begin{pmatrix} 
    y_{\mu} v_d & 0 & \lambda_E v_d \\
    \lambda_L v_d & M_L & \lambda v_d \\
    0 & \bar{\lambda} v_d & M_E 
    \end{pmatrix} U_R^e = 
    \begin{pmatrix}
    m_{\mu} e^{i \phi_{m_{\mu}}} & 0 & 0 \\
    0 & m_{e_4}e^{i\phi_{m_{e_4}}} & 0 \\
    0 & 0 & m_{e_5} e^{i\phi_{m_{e_5}}}
    \end{pmatrix}
    \label{eq:m_matrtix1}, 
\end{equation}
\begin{equation}
    U_L^{\nu \dagger}
    \begin{pmatrix} 
    0 & 0 & \kappa_N v_u \\
    0 & M_L & \kappa v_u \\
    0 & \bar{\kappa} v_u & M_N 
    \end{pmatrix} U_R^{\nu} = 
    \begin{pmatrix}
    0 & 0 & 0 \\
    0 & \ m_{\nu_4}e^{i\phi_{m_{\nu_4}}} & 0 \\
    0 & 0 & m_{\nu_5} e^{i\phi_{m_{\nu_5}}}
    \end{pmatrix}
    \label{eq:m_matrtix2}, 
\end{equation}
where $m_{\mu}, m_{e_4}, m_{e_5}, m_{\nu_4}$, and $m_{\nu_5}$ are the physical masses of the leptons. Further, we define
\begin{equation}
    \widetilde{U}_R^e \equiv U_R^e \begin{pmatrix}
    e^{-i \phi_{m_\mu}} & 0 & 0 \\
    0 & e^{-i\phi_{m_{e_4}}} & 0 \\
    0 & 0 & e^{-i\phi_{m_{e_{5}}}}
    \end{pmatrix}, 
\end{equation}
\begin{equation}
    \widetilde{U}_R^{\nu} \equiv U_R^{\nu} \begin{pmatrix}
    1 & 0 & 0 \\
    0 & e^{-i\phi_{m_{\nu_4}}} & 0 \\
    0 & 0 & e^{-i\phi_{m_{\nu_{5}}}}
    \end{pmatrix}, 
\end{equation}
so that
\begin{equation}
    U_L^{e \dagger}
    \begin{pmatrix} 
    y_{\mu} v_d & 0 & \lambda_E v_d \\
    \lambda_L v_d & M_L & \lambda v_d \\
    0 & \bar{\lambda} v_d & M_E 
    \end{pmatrix} \widetilde{U}_R^e = 
    \begin{pmatrix}
    m_{\mu} & 0 & 0 \\
    0 & m_{e_4} & 0 \\
    0 & 0 & m_{e_5}
    \end{pmatrix}, 
    \label{eq:u_mat}
\end{equation}
and
\begin{equation}
    U_L^{\nu \dagger}     \begin{pmatrix} 
    0 & 0 & \kappa_N v_u \\
    0 & M_L & \kappa v_u \\
    0 & \bar{\kappa} v_u & M_N 
    \end{pmatrix}
     \widetilde{U}_R^{\nu} = 
    \begin{pmatrix}
    0 & 0 & 0 \\
    0 & m_{\nu_4} & 0 \\
    0 & 0 & m_{\nu_5}
    \end{pmatrix}.
\end{equation}
The diagonalization matrices $U_L$ and $\widetilde{U}_{R}$ represent rotations to the mass eigenstate basis $e_L \rightarrow U_L^e \hat{e}_L$, $e_R \rightarrow \widetilde{U}_R^e \hat{e}_R$ and $\nu_L \rightarrow U_L^{\nu} \hat{\nu}_L$, $\nu_R \rightarrow \widetilde{U}_R^{\nu} \hat{\nu}_R$. This procedure will be also followed for other representations of leptons. 

In Appendix~\ref{sec:approx_form} we present approximate formulas for the diagonalization matrices and resulting couplings. Those formulas are valid only when following the above procedure.

\subsection{$\textbf{2}_{-1/2} \oplus \textbf{3}_{-1}$}

In this representation, the doublet $L$ and triplet $E$ are defined as
\begin{equation}
    L_{L,R} = \begin{pmatrix}
        L_{L,R}^0 \\ L_{L,R}^-
    \end{pmatrix}, \ \ \ \ \ \tau^a E_{L,R}^a = \begin{pmatrix}
        E_{L,R}^- & \sqrt{2} E_{L,R}^0 \\
        \sqrt{2} E_{L,R}^{--} & - E_{L,R}^-
    \end{pmatrix},
\end{equation}
and couple to $H_d$ and SM leptons via the following Lagrangian:
\begin{equation}
    \mathcal{L} \supset - \lambda_E \overline{l}_L \tau^a H_d E_R^a - \lambda_L \overline{L}_L \mu_R H_d - \lambda \overline{L}_L \tau^a H_d E_R^a - \overline{\lambda} H_d^{\dagger} \tau^a L_R \overline{E}_L^a - M_L \overline{L}_L L_R - M_E \overline{E}_L^a E_R^a + h.c.
\label{eq:L12_31}
\end{equation}
After EWSB, the mass matrices are
\begin{equation}
    \left(\overline{\mu}_L, \overline{L}_L^-, \overline{E}_L^{-} \right) \begin{pmatrix}
        y_{\mu} v_d & 0 & - \lambda_E v_d \\
        \lambda_L v_d & M_L & - \lambda v_d \\
        0 & - \overline{\lambda} v_d & M_E
    \end{pmatrix} \begin{pmatrix}
        \mu_R \\ L_R^- \\ E_R^-
    \end{pmatrix} \rightarrow \overline{\hat{e}}_L U_L^{e \dagger} \begin{pmatrix}
        y_{\mu} v_d & 0 & - \lambda_E v_d \\
        \lambda_L v_d & M_L & - \lambda v_d \\
        0 & - \overline{\lambda} v_d & M_E
    \end{pmatrix} \widetilde{U}_R^e \hat{e}_R, 
\end{equation}
\begin{equation}
    \left(\overline{\nu}_L, \overline{L}_L^0, \overline{E}_L^{0} \right) \begin{pmatrix}
        0 & 0 & \sqrt{2} \lambda_E v_d \\
        0 & M_L & \sqrt{2} \lambda v_d \\
        0 & \sqrt{2} \overline{\lambda} v_d & M_E
    \end{pmatrix} \begin{pmatrix}
        \nu_R = 0 \\ L_R^0 \\ E_R^0
    \end{pmatrix} \rightarrow \overline{\hat{\nu}}_L U_L^{\nu \dagger} \begin{pmatrix}
        0 & 0 & \sqrt{2} \lambda_E v_d \\
        0 & M_L & \sqrt{2} \lambda v_d \\
        0 & \sqrt{2} \overline{\lambda} v_d & M_E
    \end{pmatrix} \widetilde{U}_R^{\nu} \hat{\nu}_R.
\end{equation}
Additionally, there is a doubly charged mass eigenstate $E^{--}$ with mass $M_E$. Hence, 
\begin{equation}
    U_L^{e^{--}} = \widetilde{U}_R^{e^{--}} = \textbf{I}_{3 \times 3}.
\end{equation}

\subsection{$\textbf{2}_{-3/2} \oplus \textbf{1}_{-1}$}

The doublet with singly and doubly charged components is defined as
\begin{equation}
    L_{L,R} = \begin{pmatrix}
        L_{L,R}^- \\ L_{L,R}^{--}
    \end{pmatrix}. 
\end{equation}
The Lagrangian for this representation becomes
\begin{equation}
    \mathcal{L} \supset - \lambda_E \overline{l}_L E_R H_d - \lambda_L \overline{L}_L \cdot H_d^{\dagger} \mu_R - \lambda H_d^{\dagger}  \cdot \overline{L}_L E_R - \overline{\lambda} L_R \cdot H_d \overline{E}_L - M_L \overline{L}_L L_R - M_E \overline{E}_L E_R + h.c.
\label{eq:L32_11}
\end{equation}
We remind the reader that the explicit $\cdot$ corresponds to contracting $SU(2)$ indices by $\epsilon_{ij}$, such that $L_R \cdot H_d = \epsilon_{12}(L_R)_1 (H_d)_2 + \epsilon_{21} (L_R)_2 (H_d)_1 = L_R^- H_d^0 - L_R^{--} H_d^+ $, where $\epsilon_{12} = - \epsilon_{21} = +1$. Alternatively, we could continue to contract $SU(2)$ indices with $\delta_{ij}$ by introducing $\epsilon H_d^{\dagger} \equiv \widetilde{H}_d$. The difference between the two notations is an overall minus sign for the term $ \lambda \widetilde{H}_d \overline{L}_L E_R$. After EWSB, the the mass matrices are
\begin{equation}
    \left(\overline{\mu}_L, \overline{L}_L^-, \overline{E}_L^{-} \right) \begin{pmatrix}
        y_{\mu} v_d & 0 & \lambda_E v_d \\
        \lambda_L v_d & M_L & - \lambda v_d \\
        0 & \overline{\lambda} v_d & M_E
    \end{pmatrix} \begin{pmatrix}
        \mu_R \\ L_R^- \\ E_R^-
    \end{pmatrix} \rightarrow \overline{\hat{e}}_L U_L^{e \dagger} \begin{pmatrix}
        y_{\mu} v_d & 0 &  \lambda_E v_d \\
        \lambda_L v_d & M_L & - \lambda v_d \\
        0 & \overline{\lambda} v_d & M_E
    \end{pmatrix} \widetilde{U}_R^e \hat{e}_R, 
\end{equation}
which is diagonalized by $U_L^e$ and $\widetilde{U}_R^e$ as in Eq.~(\ref{eq:u_mat}), but with the replacement of $\lambda \rightarrow - \lambda$. There is one massless neutral eigenstate $\nu_{\mu}$ and one doubly charged eigenstate $L^{--}$ with mass $M_L$. Hence, 
\begin{equation}
    U_L^{\nu} = \widetilde{U}_R^{\nu} = U_L^{e^{--}} = \widetilde{U}_R^{e^{--}} = \textbf{I}_{3 \times 3}.
\end{equation}

\subsection{$\textbf{2}_{-3/2} \oplus \textbf{3}_{-1}$}

The doublet $L$ containing both a singly and doubly charged field as well as the triplet $E$ are defined as
\begin{equation}
    L_{L,R} = \begin{pmatrix}
        L_{L,R}^- \\ L_{L,R}^{--}
    \end{pmatrix}, \ \ \ \ \ \tau^a E_{L,R}^a = \begin{pmatrix}
        E_{L,R}^- & \sqrt{2} E_{L,R}^0 \\
        \sqrt{2} E_{L,R}^{--} & - E_{L,R}^-
    \end{pmatrix},
\end{equation}
which couple to the SM leptons in the following way:
\begin{equation}
    \mathcal{L} \supset - \lambda_E \overline{l}_L \tau^a H_d E_R^a - \lambda_L \overline{L}_L \cdot H_d^{\dagger} \mu_R - \lambda \overline{L}_L \tau^a \cdot H_d^{\dagger} E_R^a - \overline{\lambda} H_d \cdot \overline{E}_L^a \tau^a L_R - M_L \overline{L}_L L_R - M_E \overline{E}_L^a E_R^a + h.c.
\label{eq:L32_31}
\end{equation}
After EWSB, the mass matrices are
\begin{equation}
    \left(\overline{\mu}_L, \overline{L}_L^-, \overline{E}_L^{-} \right) \begin{pmatrix}
        y_{\mu} v_d & 0 & - \lambda_E v_d \\
        \lambda_L v_d & M_L & \lambda v_d \\
        0 & - \overline{\lambda} v_d & M_E
    \end{pmatrix} \begin{pmatrix}
        \mu_R \\ L_R^- \\ E_R^-
    \end{pmatrix} \rightarrow \overline{\hat{e}}_L U_L^{e \dagger} \begin{pmatrix}
        y_{\mu} v_d & 0 & - \lambda_E v_d \\
        \lambda_L v_d & M_L & \lambda v_d \\
        0 & - \overline{\lambda} v_d & M_E
    \end{pmatrix} \widetilde{U}_R^e \hat{e}_R, 
\end{equation}
\begin{equation}
    \left(0, \overline{L}_L^{--}, \overline{E}_L^{--} \right) \begin{pmatrix}
        0 & 0 & 0 \\
        0 & M_L & \sqrt{2} \lambda v_d \\
        0 & - \sqrt{2} \overline{\lambda} v_d & M_E
    \end{pmatrix} \begin{pmatrix}
        0 \\ L_R^{--} \\ E_R^{--}
    \end{pmatrix} \rightarrow \overline{\hat{e}}_L^{--} U_L^{e^{--} \dagger} \begin{pmatrix}
        0 & 0 & 0 \\
        0 & M_L & \sqrt{2} \lambda v_d \\
        0 & - \sqrt{2} \overline{\lambda} v_d & M_E
    \end{pmatrix} \widetilde{U}_R^{e^{--}} \hat{e}_R^{--}.
\end{equation}

\subsection{$\textbf{2}_{-1/2} \oplus \textbf{3}_{0}$}

Finally, this representation is described by the following doublet $L$ and triplet $E$:
\begin{equation}
    L_{L,R} = \begin{pmatrix}
        L_{L,R}^0 \\ L_{L,R}^-
    \end{pmatrix}, \ \ \ \ \ \tau^a E_{L,R}^a = \begin{pmatrix}
        E_{L,R}^0 & \sqrt{2} E_{L,R}^+ \\
        \sqrt{2} E_{L,R}^- & - E_{L,R}^0
    \end{pmatrix}.
\end{equation}
These fields interact with the SM leptons in the following way:
\begin{equation}
    \mathcal{L} \supset - \lambda_E \overline{l}_L \tau^a \cdot H_d^{\dagger} E_R^a - \lambda_L \overline{L}_L \mu_R H_d - \lambda \overline{L}_L \tau^a \cdot H_d^{\dagger} E_R^a - \overline{\lambda} H_d \cdot \overline{E}_L^a \tau^a L_R - M_L \overline{L}_L L_R - M_E \overline{E}_L^a E_R^a + h.c.
\label{eq:L12_30}
\end{equation}
The mass matrices after EWSB are
\begin{equation}
    \left(\overline{\mu}_L, \overline{L}_L^-, \overline{E}_L^{-} \right) \begin{pmatrix}
        y_{\mu} v_d & 0 & \sqrt{2} \lambda_E v_d \\
        \lambda_L v_d & M_L & \sqrt{2} \lambda v_d \\
        0 & - \sqrt{2} \overline{\lambda} v_d & M_E
    \end{pmatrix} \begin{pmatrix}
        \mu_R \\ L_R^- \\ E_R^-
    \end{pmatrix} \rightarrow \overline{\hat{e}}_L U_L^{e \dagger} \begin{pmatrix}
        y_{\mu} v_d & 0 & \sqrt{2} \lambda_E v_d \\
        \lambda_L v_d & M_L & \sqrt{2} \lambda v_d \\
        0 & - \sqrt{2} \overline{\lambda} v_d & M_E
    \end{pmatrix} \widetilde{U}_R^e \hat{e}_R, 
\end{equation}
and
\begin{equation}
    \left(\overline{\nu}_L, \overline{L}_L^0, \overline{E}_L^{0} \right) \begin{pmatrix}
        0 & 0 & \lambda_E v_d \\
        0 & M_L & \lambda v_d \\
        0 & -\overline{\lambda} v_d & M_E
    \end{pmatrix} \begin{pmatrix}
        \nu_R = 0 \\ L_R^0 \\ E_R^0
    \end{pmatrix} \rightarrow \overline{\hat{\nu}}_L U_L^{\nu \dagger} \begin{pmatrix}
        0 & 0 & \lambda_E v_d \\
        0 & M_L & \lambda v_d \\
        0 & -\overline{\lambda} v_d & M_E
    \end{pmatrix} \widetilde{U}_R^{\nu} \hat{\nu}_R. 
\end{equation}

\subsection{Couplings of leptons to $Z$ and $W^{\pm}$ bosons}

The couplings of leptons to the $Z$ boson come from kinetic terms in the Lagrangian, 
\begin{equation}
    \mathcal{L} \supset \overline{\hat{e}}_{La} i \slashed{D}_a \hat{e}_{La} + \overline{\hat{e}}_{Ra} i \slashed{D}_a \hat{e}_{Ra} + \overline{\hat{e}}_{La}^{--} i \slashed{D}_a \hat{e}_{La}^{--} + \overline{\hat{e}}_{Ra}^{--} i \slashed{D}_a \hat{e}_{Ra}^{--} ,
\end{equation}
whereby the covariant derivative is
\begin{equation}
    D_{\mu a} = \partial_{\mu} - i \frac{g}{\cos \theta_W} (T_a^3 - \sin^2 \theta_W Q_a) Z_{\mu}.
\end{equation}
\begin{table}[t]
\centering
\begin{tabular}{||c | c c c c c||}
 \hline
 & $\mathbf{2}_{-1/2} \oplus \mathbf{1}_{-1}$ \ \ & $\mathbf{2}_{-1/2} \oplus \mathbf{3}_{-1}$ \ \ & $\mathbf{2}_{-3/2} \oplus \mathbf{1}_{-1}$ \ \ & $\mathbf{2}_{-3/2} \oplus \mathbf{3}_{-1}$ \ \ & $\mathbf{2}_{-1/2} \oplus \mathbf{3}_{0}$ \\ [1.0ex] \hline
$\xi_L^Z$ & $0$ & $0$ & $1$ & $1$ & $0$ \\
$\chi_L^Z$ & $1$ & $1$ & $1$ & $1$ & $-1$ \\ \hline
$\xi_R^Z$ & $-1$ & $-1$ & $1$ & $1$ & $-1$ \\ 
$\chi_R^Z$ & $0$ & $0$ & $0$ & $0$ & $-1$ \\ \hline
$\xi_L^W$ & $1$ & $1$ & $-$ & $0$ & $1$ \\
$\chi_L^W$ & $0$ & $-\sqrt{2}$ & $-$ & $-\sqrt{2}$ & $\sqrt{2}$ \\ \hline
$\xi_R^W$ & $1$ & $1$ & $-$ & $0$ & $1$ \\
$\chi_R^W$ & $0$ & $-\sqrt{2}$ & $-$ & $-\sqrt{2}$ & $\sqrt{2}$ \\ \hline
$\xi_{--}^W$ & $-$ & $0$ & $1$ & $1$ & $-$ \\ 
$\chi_{--}^W$ & $-$ & $\sqrt{2}$ & $0$ & $\sqrt{2}$ & $-$ \\ [1ex] 
\hline
\end{tabular}
\caption{Representation-dependent factors for left- and right-handed couplings to $Z$ and $W^{\pm}$ gauge bosons.}
\label{tab:gauge_factors}
\end{table}
For couplings of left- and right-handed singly charged leptons to the $Z$ boson, we have
\begin{equation}
    \mathcal{L} \supset \overline{\hat{e}}_{L a} \gamma^{\mu} g_L^{Z e_a e_b} \hat{e}_{L b} Z_{\mu} + \overline{\hat{e}}_{R a} \gamma^{\mu} g_R^{Z e_a e_b} \hat{e}_{R b} Z_{\mu},
\end{equation}
whereby 
\begin{equation}
    g_L^{Z e_a e_b} = \frac{g}{\cos \theta_W} \left[\left(-\frac{1}{2} + \sin^2 \theta_W \right) \delta_{ab} + \xi_L^Z (U_L^{e \dagger})_{a4} (U_L^e)_{4b} + \frac{\chi_L^Z}{2} (U_L^{e \dagger})_{a5} (U_L^e)_{5b} \right],
\end{equation}
and
\begin{equation}
    g_R^{Z e_a e_b} = \frac{g}{\cos \theta_W} \left[ \sin^2 \theta_W \delta_{ab} + \frac{\xi_R^Z}{2} (\widetilde{U}_R^{e \dagger})_{a4} (\widetilde{U}_R^e)_{4b} + \chi_R^Z (\widetilde{U}_R^{e \dagger})_{a5} (\widetilde{U}_R^e)_{5b} \right].
\end{equation}

The couplings to $W^{\pm}$ boson are defined by
\begin{equation}
    \mathcal{L} \supset \left( \bar{\hat{\nu}}_{L a} \gamma^{\mu} g_{L}^{W \nu_a e_b }
    \hat{e}_{L b} + \bar{\hat{\nu}}_{R a} \gamma^{\mu} g_{R}^{W \nu_a e_b } \hat{e}_{R b} \right)W_{\mu}^{+} + \left( \bar{\hat{e}}_{L a}^{--} \gamma^{\mu} g_{L}^{W e_b e_a^{--}}
    \hat{e}_{L b} + \bar{\hat{e}}_{R a}^{--} \gamma^{\mu} g_{R}^{W e_b e_a^{--} } \hat{e}_{R b} \right) W_{\mu}^{-} + h.c., 
\end{equation}
where
\begin{equation}
    g_L^{W \nu_a e_b} = \frac{g}{\sqrt{2}} \left[(U_L^{\nu \dagger})_{a2} (U_L^{e})_{2b} + \xi_L^W (U_L^{\nu \dagger})_{a4} (U_L^e)_{4b} + \chi_L^W (U_L^{\nu \dagger})_{a5} (U_L^e)_{5b} \right]
\end{equation}
and
\begin{equation}
    g_R^{W \nu_a e_b} = \frac{g}{\sqrt{2}} \left[\xi_R^W (\widetilde{U}_R^{\nu \dagger})_{a4} (\widetilde{U}_R^e)_{4b} + \chi_R^W (\widetilde{U}_R^{\nu \dagger})_{a5} (\widetilde{U}_R^e)_{5b} \right].
\end{equation}
Additionally, for representations with doubly charged fermions, the left- and right-handed couplings are
\begin{equation}
    g_L^{W e_b  e_a^{--}} = \frac{g}{\sqrt{2}} \left[\xi_{--}^W (U_L^{e \dagger})_{b4} (U_L^{e^{--}})_{4a} + \chi_{--}^W (U_L^{e \dagger})_{b5} (U_L^{e^{--}})_{5a} \right]^*, 
\end{equation}
\begin{equation}
    g_R^{W e_b  e_a^{--}} = \frac{g}{\sqrt{2}} \left[\xi_{--}^W (\widetilde{U}_R^{e \dagger})_{b4} (\widetilde{U}_R^{e^{--}})_{4a} + \chi_{--}^W (\widetilde{U}_R^{e \dagger})_{b5} (\widetilde{U}_R^{e^{--}})_{5a} \right]^*.
\end{equation}
The representation-dependent $\xi_i$ and $\chi_i$ factors for each set of left- and right-handed $Z$ and $W^{\pm}$ gauge couplings are collected in Table~\ref{tab:gauge_factors}.

\subsection{Couplings to 
Higgs bosons} 

The softly broken 2HDM-II scalar potential is given by Eq.~(\ref{eq:potential}) and is diagonalized to the mass eigenstate basis through Eqs.~(\ref{eq:hd0})-(\ref{eq:huc}). We also work in the alignment limit where $\beta - \alpha = \pi / 2$ such that couplings to the light eigenstate $h$ are SM-like. For singly charged leptons in a given model, the Lagrangian for Yukawa couplings to neutral Higgs bosons is defined by 
\begin{equation}
    \mathcal{L}_{h, H, A} = - \frac{1}{\sqrt{2}}  \bar{\hat{e}}_{L a} \lambda_{e_a e_b}^h \hat{e}_{R b} h - \frac{1}{\sqrt{2}}  \bar{\hat{e}}_{L a} \lambda_{e_a e_b}^H \hat{e}_{R b} H - \frac{1}{\sqrt{2}}  \bar{\hat{e}}_{L a} \lambda_{e_a e_b}^A\hat{e}_{R b} A + h.c., 
\end{equation}
where 
\begin{equation}
\begin{split}
    & \lambda_{e_a e_b}^h = \cos \beta (U_L^{e \dagger} Y_E \widetilde{U}_R^e)_{ab}, \\
    & \lambda_{e_a e_b}^H = \sin \beta (U_L^{e \dagger} Y_E \widetilde{U}_R^e)_{ab}, \\
    & \lambda_{e_a e_b}^A = -i \sin \beta (U_L^{e \dagger} Y_E^A \widetilde{U}_R^e)_{ab},
\label{eq:yukawa_c}
\end{split}
\end{equation}
where $Y_E$ and $Y_E^A$ are Yukawa matrices defined per representation for $CP$-even $(h,H)$ and $CP$-odd $(A)$ bosons, respectively. The notations of Eq.~(\ref{eq:yukawa_c}) are the same for all representations except for their respective diagonalization and Yukawa matrices.

Yukawa couplings of the charged Higgs boson $H^{\pm}$ to leptons are defined by
\begin{equation}
    \mathcal{L}_{H^{\pm}} = -\bar{\hat{\nu}}_{L a} \lambda_{\nu_a e_b}^{H^{\pm}} \hat{e}_{R b} H^+ - \bar{\hat{e}}_{L a} \lambda_{e_a \nu_b}^{H^{\pm}} \hat{\nu}_{R b} H^- -\bar{\hat{e}}_{L a} \lambda_{e_a e_b^{--}}^{H^{\pm}} \hat{e}_{R b}^{--} H^+ - \bar{\hat{e}}_{L a}^{--} \lambda_{e_a^{--} e_b}^{H^{\pm}} \hat{e}_{R b} H^- + h.c.
\end{equation}
The last two terms are relevant for representations with doubly charged leptons. The couplings are defined as
\begin{equation}
\begin{split}
    & \lambda_{\nu_a e_b}^{H^{\pm}} = (U_L^{\nu \dagger} Y_N^{H^{\pm}} \widetilde{U}_R^e)_{ab}, \\
    & \lambda_{e_a \nu_b}^{H^{\pm}} = (U_L^{e \dagger} Y_E^{H^{\pm}} \widetilde{U}_R^{\nu})_{ab}, \\
    & \lambda_{e_a^{--} e_b}^{H^{\pm}} = (U_L^{e^{--} \dagger} Y_{N'}^{H^{\pm}} \widetilde{U}_R^e)_{ab}, \\
    & \lambda_{e_a e_b^{--}}^{H^{\pm}} = (U_L^{e \dagger} Y_{E'}^{H^{\pm}} \widetilde{U}_R^{e^{--}})_{ab}. 
\end{split} 
\end{equation}
The matrices $Y_{E', N'}^{H^{\pm}}$ are the Yukawa matrices present for doubly charged fermions coupling to singly charged ones. 

For the $\textbf{2}_{-1/2} \oplus \textbf{1}_{-1}$ model, the Yukawa matrices for neutral Higgs bosons are 
\begin{equation}
    \begin{split}
        Y_E = \begin{pmatrix}
            y_{\mu} & 0 & \lambda_E \\
            \lambda_L & 0 & \lambda \\
            0 & \overline{\lambda} & 0
        \end{pmatrix}, \ \ \ \ \ Y_E^A = \begin{pmatrix}
            y_{\mu} & 0 & \lambda_E \\
            \lambda_L & 0 & \lambda \\
            0 & - \overline{\lambda} & 0
        \end{pmatrix}. 
    \end{split}
\end{equation}
The Yukawa matrices for charged Higgs bosons are 
\begin{equation}
    \begin{split}
        Y_N^{H^{\pm}} = - \sin \beta \begin{pmatrix}
            y_{\mu} & 0 & \lambda_E \\
            \lambda_L & 0 & \lambda \\
            0 & \overline{\kappa} / \tan \beta & 0
        \end{pmatrix}, \ \ \ \ \ Y_E^{H^{\pm}} = - \cos \beta \begin{pmatrix}
            0 & 0 & \kappa_N \\
            0 & 0 & \kappa \\
            0 & \overline{\lambda}  \tan \beta & 0
        \end{pmatrix}. 
    \end{split}
\end{equation}

In the $\textbf{2}_{-1/2} \oplus \textbf{3}_{-1}$ representation, the Yukawa matrices for neutral Higgs bosons are 
\begin{equation}
    \begin{split}
        Y_E = \begin{pmatrix}
            y_{\mu} & 0 & -\lambda_E \\
            \lambda_L & 0 & -\lambda \\
            0 & -\overline{\lambda} & 0
        \end{pmatrix}, \ \ \ \ \ Y_E^A = \begin{pmatrix}
            y_{\mu} & 0 & -\lambda_E \\
            \lambda_L & 0 & -\lambda \\
            0 & \overline{\lambda} & 0
        \end{pmatrix}. 
    \end{split}
\end{equation}
The Yukawa matrices for charged Higgs bosons are 
\begin{equation}
    \begin{split}
        & Y_N^{H^{\pm}} = - \sin \beta \begin{pmatrix}
            y_{\mu} & 0 & \lambda_E \\
            \lambda_L & 0 & \lambda \\
            0 & 0 & 0
        \end{pmatrix}, \ \ \ \ \ Y_E^{H^{\pm}} = - \cos \beta \begin{pmatrix}
            0 & 0 & 0 \\
            0 & 0 & 0 \\
            0 & \overline{\lambda}  \tan \beta & 0
        \end{pmatrix}, \\
        & Y_{N'}^{H^{\pm}} = - \sin \beta \begin{pmatrix} 0 & 0 & 0 \\
        0 & 0 & 0 \\
        0 & \sqrt{2} \overline{\lambda} & 0
        \end{pmatrix}, \ \ \ \ \ Y_{E'}^{H^{\pm}} = - \sin \beta \begin{pmatrix}
            0 & 0 & \sqrt{2} \lambda_E \\
            0 & 0 & \sqrt{2} \lambda \\ 
            0 & 0 & 0
        \end{pmatrix}.
    \end{split}
\end{equation} 

Then, for the $\textbf{2}_{-3/2} \oplus \textbf{1}_{-1}$, we have the following Yukawa matrices for neutral Higgs bosons: 
\begin{equation}
    \begin{split}
        Y_E = \begin{pmatrix}
            y_{\mu} & 0 & \lambda_E \\
            \lambda_L & 0 & - \lambda \\
            0 & \overline{\lambda} & 0
        \end{pmatrix}, \ \ \ \ \ Y_E^A = \begin{pmatrix}
            y_{\mu} & 0 & \lambda_E \\
            - \lambda_L & 0 & \lambda \\
            0 & \overline{\lambda} & 0
        \end{pmatrix}. 
    \end{split}
\end{equation}
The Yukawa matrices for charged Higgs bosons are 
\begin{equation}
    \begin{split}
        & Y_N^{H^{\pm}} = - \sin \beta \begin{pmatrix}
            y_{\mu} & 0 & \lambda_E \\
            0 & 0 & 0\\
            0 & 0 & 0
        \end{pmatrix}, \ \ \ \ \ Y_E^{H^{\pm}} = \textbf{0}_{3 \times 3} \\ 
        & Y_{N'}^{H^{\pm}} = - \sin \beta \begin{pmatrix} 0 & 0 & 0 \\
        - \lambda_L & 0 & \lambda \\
        0 & 0 & 0
        \end{pmatrix}, \ \ \ \ \ Y_{E'}^{H^{\pm}} = - \sin \beta \begin{pmatrix}
            0 & 0 & 0 \\
            0 & 0 & 0\\ 
            0 & - \overline{\lambda} & 0
        \end{pmatrix}.
    \end{split}
\end{equation}

The Yukawa matrices for neutral Higgs bosons in the $\textbf{2}_{-3/2} \oplus \textbf{3}_{-1}$ model are 
\begin{equation}
    \begin{split}
        Y_E = \begin{pmatrix}
            y_{\mu} & 0 & - \lambda_E \\
            \lambda_L & 0 & \lambda \\
            0 & -\overline{\lambda} & 0
        \end{pmatrix}, \ \ \ \ \ Y_E^A = \begin{pmatrix}
            y_{\mu} & 0 & - \lambda_E \\
            - \lambda_L & 0 & -\lambda \\
            0 & - \overline{\lambda} & 0
        \end{pmatrix}. 
    \end{split}
\end{equation}
The Yukawa matrices for charged Higgs bosons are
\begin{equation}
    \begin{split}
        & Y_N^{H^{\pm}} = - \sin \beta \begin{pmatrix}
            y_{\mu} & 0 & \lambda_E \\
            0 & 0 & 0\\
            0 & \sqrt{2} \overline{\lambda} & 0
        \end{pmatrix}, \ \ \ \ \ Y_E^{H^{\pm}} = - \sin \beta \begin{pmatrix}
            0 & 0 & 0 \\
            0 & 0 & - \sqrt{2} \lambda \\
            0 & 0 & 0
        \end{pmatrix} \\ 
        & Y_{N'}^{H^{\pm}} = - \sin \beta \begin{pmatrix} 0 & 0 & 0 \\
        - \lambda_L & 0 & \lambda \\
        0 & 0 & 0
        \end{pmatrix}, \ \ \ \ \ Y_{E'}^{H^{\pm}} = - \sin \beta \begin{pmatrix}
            0 & 0 & \sqrt{2} \lambda_E \\
            0 & 0 & 0\\ 
            0 & - \overline{\lambda} & 0
        \end{pmatrix}.
    \end{split}
\end{equation}

For the last model we consider, $\textbf{2}_{-1/2} \oplus \textbf{3}_{0}$, we find that the Yukawa matrices involving neutral Higgs bosons are
\begin{equation}
    Y_E = \begin{pmatrix}
        y_{\mu} & 0 & \sqrt{2} \lambda_E \\
        \lambda_L & 0 & \sqrt{2} \lambda \\
        0 & - \sqrt{2} \overline{\lambda} & 0
    \end{pmatrix}, \ \ \ \ \
        Y_E^A = \begin{pmatrix}
            y_{\mu} & 0 & - \sqrt{2} \lambda_E \\
            \lambda_L & 0 & - \sqrt{2} \lambda \\
            0 & - \sqrt{2} \overline{\lambda} & 0
            \end{pmatrix}.
\end{equation}
The Yukawa matrices for the charged Higgs are
\begin{equation}
    \begin{split}
        & Y_N^{H^{\pm}} = - \sin \beta \begin{pmatrix}
        y_{\mu} & 0 & 0 \\
        \lambda_L & 0 & 0 \\
        0 & - \overline{\lambda} & 0
        \end{pmatrix}, \ \ \ \ \ Y_E^{H^{\pm}} = - \sin \beta \begin{pmatrix}
            0 & 0 & \lambda_E \\
            0 & 0 & \lambda \\
            0 & 0 & 0
        \end{pmatrix}.
    \end{split}
\end{equation}
 
\section{Approximate formulas for diagonalization matrices and couplings}
\label{sec:approx_form}

We present in full detail approximate diagonalization matrices for our main representation, as defined in Eqs.~(\ref{eq:m_matrtix1}) and~(\ref{eq:m_matrtix2}), and the other four models. We also give a complete list of approximate couplings to bosons in our representation. Computing approximate couplings in other representations is straightforward and follows similar arguments as those presented for our model.

\subsection{$\textbf{2}_{-1} \oplus \textbf{1}_{-1}$}

In the limit when the mass eigenstates $e_4$ and $\nu_4$ are mostly doublet-like, while $e_5$ and $\nu_5$ are singlet-like, the masses approximately become $m_{e_4} e^{i \phi_{m_{e_4}}} \simeq m_{\nu_4} e^{i \phi_{m_{\nu_4}}} \simeq |M_L| e^{i \phi_{M_L}}$ and $m_{e_5} e^{i \phi_{m_{e_5}}} \simeq |M_E| e^{i \phi_{M_E}}, \ \ m_{\nu_5} e^{i \phi_{\nu_5}} \simeq |M_N| e^{i \phi_{M_N}}$, respectively. From now on, eigenstates corresponding to the $SU(2)$ vector-like doublet are labeled with index $L$, whereas the $SU(2)$ charged and neutral singlet states are indexed with labels $E$ and $N$. In this approximation, one can also expand the diagonalization matrices $U_{L,R}^e$ $U_{L,R}^{\nu}$ in terms of $\epsilon_E = v_d \times (\lambda_L, \lambda_E, \lambda, \bar{\lambda}) / M_{L,E}$ and $ \epsilon_N = v_u \times (\kappa_N, \kappa, \bar{\kappa}) / M_{L,N}$ for $ |\epsilon_{E,N}| \ll 1$. Keeping terms up to $\mathcal{O} (\epsilon_{E,N}^2)$, we find for the diagonalization matrices
\begin{equation}
    U_L^e = \begin{pmatrix}
    1-v_d^2\frac{|\lambda_E|^2}{2|M_E|^2} & - v_d^2 \left(\frac{\lambda_E (\bar{\lambda} M_E^* + \lambda^* M_L)}{M_L (|M_E|^2 - |M_L|^2)} - \frac{y_{\mu} \lambda_L^*}{|M_L|^2}\right) & v_d\frac{\lambda_E}{M_E} \\
    v_d^2 \frac{(\bar{\lambda}^* \lambda_E^* M_L - y_{\mu}^* \lambda_L M_E^*)
    }{|M_L|^2 M_E^*} & 1- v_d^2 \frac{|\lambda^* M_E + \bar{\lambda} M_L^*|^2}{2 (|M_E|^2 - |M_L|^2)^2} & v_d \frac{(\bar{\lambda}^* M_L + \lambda M_E^*)}{|M_E|^2 - |M_L|^2} \\
    - v_d \frac{\lambda_E^*}{M_E^*} & - v_d \frac{(\bar{\lambda} M_L^* + \lambda^* M_E)}{|M_E|^2 - |M_L|^2} & 1 - v_d^2 
    \frac{|\lambda_E|^2}{2 |M_E|^2} - v_d^2 \frac{|\bar{\lambda}^* M_L + \lambda M_E^*|^2}{2(|M_E|^2 - |M_L|^2)^2}
    \end{pmatrix},
\label{eq:ULe}
\end{equation}
\begin{equation}
    U_R^e = \begin{pmatrix}
    1 - v_d^2 \frac{|\lambda_L|^2}{2 |M_L|^2} & v_d \frac{\lambda_L^*}{M_L^*} & v_d^2 \left( \frac{\lambda_L^* (\bar{\lambda}^* M_L + \lambda M_E^*)}{M_E^* (|M_E|^2 - |M_L|^2)} + \frac{y_{\mu}^* \lambda_E}{|M_E|^2} \right) \\
    - v_d \frac{\lambda_L}{M_L} & 1 - v_d^2 
    \frac{|\lambda_L|^2}{2 |M_L|^2} - v_d^2 \frac{|\lambda^* M_L + \bar{\lambda} M_E^*|^2}{2(|M_L|^2 - |M_E|^2)^2} & v_d \frac{(\bar{\lambda}^* M_E + \lambda M_L^*)}{|M_E|^2 - |M_L|^2} \\
    v_d^2 \frac{(\lambda_L \bar{\lambda} M_E^* - y_{\mu} \lambda_E^* M_L)}{M_L |M_E|^2} & -v_d \frac{(\bar{\lambda} M_E^* + \lambda^* M_L)}{|M_E|^2 - |M_L|^2} & 1 - v_d^2 \frac{|\lambda M_L^* + \bar{\lambda}^* M_E|^2}{2 (|M_E|^2 - |M_L|^2)^2}
    \end{pmatrix}, 
\label{eq:URe}
\end{equation}
\begin{equation}
    \widetilde{U}_R^e = U_R^e \begin{pmatrix}
    e^{-i \phi_{m_\mu}} & 0 & 0 \\
    0 & e^{-i\phi_{M_L}} & 0 \\
    0 & 0 & e^{-i\phi_{M_E}}
    \end{pmatrix}, 
\label{eq:m1_phase}
\end{equation}
\begin{equation}
    U_L^{\nu} = \begin{pmatrix}
    1 - v_u^2 \frac{|\kappa_N|^2}{2|M_N|^2} & - v_u^2 \frac{\kappa_N (\kappa^* M_L + \bar{\kappa} M_N^*)}{M_L (|M_N|^2 - |M_L|^2)} & v_u \frac{\kappa_N}{M_N} \\
    v_u^2 \frac{\kappa_N^* \bar{\kappa}^*}{M_L^* M_N^*} & 1 - v_u^2 \frac{|\bar{\kappa}M_L^* + \kappa^* M_N|^2}{2 (|M_N|^2 - |M_L|^2)^2} & v_u \frac{(\bar{\kappa}^* M_L + \kappa M_N^*)}{|M_N|^2 - |M_L|^2} \\
    - v_u \frac{\kappa_N^*}{M_N^*} & - v_u \frac{(\bar{\kappa} M_L^* + \kappa^* M_N)}{|M_N|^2 - |M_L|^2} & 1 - v_u^2 \frac{|\kappa_N|^2}{2|M_N|^2} - v_u^2 \frac{|\bar{\kappa}^* M_L + \kappa M_N^*|^2}{2 (|M_N|^2 - |M_L|^2)^2}
    \end{pmatrix},
\end{equation}
\begin{equation}
    U_R^{\nu} = \begin{pmatrix}
    1 & 0 & 0 \\
    0 & 1 - v_u^2 \frac{|\kappa^* M_L + \bar{\kappa} M_N^*|^2}{2 (|M_N|^2 - |M_L|^2)^2} & v_u \frac{(\kappa M_L^* + \bar{\kappa}^* M_N)}{|M_N|^2 - |M_L|^2} \\
    0 & - v_u \frac{(\kappa^* M_L + \bar{\kappa} M_N^*)}{|M_N|^2 - |M_L|^2} & 1 - v_u^2 \frac{|\kappa M_L^* + \bar{\kappa}^* M_N|^2}{2 (|M_N|^2 - |M_L|^2)^2}
    \end{pmatrix},
\end{equation}
and
\begin{equation}
    \widetilde{U}_R^{\nu} = U_R^{\nu} \begin{pmatrix}
    1 & 0 & 0 \\
    0 & e^{-i\phi_{M_L}} & 0 \\
    0 & 0 & e^{-i\phi_{M_N}}
    \end{pmatrix}.
\label{eq:m2_phase}
\end{equation}

Note that exact couplings of fermions to $Z,W^{\pm},H,h,A,H,$ and $H^{\pm}$ bosons in the mass eigenstate basis were defined first in Appendices A1 and A2 of \cite{Dermisek:2021ajd}. Those definitions apply identically in this paper, except that one must replace the diagonalization matrices with their complex versions $U_{L}^{e, \nu}$ and $\widetilde{U}_R^{e, \nu}$.

In this limit, couplings of charged fermions to the $Z$ boson are approximated by
\begin{flalign}
    g_L^{Z \mu L} = \frac{g}{2 \cos \theta_W} \left( \frac{v_d^2 \lambda_E (\lambda^* M_E + \bar{\lambda} M_L^*)}{M_E (|M_E|^2 - |M_L|^2)} \right), &&
\end{flalign}
\begin{flalign}
    \begin{split}
        g_R^{Z \mu L} & = \frac{g}{2 \cos \theta_W} \frac{v_d \lambda_L^* }{M_L^*} e^{i \phi_{m_{\mu}}} \left(1 - v_d^2 \frac{|\lambda_L|^2}{2 |M_L|^2} - v_d^2 \frac{|\lambda^* M_L + \bar{\lambda} M_E^*|^2}{2 (|M_L|^2 - |M_E|^2)^2} \right) e^{- i \phi_{M_L}} \\ 
        & \simeq \frac{g}{2 \cos \theta_W} \frac{v_d \lambda_L^*}{|M_L|} e^{i \phi_{m_{\mu}}},
    \end{split} &&
\end{flalign}
\begin{flalign}
    \begin{split}
        g_L^{Z \mu E} & = - \frac{g}{2 \cos \theta_W} \frac{v_d \lambda_E}{M_E} \left(1 - v_d^2 \frac{|\lambda_E|^2}{2 |M_E|^2} - v_d^2 \frac{|\bar{\lambda}^* M_L + \lambda M_E^*|^2}{2 (|M_E|^2 - |M_L|^2)^2} \right) \\ 
        & \simeq -\frac{g}{2 \cos \theta_W} \frac{v_d \lambda_E}{M_E},
    \end{split} &&
\end{flalign}
\begin{flalign}
    g_R^{Z \mu E} = \frac{g}{2 \cos \theta_W} \frac{v_d^2 \lambda_L^* (\bar{\lambda}^* M_E + \lambda M_L^*)}{M_L^* (|M_E|^2 - |M_L|^2)} e ^{i \phi_{m_{\mu}}} e^{-i \phi_{M_E}}. &&
\end{flalign}

Couplings of charged and neutral fermions to the $W^{\pm}$ boson are approximated below:
\begin{flalign}
    \begin{split}
      g_L^{W L \mu} = & \frac{g}{\sqrt{2}} \left[ v_d^2\left(1 - v_u^2 \frac{|\bar{\kappa} M_L^* - \kappa^* M_N|^2}{2 (|M_N|^2 -|M_L|^2)^2} \right) \left(\frac{(\bar{\lambda}^* \lambda_E^* M_L - y_{\mu}^* \lambda_L M_E^*)}{|M_L|^2 M_E^*} \right) \right. \\
      & \left. - v_u^2 \left(\frac{\kappa_N^* (\kappa M_L^* + \bar{\kappa}^* M_N)}{M_L^* (|M_N|^2 - |M_L|^2)^2} \right) \left(1 -v_d^2 \frac{|\lambda_E^2|}{2 |M_E|^2} \right) \right] \\
      & \simeq \frac{g}{\sqrt{2}} \left[- v_u^2 \frac{\kappa_N^*}{M_L^*} \left( \frac{\kappa M_L^* + \bar{\kappa}^* M_N}{|M_N|^2 - |M_L|^2}\right) + v_d^2 \left( \frac{\bar{\lambda}^* \lambda_E^*}{M_L^* M_E^*}\right) \right],
    \end{split} &&
\end{flalign}
\begin{flalign}
    \begin{split}
        g_R^{W L \mu} & = - \frac{g}{\sqrt{2}} \frac{v_d \lambda_L}{M_L} e^{-i\phi_{m_{\mu}}} \left(1 - v_u^2 \frac{|\kappa^* M_L + \bar{\kappa} M_N^*|^2}{2 (|M_N|^2 - M_L^2)^2} \right) e^{i \phi_{M_L}} \\
        & \simeq - \frac{g}{\sqrt{2}} \frac{v_d \lambda_L}{|M_L|} e^{-i \phi_{m_{\mu}}},
    \end{split} &&
\end{flalign}
\begin{flalign}
\begin{split}
    g_L^{W N \mu} & = \frac{g}{\sqrt{2}} \left[\frac{v_u \kappa_N^*}{M_N^*} \left(1 - v_d^2 \frac{|\lambda_E|^2}{2 |M_E|^2} \right) + v_u v_d^2 \left( \frac{(\bar{\kappa} M_L^* + \kappa^* M_N)}{|M_N|^2 - |M_L|^2} \right) \left( \frac{(\bar{\lambda}^* \lambda_E^* M_L - y_{\mu}^* \lambda_L M_E^*)}{|M_L|^2 M_E^*} \right) \right] \\
    & \simeq \frac{g}{\sqrt{2}} \frac{v_u \kappa_N^*}{M_N^*},
\end{split} &&
\end{flalign}
\begin{flalign}
    \begin{split}
        g_R^{W N \mu} = - \frac{g}{\sqrt{2}} \frac{v_u v_d \lambda_L}{M_L} \frac{(\kappa^* M_L + \bar{\kappa} M_N^*)}{|M_N|^2 - |M_L|^2} e^{-i \phi_{m_{\mu}}} e^{i \phi_{M_N}}.
    \end{split} &&
\end{flalign}

Approximate couplings of charged fermions to the light SM Higgs boson $h$ are given below:
\begin{flalign}
    \lambda_{\mu \mu}^h = \left[ y_{\mu} \cos \beta \left(1 - \frac{3 v_d^2 |\lambda_E|^2}{2|M_E|^2} - \frac{3 v_d^2 |\lambda_L|^2}{2 |M_L^2|} \right) + \frac{3 \cos \beta v_d^2 \lambda_L \lambda_E \bar{\lambda}}{M_L M_E} \right] e^{- \phi_{m_{\mu}}}, &&
\end{flalign}
\begin{flalign}
    \lambda_{\mu L}^h = \left[- \frac{\cos \beta v_d \lambda_E \bar{\lambda}}{M_E} - \frac{\cos \beta v_d (\lambda_E \bar{\lambda} M_E^* + \lambda_E \lambda^* M_L)}{|M_E|^2 - |M_L|^2} \right] e^{-i \phi_{M_L}}, &&
\end{flalign}
\begin{flalign}
    \lambda_{\mu E}^h = \cos \beta \lambda_E e^{-i \phi_{M_E}}, &&
\end{flalign}
\begin{flalign}
    \lambda_{L \mu}^h = \cos \beta \lambda_L e^{-i \phi_{m_{\mu}}}, &&
\end{flalign}
\begin{flalign}
    \lambda_{E \mu}^h = \left[- \frac{\cos \beta v_d \lambda_L \bar{\lambda}}{M_L} + \frac{\cos \beta v_d (\lambda_L \bar{\lambda} M_L^* + \lambda_L \lambda^* M_E)}{|M_E|^2 - |M_L|^2} \right] e^{-i \phi_{m_{\mu}}}, &&
\end{flalign}
\begin{flalign}
    \lambda_{LL}^h = \cos \beta \left[\frac{v_d |\lambda_L|^2}{M_L^*} - v_d \frac{\lambda (\overline{\lambda} M_E^* + \lambda^* M_L)}{|M_E|^2 - |M_L|^2} - v_d \frac{\overline{\lambda} (\overline{\lambda}^* M_L + \lambda M_E^*)}{|M_E|^2 - |M_L|^2} \right] e^{-i \phi_{M_L}}, &&
\end{flalign}
\begin{flalign}
    \lambda_{LE}^h = \cos \beta \lambda e^{-i \phi_{M_E}}, &&
\end{flalign}
\begin{flalign}
    \lambda_{EE}^h = \cos \beta \left[\frac{v_d |\lambda_E|^2}{M_E^*} + v_d \frac{\lambda (\overline{\lambda} M_L^* + \lambda^* M_E)}{|M_E|^2 - |M_L|^2} + v_d \frac{\overline{\lambda} (\overline{\lambda}^* M_E + \lambda M_L^*)}{|M_E|^2 - |M_L|^2} \right]  e^{-i \phi_{M_E}}, &&
\end{flalign}
\begin{flalign}
    \lambda_{EL}^h = \cos \beta \overline{\lambda} e^{-i \phi_{M_L}}. &&
\end{flalign}

Couplings to the $CP$-even heavy Higgs field $H$ can be found by simply replacing $\cos \beta \rightarrow \sin \beta$ in the above couplings to $h$. Couplings to the $CP$-odd heavy Higgs field $A$ are given below:
\begin{flalign}
    i \lambda_{\mu L}^A = \left[\frac{\sin \beta v_d \lambda_E \bar{\lambda}}{M_E} - \frac{\sin \beta v_d (\lambda_E \bar{\lambda} M_E^* + \lambda_E \lambda^* M_L)}{|M_E|^2 - |M_L|^2} \right] e^{-i \phi_{M_L}},&&
\end{flalign} 
\begin{flalign}
    i \lambda_{\mu E}^A = \sin \beta \lambda_E e^{-i \phi_{M_E}}, &&
\end{flalign}
\begin{flalign}
    i \lambda_{L \mu}^A = \sin \beta \lambda_L e^{-i \phi_{m_{\mu}}}, &&
\end{flalign}
\begin{flalign}
    i \lambda_{E \mu}^A =  \left[\frac{\sin \beta v_d \lambda_L \bar{\lambda}}{M_L} + \frac{\sin \beta v_d (\lambda_L \bar{\lambda} M_L^* + \lambda_L \lambda^* M_E)}{|M_E|^2 - |M_L|^2} \right] e^{-i \phi_{m_{\mu}}}, &&
\end{flalign}
\begin{flalign}
    i \lambda_{LL}^A = \sin \beta \left[\frac{v_d |\lambda_L|^2}{M_L^*} - v_d \frac{\lambda (\overline{\lambda} M_E^* + \lambda^* M_L)}{|M_E|^2 - |M_L|^2} + v_d \frac{\overline{\lambda} (\overline{\lambda}^* M_L + \lambda M_E^*)}{|M_E|^2 - |M_L|^2} \right] e^{-i \phi_{M_L}}, &&
\end{flalign}
\begin{flalign}
    i\lambda_{LE}^A = \sin \beta \lambda e^{-i \phi_{M_E}}, &&
\end{flalign}
\begin{flalign}
    i\lambda_{EE}^A = \sin \beta \left[\frac{v_d |\lambda_E|^2}{M_E^*} + v_d \frac{\lambda (\overline{\lambda} M_L^* + \lambda^* M_E)}{|M_E|^2 - |M_L|^2} - v_d \frac{\overline{\lambda} (\overline{\lambda}^* M_E + \lambda M_L^*)}{|M_E|^2 - |M_L|^2} \right]  e^{-i \phi_{M_E}}, &&
\end{flalign}
\begin{flalign}
    i\lambda_{EL}^A = - \sin \beta \overline{\lambda} e^{-i \phi_{M_L}}. &&
\end{flalign}

Finally, the approximate couplings of fermions to the charged Higgs field $H^{\pm}$ are given by
\begin{flalign}
    \lambda_{L \mu}^{H^{\pm}} = - \sin \beta \lambda_L e^{-i \phi_{m_{\mu}}}, &&
\end{flalign}
\begin{flalign}
    \lambda_{N \mu}^{H^{\pm}} = \left[\frac{\cos \beta v_d \lambda_L \bar{\kappa}}{M_L} - \frac{\sin \beta v_u (\lambda_L \bar{\kappa} M_L^* + \lambda_L \kappa^* M_N)}{|M_N|^2 - |M_L|^2} \right] e^{-i \phi_{m_{\mu}}}, &&
\end{flalign}
\begin{flalign}
    \lambda_{LL}^{H^{\pm}} = - \sin \beta \left[\frac{v_d |\lambda_L|^2}{M_L^*} - \frac{v_d \lambda (\overline{\lambda}M_E^* + \lambda^* M_L)}{|M_E|^2 - |M_L|^2} + \frac{v_u \overline{\kappa}(\overline{\kappa}^* M_L + \kappa M_N^*)}{\tan \beta (|M_N|^2 - |M_L|^2)} \right] e^{-i \phi_{M_L}}, &&
\end{flalign}
\begin{flalign}
    \lambda_{LE}^{H^{\pm}} = -\sin \beta \lambda e^{-i \phi_{M_E}}, &&
\end{flalign}
\begin{flalign}
    \lambda_{N E}^{H^{\pm}} = - \sin \beta \left[ \frac{v_u \lambda_E \kappa_N^*}{M_N^*} + \frac{v_u \lambda(\overline{\kappa} M_L^* + \kappa^* M_N)}{|M_N|^2 - |M_L|^2} + \frac{v_d \overline{\kappa}(\overline{\lambda}^* M_E + \lambda M_L^*)}{\tan \beta (|M_E|^2 - |M_L|^2)} \right] e^{-i \phi_{M_E}}, &&
\end{flalign}
\begin{flalign}
    \lambda_{N L}^{H^{\pm}} = - \cos \beta \overline{\kappa} e^{-i \phi_{M_L}}, &&
\end{flalign}
\begin{flalign}
    \lambda_{\mu L}^{H^{\pm}} = \left[\frac{\sin \beta v_d \lambda_E \bar{\lambda}}{M_E} + \frac{ \cos \beta v_u (\kappa_N \bar{\kappa} M_N^* + \kappa_N \kappa^* M_L)}{|M_N|^2 - |M_L|^2} \right] e^{-i \phi_{M_L}}, &&
\end{flalign}
\begin{flalign}
    \lambda_{\mu N}^{H^{\pm}} = - \cos \beta \kappa_N e^{-i \phi_{M_N}}, &&
\end{flalign}
\begin{flalign}
    \lambda_{L L}^{H^{\pm}} = - \cos \beta \left[ -\frac{v_u \kappa(\kappa^* M_L + \overline{\kappa} M_N^*)}{|M_N|^2 - |M_L|^2} - \frac{v_d \tan \beta \overline{\lambda}(\overline{\lambda}^* M_L + \lambda M_E^*)}{|M_E|^2 - |M_L|^2} \right] e^{-i \phi_{M_L}}, &&
\end{flalign}
\begin{flalign}
    \lambda_{E L}^{H^{\pm}} = - \sin \beta \overline{\lambda} e^{- \phi_{M_L}}, &&
\end{flalign}
\begin{flalign}
    \lambda_{E N}^{H^{\pm}} = - \cos \beta \left[\frac{v_d \lambda_E^* \kappa_N}{M_E^*} + \frac{v_d \kappa (\overline{\lambda} M_L^* + \lambda^* M_E)}{|M_E|^2 - |M_L|^2} + \frac{v_u \tan 
    \beta \overline{\lambda}(\kappa M_L^* + \overline{\kappa}^* M_N)}{|M_N|^2 - |M_L|^2}\right] e^{-i\phi_{M_N}}, &&
\end{flalign}
\begin{flalign}
    \lambda_{L N}^{H^{\pm}} = - \cos \beta \kappa e^{-i \phi_{M_N}}. &&
\end{flalign}

\subsection{$\textbf{2}_{-1/2} \oplus \textbf{3}_{-1}$}

In the same expansion limit as for the main model above, the diagonalization matrices become
\begin{equation}
    U_L^e = \begin{pmatrix}
        1 - v_d^2 \frac{|\lambda_E|^2}{2 |M_E|^2} & v_d^2\left( \frac{\lambda_E \lambda^* M_L + \lambda_E \overline{\lambda} M_E^*}{M_L (|M_L|^2 - |M_E|^2)} + \frac{y_{\mu} \lambda_L^*}{|M_L|^2} \right) & - v_d \frac{\lambda_E}{M_E} \\ 
        - v_d^2 \frac{y_{\mu}^* \lambda_L}{|M_L|^2} + v_d^2 \frac{\lambda_E^* \overline{\lambda}^* }{M_L^* M_E^*} & 1 - v_d^2 \frac{|\overline{\lambda} M_L^* + \lambda^* M_E|^2}{2 (|M_L|^2 - |M_E|^2)^2} & v_d \frac{(\overline{\lambda}^* M_L + \lambda M_E^*)}{|M_L|^2 - |M_E|^2} \\
        v_d \frac{\lambda_E^*}{M_E^*} & - v_d \frac{(\overline{\lambda} M_L^* + \lambda^* M_E)}{|M_L|^2 - |M_E|^2} & 1 - v_d^2 \frac{|\lambda_E|^2}{2|M_E|^2} - v_d^2 \frac{|\overline{\lambda}^* M_L + \lambda M_E^*|^2}{2 (|M_L|^2 - |M_E|^2)^2}
    \end{pmatrix},
\end{equation}
\begin{equation}
    U_R^e = \begin{pmatrix}
        1 - v_d^2\frac{|\lambda_L|^2}{2 |M_L|^2} & v_d \frac{\lambda_L^*}{M_L^*} & v_d^2 \left( \frac{\lambda_L^* \lambda M_E^* + \lambda_L^* \overline{\lambda}^* M_L}{M_E^* (|M_L|^2 - |M_E|^2)} - \frac{y_{\mu}^* \lambda_E}{|M_E|^2}\right) \\
        - v_d \frac{\lambda_L}{M_L} & 1 - v_d^2 \frac{|\lambda_L|^2}{2 |M_L|^2} - v_d^2 \frac{|\overline{\lambda} M_E^* + \lambda^* M_L|^2}{2 (|M_L|^2 - |M_E|^2)^2} & v_d \frac{(\lambda M_L^* + \overline{\lambda}^* M_E)}{|M_L|^2 - |M_E|^2} \\
        v_d^2 \frac{y_{\mu} \lambda_E^*}{|M_E|^2} - v_d^2 \frac{\lambda_L \overline{\lambda}}{M_L M_E} & - v_d \frac{(\lambda^* M_L + \overline{\lambda} M_E^*)}{|M_L|^2 - |M_E|^2} & 1 - v_d^2 \frac{|\lambda M_L^* + \overline{\lambda}^* M_E|^2}{2(|M_L|^2 - |M_E|^2)^2}
    \end{pmatrix},
\end{equation}
\begin{equation}
    \widetilde{U}_R^e = U_R^e  \begin{pmatrix}
    e^{-i \phi_{m_{\mu}}} & 0 & 0 \\
    0 & e^{-i\phi_{M_L}} & 0 \\
    0 & 0 & e^{-i\phi_{M_E}}
    \end{pmatrix}
\end{equation}
\begin{equation}
    U_L^{\nu} = \begin{pmatrix}
        1 - v_d^2\frac{|\lambda_E|^2}{|M_E|^2} & 2 v_d^2 \left( \frac{\lambda_E \lambda^* M_L + \lambda_E \overline{\lambda} M_E^* }{M_L (|M_L|^2 - |M_E|^2)}\right)  & \sqrt{2} v_d \frac{\lambda_E}{M_E} \\ 
        2 v_d^2 \frac{\lambda_E^* \overline{\lambda}^*}{M_L^* M_E^*} & 1 - v_d^2\frac{|\overline{\lambda} M_L^* + \lambda^* M_E|^2}{(|M_L|^2 - |M_E|^2)^2} & -\sqrt{2} v_d \frac{(\overline{\lambda}^* M_L + \lambda M_E^*)}{|M_L|^2 - |M_E|^2} \\
        - \sqrt{2} v_d \frac{\lambda_E^*}{M_E^*} & \sqrt{2} v_d \frac{(\overline{\lambda} M_L^* + \lambda^* M_E)}{|M_L|^2 - |M_E|^2} & 1 - v_d^2 \frac{|\lambda_E|^2}{|M_E|^2} - v_d^2 \frac{|\overline{\lambda} M_L^* + \lambda^* M_E|^2}{(|M_L|^2 - |M_E|^2)^2}
    \end{pmatrix}, 
\end{equation}
\begin{equation}
    U_R^{\nu} = \begin{pmatrix}
        1 & 0 & 0 \\
        0 & 1 - v_d^2 \frac{|\lambda^* M_L + \overline{\lambda} M_E^*|^2}{(|M_L|^2 - |M_E|^2)^2} & -\sqrt{2} v_d \frac{(\lambda M_L^* + \overline{\lambda}^* M_E)}{|M_L|^2 - |M_E|^2} \\
        0 & \sqrt{2} v_d \frac{(\lambda^* M_L + \overline{\lambda} M_E^*)}{|M_L|^2 - |M_E|^2} & 1 - v_d^2 \frac{|\lambda M_L^* + \overline{\lambda}^* M_E|^2}{(|M_L|^2 - |M_E|^2)^2}
    \end{pmatrix},
\end{equation}
and
\begin{equation}
    \widetilde{U}_R^{\nu} = U_R^{\nu} \begin{pmatrix}
    1 & 0 & 0 \\
    0 & e^{-i\phi_{M_L}} & 0 \\
    0 & 0 & e^{-i\phi_{M_E}}
    \end{pmatrix}.
\end{equation}

\subsection{$\textbf{2}_{-3/2} \oplus \textbf{1}_{-1}$}

The diagonalization matrices for this model are the same as in the $\textbf{2}_{-1/2} \oplus \textbf{1}_{-1}$ model, Eqs.~(\ref{eq:ULe})-(\ref{eq:m1_phase}), with the exception that $\lambda \rightarrow - \lambda$.

\subsection{$\textbf{2}_{-3/2} \oplus \textbf{3}_{-1}$}

In the same mass limit, the diagonalization matrices for this representation are
\begin{equation}
    U_L^e = \begin{pmatrix}
        1 - v_d^2 \frac{|\lambda_E|^2}{2|M_E|^2} & v_d^2 \left(\frac{y_{\mu} \lambda_L^*}{|M_L|^2} - \frac{\lambda^* \lambda_E M_L - \lambda_E \overline{\lambda} M_E^*}{M_L (|M_L|^2 - |M_E|^2)} \right) & - v_d \frac{\lambda_E}{M_E} \\ 
        - v_d^2\frac{y_{\mu}^* \lambda_L}{|M_L|^2} + v_d^2\frac{\overline{\lambda}^* \lambda_E^*}{M_L^* M_E^*} & 1 - v_d^2 \frac{|\overline{\lambda} M_L^* - \lambda^* M_E|^2}{2(|M_L|^2 - |M_E|^2)^2} &  v_d \frac{(\overline{\lambda} M_L - \lambda M_E^* )}{|M_L|^2 - |M_E|^2} \\
        v_d \frac{\lambda_E^*}{M_E^*} & - v_d \frac{(\overline{\lambda} M_L^* - \lambda^* M_E)}{|M_L|^2 - |M_E|^2} & 1 - v_d^2 \frac{|\lambda_E|^2}{2|M_E|^2} - v_d^2 \frac{|\overline{\lambda}^* M_L - \lambda M_E^*|^2}{2(|M_L|^2 - |M_E|^2)^2}
    \end{pmatrix}, 
\end{equation}
\begin{equation}
    U_R^e = \begin{pmatrix}
        1 - v_d^2 \frac{|\lambda_L|^2}{2|M_L|^2} & v_d \frac{\lambda_L^*}{M_L^*} & -v_d^2 \left(\frac{y_{\mu}^* \lambda_E}{|M_E|^2} + \frac{\lambda \lambda_L^* M_E^* - \lambda_L^* \overline{\lambda}^* M_L}{M_E^*(|M_L|^2 - |M_E|^2)} \right) \\
        - v_d \frac{\lambda_L}{M_L} & 1 - v_d^2 \frac{|\lambda_L|^2}{2 |M_L|^2} - v_d^2 \frac{|\lambda^* M_L - \overline{\lambda} M_E^*|^2}{2(|M_L|^2 - |M_E|^2)^2} & -v_d \frac{(\lambda M_L^* - \overline{\lambda}^* M_E)}{|M_L|^2 - |M_E|^2} \\
        v_d^2 \frac{y_{\mu} \lambda_E^*}{|M_E|^2} - v_d^2 \frac{\overline{\lambda} \lambda_L}{M_L M_E} & v_d \frac{(\lambda^* M_L - \overline{\lambda} M_E^*)}{|M_L|^2 - |M_E|^2} & 1 - v_d^2 \frac{|\lambda M_L^* - \overline{\lambda}^* M_E|^2}{2 (|M_L|^2 - |M_E|^2)^2}
    \end{pmatrix}, 
\end{equation}
and
\begin{equation}
    \widetilde{U}_R^e = U_R^e \begin{pmatrix}
    e^{-i \phi_{m_{\mu}}} & 0 & 0 \\
    0 & e^{-i\phi_{M_L}} & 0 \\
    0 & 0 & e^{-i\phi_{M_E}}
    \end{pmatrix}.
\end{equation}
The diagonalization matrices for doubly charged fermions are
\begin{equation}
    U_L^{e^{--}} = \begin{pmatrix}
        1 & 0 & 0 \\
        0 & 1 - v_d^2 \frac{|\lambda^* M_L - \overline{\lambda} M_E^*|^2}{(|M_L|^2 - |M_E|^2)^2} & \sqrt{2}v_d \frac{(\overline{\lambda}^* M_L - \lambda M_E^*)}{|M_L|^2 - |M_E|^2} \\
        0 & - \sqrt{2} v_d \frac{(\overline{\lambda} M_L^* - \lambda^* M_E)}{|M_L|^2 - |M_E|^2} & 1 - v_d^2 \frac{|\lambda M_L^* - \overline{\lambda}^* M_E|^2}{(|M_L|^2 - |M_E|^2)^2}
    \end{pmatrix}, 
\end{equation}
\begin{equation}
    U_R^{e^{--}} = \begin{pmatrix}
        1 & 0 & 0 \\
        0 & 1 - v_d^2 \frac{|\lambda^* M_L - \overline{\lambda} M_E^*|^2}{(|M_L|^2 - |M_E|^2)^2} & - \sqrt{2}v_d \frac{(\lambda M_L^* - \overline{\lambda}^* M_E)}{|M_L|^2 - |M_E|^2} \\
        0 & \sqrt{2} v_d \frac{(\lambda^* M_L - \overline{\lambda} M_E^*)}{|M_L|^2 - |M_E|^2} & 1 - v_d^2 \frac{|\lambda M_L^* - \overline{\lambda}^* M_E|^2}{(|M_L|^2 - |M_E|^2)^2}
    \end{pmatrix}, 
\end{equation}
as well as 
\begin{equation}
    \widetilde{U}_R^{e^{--}} = U_R^{e^{--}} \begin{pmatrix}
        1 & 0 & 0 \\
        0 & e^{-i \phi_{M_L}} & 0 \\ 0 & 0 & e^{-i \phi_{M_E}}
    \end{pmatrix}.
\end{equation}

\subsection{$\textbf{2}_{-1/2} \oplus \textbf{3}_{0}$}

Evaluating the same mass limit as above, the diagonalization matrices are 
\begin{equation}
    U_L^e = \begin{pmatrix}
        1 - v_d^2 \frac{|\lambda_E|^2}{|M_E|^2} & v_d^2 \left(\frac{2( \lambda_E \lambda^* M_L - \lambda_E \overline{\lambda} M_E^*)}{M_L (|M_L|^2 - |M_E|^2)} + \frac{y_{\mu} \lambda_L^*}{|M_L|^2} \right) & \sqrt{2} v_d \frac{\lambda_E}{|M_E} \\
        - 2 v_d^2 \frac{\overline{\lambda}^* \lambda_E^*}{M_L^* M_E^*} - v_d^2 \frac{y_{\mu}^* \lambda_L}{|M_L|^2} & 1 - v_d^2 \frac{|\lambda^* M_E - \overline{\lambda} M_L^*|^2}{(|M_L|^2 - |M_E|^2)^2} & \sqrt{2} v_d  \frac{(\overline{\lambda}^* M_L - \lambda M_E^*)}{|M_L|^2 - |M_E|^2} \\
        - \sqrt{2} v_d \frac{\lambda_E^*}{M_E^*} & - \sqrt{2} v_d \frac{(\overline{\lambda} M_L^* - \lambda^* M_E)}{|M_L|^2 - |M_E|^2} & 1 - v_d^2 \frac{|\lambda_E|^2}{|M_E|^2} - v_d^2 \frac{|\overline{\lambda}^* M_L - \lambda M_E^*|^2}{(|M_L|^2 - |M_E|^2)^2} 
    \end{pmatrix},
\end{equation}
\begin{equation}
    U_R^e = \begin{pmatrix}
        1 - v_d^2 \frac{|\lambda_L|^2}{2 |M_L|^2} & v_d \frac{\lambda_L^*}{M_L^*} & \sqrt{2} v_d^2 \left( \frac{y_{\mu}^* \lambda_E}{|M_E|^2} + \frac{(\lambda_L^*\overline{\lambda}^* M_L - \lambda_L^* \lambda M_E^*)}{M_E^* (|M_L|^2 - |M_E|^2)} \right) \\
        - v_d \frac{\lambda_L}{M_L} & 1 - v_d^2 \frac{|\lambda_L|^2}{2 |M_L|^2} - v_d^2 \frac{|\lambda^* M_L - \overline{\lambda} M_E^*|^2}{(|M_L|^2 - |M_E|^2)^2} & \sqrt{2} v_d \frac{(\overline{\lambda}^* M_E - \lambda M_L^*)}{|M_L|^2 - |M_E|^2} \\
        - \sqrt{2} v_d^2 \frac{\lambda_L \overline{\lambda}}{M_L M_E} - \sqrt{2} v_d^2 \frac{y_{\mu} \lambda_E^*}{|M_E|^2} & - \sqrt{2} v_d \frac{(\overline{\lambda} M_E^* - \lambda^* M_L)}{|M_L|^2 - |M_E|^2} & 1 - v_d^2 \frac{|\lambda M_L^* - \overline{\lambda}^* M_E|^2}{(|M_L|^2 - |M_E|^2)^2}
    \end{pmatrix},
\end{equation}
\begin{equation}
    \widetilde{U}_R^e = U_R^e \begin{pmatrix}
        e^{-i\phi_{m_{\mu}}} & 0 & 0 \\
        0 & e^{-i\phi_{M_L}} & 0 \\
        0 & 0 & e^{-i\phi_{M_E}}
    \end{pmatrix}.
\end{equation}
The diagonalization matrices for the neutral sector are
\begin{equation}
    U_L^{\nu} = \begin{pmatrix}
        1 - v_d^2 \frac{|\lambda_E|^2}{2 |M_E|^2} & v_d^2 \frac{\lambda_E (\lambda^* M_L - \overline{\lambda} M_E^*)}{M_L(|M_L|^2 - |M_E|^2)} & v_d \frac{\lambda_E}{M_E} \\
        - v_d^2 \frac{\lambda_E^* \overline{\lambda}^*}{M_L^* M_E^*} & 1 - v_d^2 \frac{|\overline{\lambda} M_L^* - \lambda^* M_E|^2}{2(|M_L|^2 - |M_E|^2)^2} & v_d \frac{(\overline{\lambda}^* M_L - \lambda M_E^*)}{|M_L|^2 - |M_E|^2} \\ 
        - v_d \frac{\lambda_E^*}{M_E^*} & - v_d \frac{(\overline{\lambda} M_L^* - \lambda^* M_E)}{|M_L|^2 - |M_E|^2} & 1 - v_d^2 \frac{|\lambda_E|}{2 |M_E|^2} - v_d^2 \frac{|\overline{\lambda}^* M_L - \lambda M_E^*|^2}{2(|M_L|^2 - |M_E|^2)^2}
    \end{pmatrix},
\end{equation}
\begin{equation}
    U_R^{\nu} = \begin{pmatrix}
        1 & 0 & 0 \\
        0 & 1 - v_d^2 \frac{|\lambda^* M_L - \overline{\lambda} M_E^*|^2}{2(|M_L|^2 - |M_E|^2)^2} & -v_d \frac{(\lambda M_L^* - \overline{\lambda}^* M_E)}{|M_L|^2 - |M_E|^2} \\
        0 & v_d \frac{(\lambda^* M_L - \overline{\lambda} M_E^*)}{|M_L|^2 - |M_E|^2} & 1 - v_d^2 \frac{|\lambda M_L^* - \overline{\lambda}^* M_E|^2}{2(|M_L|^2 - |M_E|^2)^2}
    \end{pmatrix}, 
\end{equation}
\begin{equation}
    \widetilde{U}_R^{\nu} = U_R^{\nu} \begin{pmatrix}
        1 & 0 & 0 \\
        0 & e^{-i \phi_{M_L}} & 0 \\
        0 & 0 & e^{-i \phi_{M_E}}
    \end{pmatrix}.
\end{equation}

\section{Contributions to the dipole moments in the Higgs basis}
\label{sec:tree_contributions}

In this appendix, we list the full contributions of $C_{\mu B}$ and $C_{\mu W}$ in the context of $SU(2)$ singlets, doublets, and triplets discussed in \cite{Kannike:2011ng}. For compactness, we use the shorthand notation $\cos \beta \equiv c_{\beta}$ and $\sin \beta \equiv s_{\beta}$. Loop functions can be found in Eqs.~(\ref{eq:F_fun})-(\ref{eq:J_fun}) in this appendix and their useful limits can be found in Eqs~(\ref{eq:F_lim})-(\ref{eq:J_lim}).

For the main representation of the paper, $\textbf{2}_{-1/2} \oplus \textbf{1}_{-1}$, and translating the Lagrangian of Eq.~(\ref{eq:vll_lag}) to the Higgs basis while omitting up-type couplings, the Wilson coefficients are
\begin{equation}
    \begin{split}
        C_{\mu B} = &- \left( \frac{\lambda_L \lambda_E \bar{\lambda}}{64 \pi^2 M_L M_E }\right) g' c^2_{\beta} \left[Y_E F(x_E^{(1)}) + 2 Y_L F(x_L^{(1)}) - \frac{1}{2} G(x_E^{(1)}) + G(x_L^{(1)}) \right] \\
        & - \left( \frac{\lambda_L \lambda_E \bar{\lambda}}{64 \pi^2 M_L M_E }\right) g' s^2_{\beta} \left[Y_E F(x_E^{(2)}) + 2 Y_L F(x_L^{(2)}) - \frac{1}{2} G(x_E^{(2)}) + G(x_L^{(2)}) \right],
    \end{split}
\end{equation}
and 
\begin{equation}
    C_{\mu W} = \left( \frac{\lambda_L \lambda_E \bar{\lambda}}{128 \pi^2 M_L M_E } \right) g c^2_{\beta} G(x_E^{(1)}) + \left( \frac{\lambda_L \lambda_E \bar{\lambda}}{128 \pi^2 M_L M_E } \right) g s^2_{\beta} G(x_E^{(2)}),
\end{equation}
where $Y_{L,E}$ are the hypercharges of the new fermions and $x_{L,E}^{(1,2)} = M_{L,E}^2 / M_{1,2}^2$. In the limit where $M_{L,E}^2 \gg M_1^2$ and $M_{L,E}^2 \simeq M_2^2$ while inserting their respective hypercharges, we find
\begin{equation}
    C_{\mu \gamma} = \left( \frac{\lambda_L \lambda_E \bar{\lambda}}{64 \pi^2 M_L M_E } \right) e c^2_{\beta} \left(1 + \tan^2 \beta \right).
\end{equation}
If we consider the limit where $M_{L,E}^2 \gg M_1^2$ and $M_2^2$, we find the same expression.

Next, the Wilson coefficients for the $\textbf{2}_{-1/2} \oplus \textbf{3}_{-1}$ representation, using the Lagrangian of Eq.~(\ref{eq:L12_31}) in the Higgs basis, are
\begin{equation}
    \begin{split}
        C_{\mu B} = & - \left(\frac{3 \lambda_L \lambda_E \bar{\lambda}}{64 \pi^2 M_L M_E} \right) g' c^2_{\beta} \left[ Y_E F(x_E^{(1)}) - \frac{1}{2} G(x_E^{(1)}) \right] \\
        & -\left(\frac{3 \lambda_L \lambda_E \bar{\lambda}}{64 \pi^2 M_L M_E} \right) g' s^2_{\beta} \left[ Y_E F(x_E^{(2)}) - \frac{1}{2} G(x_E^{(2)}) \right],
    \end{split}
\end{equation}
and 
\begin{equation}
    \begin{split}
        C_{\mu W} = & - \left(\frac{\lambda_L \lambda_E \bar{\lambda}}{64 \pi^2 M_L M_E} \right) g c^2_{\beta} \left[ 2 F(x_E^{(1)}) + F(x_L^{(1)}) + \frac{1}{2} G(x_E^{(1)}) + G(x_L^{(1)}) \right] \\
        & -\left(\frac{\lambda_L \lambda_E \bar{\lambda}}{64 \pi^2 M_L M_E} \right) g s^2_{\beta} \left[ 2 F(x_E^{(2)}) + F(x_L^{(2)}) + \frac{1}{2} G(x_E^{(2)}) + G(x_L^{(2)}) \right].
    \end{split}
\end{equation}
Taking the same mass limits $x_{L,E}^{(1)} \rightarrow \infty$ and $x_{L,E}^{(2)} \rightarrow 1$ while inserting respective hypercharges, we find 
\begin{equation}
    C_{\mu \gamma} = \left( \frac{\lambda_L \lambda_E \bar{\lambda}}{64 \pi^2 M_L M_E } \right) e c^2_{\beta} \left(9 + 5\tan^2 \beta \right).
\end{equation}
Considering the other limit where $x_{L,E}^{(1,2)} \rightarrow \infty$, we have
\begin{equation}
    C_{\mu \gamma} = 9 \left( \frac{\lambda_L \lambda_E \bar{\lambda}}{64 \pi^2 M_L M_E } \right) e c^2_{\beta} \left(1 + \tan^2 \beta \right).
\end{equation}

For the $\textbf{2}_{-3/2} \oplus \textbf{1}_{-1}$ representation, using the Lagrangian of Eq.~(\ref{eq:L32_11}) in the Higgs basis, the Wilson coefficients are
\begin{equation}
    \begin{split}
            C_{\mu B} = - \left( \frac{\lambda_L \lambda_E \bar{\lambda}}{64 \pi^2 M_L M_E} \right) g' c_{\beta}^2 & \left[ Y_E \left(A(x_E^{(1)}, x_L^{(1)}) + B(x_E^{(1)}, x_L^{(1)}) \right) \right. \\ 
            & \left. + Y_L \left( A(x_L^{(1)}, x_E^{(1)}) + B(x_L^{(1)}, x_E^{(1)}) + 2 F(x_L^{(1)}) \right) \right. \\
            & \left. - C(x_L^{(1)}, x_E^{(1)}) - G(x_L^{(1)})\right] \\
            - \left( \frac{\lambda_L \lambda_E \bar{\lambda}}{64 \pi^2 M_L M_E} \right) g' s_{\beta}^2 & \left[ Y_E \left(A(x_E^{(2)}, x_L^{(2)}) + B(x_E^{(2)}, x_L^{(2)}) \right) \right. \\ 
            & \left. + Y_L \left( A(x_L^{(2)}, x_E^{(2)}) + B(x_L^{(2)}, x_E^{(2)}) + 2 F(x_L^{(2)}) \right) \right. \\
            & \left. - C(x_L^{(2)}, x_E^{(2)}) - G(x_L^{(2)})\right]
            \\ + \left( \frac{\lambda_L \lambda_E \lambda^*}{64 \pi^2 M_L M_E} \right) g' c_{\beta}^2 & \left[ Y_E \left( 2 I(x_E^{(1)}, x_L^{(1)}) + \left(\frac{M_E}{M_L}  \right) A(x_E^{(1)}, x_L^{(1)}) \right) \right. \\
            & + \left. Y_L \left( 2 I(x_L^{(1)}, x_E^{(1)}) + \left(\frac{M_L}{M_E}  \right) A(x_L^{(1)}, x_E^{(1)}) \right) + J(x_L^{(1)}, x_E^{(1)}) \right] \\
            + \left( \frac{\lambda_L \lambda_E \lambda^*}{64 \pi^2 M_L M_E} \right) g' s_{\beta}^2 & \left[ Y_E \left( 2 I(x_E^{(2)}, x_L^{(2)}) + \left(\frac{M_E}{M_L}  \right) A(x_E^{(2)}, x_L^{(2)}) \right) \right. \\
            & + \left. Y_L \left( 2 I(x_L^{(2)}, x_E^{(2)}) + \left(\frac{M_L}{M_E}  \right) A(x_L^{(2)}, x_E^{(2)}) \right) + J(x_L^{(2)}, x_E^{(2)}) \right],
    \end{split}
\end{equation}
and 
\begin{equation}
    \begin{split}
            C_{\mu W} =  \left( \frac{\lambda_L \lambda_E \bar{\lambda}}{64 \pi^2 M_L M_E} \right) g c_{\beta}^2 & \left[  \frac{1}{2}\left(A(x_L^{(1)}, x_E^{(1)}) + B(x_L^{(1)}, x_E^{(1)}) \right) + C(x_L^{(1)},x_E^{(1)}) \right] \\
            + \left( \frac{\lambda_L \lambda_E \bar{\lambda}}{64 \pi^2 M_L M_E} \right) g s_{\beta}^2 & \left[ \frac{1}{2}\left(A(x_L^{(2)}, x_E^{(2)}) + B(x_L^{(2)}, x_E^{(2)}) \right) + C(x_L^{(2)},x_E^{(2)}) \right]
            \\ - \left( \frac{\lambda_L \lambda_E \lambda^*}{64 \pi^2 M_L M_E} \right) g c_{\beta}^2 & \left[ \frac{1}{2} \left(2 I(x_L^{(1)}, x_E^{(1)}) + \left( \frac{M_L}{M_E} \right) A(x_{L}^{(1)}, x_E^{(1)}) \right) - J(x_L^{(1)}, x_E^{(1)}) \right] \\
            - \left( \frac{\lambda_L \lambda_E \lambda^*}{64 \pi^2 M_L M_E} \right) g s_{\beta}^2 & \left[ \frac{1}{2} \left(2 I(x_L^{(2)}, x_E^{(2)}) + \left( \frac{M_L}{M_E} \right) A(x_{L}^{(2)}, x_E^{(2)}) \right) - J(x_L^{(2)}, x_E^{(2)}) \right].
    \end{split}
\end{equation}
Note that the combination $2 I(x,y) + \sqrt{x/ y} A(x,y)$ multiplying the $\lambda$ coupling vanishes in the limit $x,y \rightarrow \infty$. In the limit $x_{L,E}^{(1)} \rightarrow \infty$ and $x_{L,E}^{(2)} \rightarrow 1$,
\begin{equation}
        C_{\mu \gamma} = \left( \frac{\lambda_L \lambda_E \bar{\lambda}}{64 \pi^2 M_L M_E} \right) e c_{\beta}^2 \left(5 + \frac{17}{6} \tan^2 \beta \right) + \left( \frac{\lambda_L \lambda_E \lambda^*}{64 \pi^2 M_L M_E} \right) e c_{\beta}^2 \left(\frac{1}{6}\tan^2 \beta \right),
\label{eq:cmugamma_lam1}
\end{equation}
and in the limit where $x_{L,E}^{(1,2)} \rightarrow \infty$, we find
\begin{equation}
        C_{\mu \gamma} = 5 \left( \frac{\lambda_L \lambda_E \bar{\lambda}}{64 \pi^2 M_L M_E} \right) e c_{\beta}^2 \left(1 + \tan^2 \beta \right).
\end{equation}

Likewise, using the Lagrangian of Eq.~(\ref{eq:L32_31}) in the Higgs basis for the $\textbf{2}_{-3/2} \oplus \textbf{3}_{-1}$ representation, the Wilson coefficients are
\begin{equation}
    \begin{split}
            C_{\mu B} = - \left( \frac{3 \lambda_L \lambda_E \bar{\lambda}}{64 \pi^2 M_L M_E} \right) g' c_{\beta}^2 & \left[ Y_E \left(A(x_E^{(1)}, x_L^{(1)}) + B(x_E^{(1)}, x_L^{(1)}) \right) \right. \\ 
            & \left. + Y_L \left( A(x_L^{(1)}, x_E^{(1)}) + B(x_L^{(1)}, x_E^{(1)}) \right) - C(x_L^{(1)}, x_E^{(1)}) \right] \\
            - \left( \frac{3 \lambda_L \lambda_E \bar{\lambda}}{64 \pi^2 M_L M_E} \right) g' s_{\beta}^2 & \left[ Y_E \left(A(x_E^{(2)}, x_L^{(2)}) + B(x_E^{(2)}, x_L^{(2)}) \right) \right. \\ 
            & \left. + Y_L \left( A(x_L^{(2)}, x_E^{(2)}) + B(x_L^{(2)}, x_E^{(2)}) \right) - C(x_L^{(2)}, x_E^{(2)}) \right]
            \\ + \left( \frac{3 \lambda_L \lambda_E \lambda^*}{64 \pi^2 M_L M_E} \right) g' c_{\beta}^2 & \left[ Y_E \left( 2 I(x_E^{(1)}, x_L^{(1)}) + \left(\frac{M_E}{M_L}  \right) A(x_E^{(1)}, x_L^{(1)}) \right) \right. \\
            & + \left. Y_L \left( 2 I(x_L^{(1)}, x_E^{(1)}) + \left(\frac{M_L}{M_E}  \right) A(x_L^{(1)}, x_E^{(1)}) \right) + J(x_L^{(1)}, x_E^{(1)}) \right] \\
            + \left( \frac{3 \lambda_L \lambda_E \lambda^*}{64 \pi^2 M_L M_E} \right) g' s_{\beta}^2 & \left[ Y_E \left( 2 I(x_E^{(2)}, x_L^{(2)}) + \left(\frac{M_E}{M_L}  \right) A(x_E^{(2)}, x_L^{(2)}) \right) \right. \\
            & + \left. Y_L \left( 2 I(x_L^{(2)}, x_E^{(2)}) + \left(\frac{M_L}{M_E}  \right) A(x_L^{(2)}, x_E^{(2)}) \right) + J(x_L^{(2)}, x_E^{(2)}) \right],
    \end{split}
\end{equation}
\begin{equation}
    \begin{split}
            C_{\mu W} = - \left( \frac{\lambda_L \lambda_E \bar{\lambda}}{64 \pi^2 M_L M_E} \right) g c_{\beta}^2 & \left[ - F(x_L^{(1)}) - G(x_L^{(1)}) + 2 \left(A(x_E^{(1)}, x_L^{(1)}) + B(x_E^{(1)}, x_L^{(1)} \right) \right. \\ 
            & \left. + \frac{1}{2}\left(A(x_L^{(1)}, x_E^{(1)}) + B(x_L^{(1)}, x_E^{(1)}) \right) + C(x_L^{(1)},x_E^{(1)}) \right] \\
            - \left( \frac{\lambda_L \lambda_E \bar{\lambda}}{64 \pi^2 M_L M_E} \right) g s_{\beta}^2 & \left[ - F(x_L^{(2)}) - G(x_L^{(2)}) + 2 \left(A(x_E^{(2)}, x_L^{(2)}) + B(x_E^{(2)}, x_L^{(2)}) \right) \right. \\ 
            & \left. + \frac{1}{2}\left(A(x_L^{(2)}, x_E^{(2)}) + B(x_L^{(2)}, x_E^{(2)}) \right) + C(x_L^{(1)},x_E^{(1)}) \right]
            \\ + \left( \frac{\lambda_L \lambda_E \lambda^*}{64 \pi^2 M_L M_E} \right) g c_{\beta}^2 & \left[ 2 \left(2 I(x_E^{(1)}, x_L^{(1)}) + \left( \frac{M_E}{M_L} \right) A(x_{E}^{(1)}, x_L^{(1)}) \right) \right. \\ 
            & \left. + \frac{1}{2} \left(2 I(x_L^{(1)}, x_E^{(1)}) + \left( \frac{M_L}{M_E} \right) A(x_{L}^{(1)}, x_E^{(1)}) \right) - J(x_L^{(1)}, x_E^{(1)}) \right] \\
            + \left( \frac{\lambda_L \lambda_E \lambda^*}{64 \pi^2 M_L M_E} \right) g s_{\beta}^2 & \left[ 2 \left(2 I(x_E^{(2)}, x_L^{(2)}) + \left( \frac{M_E}{M_L} \right) A(x_{E}^{(2)}, x_L^{(2)}) \right) \right. \\ 
            & \left. + \frac{1}{2} \left(2 I(x_L^{(2)}, x_E^{(2)}) + \left( \frac{M_L}{M_E} \right) A(x_{L}^{(2)}, x_E^{(2)}) \right) - J(x_L^{(2)}, x_E^{(2)}) \right].
    \end{split}
\end{equation}
Taking the limits $x_{L,E}^{(1)} \rightarrow \infty$ and $x_{L,E}^{(2)} \rightarrow 1$, we find
\begin{equation}
        C_{\mu \gamma} = \left( \frac{\lambda_L \lambda_E \bar{\lambda}}{64 \pi^2 M_L M_E} \right) e c_{\beta}^2 \left(5 + \frac{11}{6} \tan^2 \beta \right) + \left( \frac{\lambda_L \lambda_E \lambda^*}{64 \pi^2 M_L M_E} \right) e c_{\beta}^2 \left(\frac{7}{6}\tan^2 \beta \right).
\label{eq:cmugamma_lam2}
\end{equation}
In the limit where $x_{L,E}^{(1,2)} \rightarrow \infty$, we have
\begin{equation}
        C_{\mu \gamma} = 5 \left( \frac{\lambda_L \lambda_E \bar{\lambda}}{64 \pi^2 M_L M_E} \right) e c_{\beta}^2 \left(1 + \tan^2 \beta \right).
\end{equation}

Finally, for the last representation $\textbf{2}_{-1/2} \oplus \textbf{3}_{0}$, we translate the Lagrangian of Eq.~(\ref{eq:L12_30}) to the Higgs basis and compute the following Wilson coefficients:
\begin{equation}
    \begin{split}
            C_{\mu B} =  \left( \frac{3 \lambda_L \lambda_E \bar{\lambda}}{64 \pi^2 M_L M_E} \right) g' c_{\beta}^2 & \left[ Y_E \left(A(x_E^{(1)}, x_L^{(1)}) + B(x_E^{(1)}, x_L^{(1)}) + F(x_E^{(1)}) \right) \right. \\ 
            & \left. + Y_L \left( A(x_L^{(1)}, x_E^{(1)}) + B(x_L^{(1)}, x_E^{(1)}) \right) + C(x_L^{(1)}, x_E^{(1)}) + \frac{1}{2} G(x_E^{(1)}) \right] \\
            + \left( \frac{3 \lambda_L \lambda_E \bar{\lambda}}{64 \pi^2 M_L M_E} \right) g' s_{\beta}^2 & \left[ Y_E \left(A(x_E^{(2)}, x_L^{(2)}) + B(x_E^{(2)}, x_L^{(2)}) + F(x_E^{(2)}) \right) \right. \\ 
            & \left. + Y_L \left( A(x_L^{(2)}, x_E^{(2)}) + B(x_L^{(2)}, x_E^{(2)}) \right) + C(x_L^{(2)}, x_E^{(2)}) + \frac{1}{2} G(x_E^{(2)}) \right]
            \\ - \left( \frac{3 \lambda_L \lambda_E \lambda^*}{64 \pi^2 M_L M_E} \right) g' c_{\beta}^2 & \left[ Y_E \left( 2 I(x_E^{(1)}, x_L^{(1)}) + \left(\frac{M_E}{M_L}  \right) A(x_E^{(1)}, x_L^{(1)}) \right) \right. \\
            & + \left. Y_L \left( 2 I(x_L^{(1)}, x_E^{(1)}) + \left(\frac{M_L}{M_E}  \right) A(x_L^{(1)}, x_E^{(1)}) \right) - J(x_L^{(1)}, x_E^{(1)}) \right] \\
            - \left( \frac{3 \lambda_L \lambda_E \lambda^*}{64 \pi^2 M_L M_E} \right) g' s_{\beta}^2 & \left[ Y_E \left( 2 I(x_E^{(2)}, x_L^{(2)}) + \left(\frac{M_E}{M_L}  \right) A(x_E^{(2)}, x_L^{(2)}) \right) \right. \\
            & + \left. Y_L \left( 2 I(x_L^{(2)}, x_E^{(2)}) + \left(\frac{M_L}{M_E}  \right) A(x_L^{(2)}, x_E^{(2)}) \right) - J(x_L^{(2)}, x_E^{(2)}) \right],
    \end{split}
\end{equation}
\begin{equation}
    \begin{split}
            C_{\mu W} = \left( \frac{\lambda_L \lambda_E \bar{\lambda}}{64 \pi^2 M_L M_E} \right) g c_{\beta}^2 & \left[ 2 \left(A(x_E^{(1)}, x_L^{(1)}) + B(x_E^{(1)}, x_L^{(1)}) + F(x_E^{(1)}) \right) \right. \\ 
            & \left. + \frac{1}{2}\left(A(x_L^{(1)}, x_E^{(1)}) + B(x_L^{(1)}, x_E^{(1)}) \right) + C(x_L^{(1)},x_E^{(1)}) + \frac{1}{2} G(x_E^{(1)}) \right] \\
            + \left( \frac{\lambda_L \lambda_E \bar{\lambda}}{64 \pi^2 M_L M_E} \right) g s_{\beta}^2 & \left[ 2 \left(A(x_E^{(2)}, x_L^{(2)}) + B(x_E^{(2)}, x_L^{(2)}) + F(x_E^{(2)}) \right) \right. \\ 
            & \left. + \frac{1}{2}\left(A(x_L^{(2)}, x_E^{(2)}) + B(x_L^{(2)}, x_E^{(2)}) \right) + C(x_L^{(2)},x_E^{(2)}) + \frac{1}{2} G(x_E^{(2)}) \right]
            \\ - \left( \frac{\lambda_L \lambda_E \lambda^*}{64 \pi^2 M_L M_E} \right) g c_{\beta}^2 & \left[ 2 \left(2 I(x_E^{(1)}, x_L^{(1)}) + \left( \frac{M_E}{M_L} \right) A(x_{E}^{(1)}, x_L^{(1)}) \right) \right. \\ 
            & \left. + \frac{1}{2} \left(2 I(x_L^{(1)}, x_E^{(1)}) + \left( \frac{M_L}{M_E} \right) A(x_{L}^{(1)}, x_E^{(1)}) \right) - J(x_L^{(1)}, x_E^{(1)}) \right] \\
            - \left( \frac{\lambda_L \lambda_E \lambda^*}{64 \pi^2 M_L M_E} \right) g s_{\beta}^2 & \left[ 2 \left(2 I(x_E^{(2)}, x_L^{(2)}) + \left( \frac{M_E}{M_L} \right) A(x_{E}^{(2)}, x_L^{(2)}) \right) \right. \\ 
            & \left. + \frac{1}{2} \left(2 I(x_L^{(2)}, x_E^{(2)}) + \left( \frac{M_L}{M_E} \right) A(x_{L}^{(2)}, x_E^{(2)}) \right) - J(x_L^{(2)}, x_E^{(2)}) \right].
    \end{split}
\end{equation}
Taking $x_{L,E}^{(1)} \rightarrow \infty$ and $x_{L,E}^{(2)} \rightarrow 1$, we have 
\begin{equation}
    C_{\mu \gamma} = - \left( \frac{\lambda_L \lambda_E \bar{\lambda}}{64 \pi^2 M_L M_E} \right) e c_{\beta}^2 \left(2 + \frac{11}{6} \tan^2 \beta \right) - \left( \frac{\lambda_L \lambda_E \lambda^*}{64 \pi^2 M_L M_E} \right) e c_{\beta}^2 \left(\frac{1}{6}\tan^2 \beta \right),
    \label{eq:cmugamma_lam3}
\end{equation}
and in the limit $x_{L,E}^{(1,2)} \rightarrow \infty$, we find
\begin{equation}
    C_{\mu \gamma} = - 2 \left( \frac{\lambda_L \lambda_E \bar{\lambda}}{64 \pi^2 M_L M_E} \right) e c_{\beta}^2 \left( 1+ \tan^2 \beta \right).
\end{equation}
Note that for this representation $m_{\mu}^{LE} = - 2 \times \lambda_L \lambda_E \bar{\lambda} v_d^3 / M_L M_E$ when rewriting $C_{\mu \gamma}$ in terms of $C_{\mu H_1}$.

The loop functions relevant for contributions to tree models are given by
\begin{equation}
    F(x) = x G_{\phi}(x) = - \frac{x^3 - 4x^2 + 3x + 2x \ \textrm{ln}(x)}{(1-x)^3}
\label{eq:F_fun},
\end{equation}

\begin{equation}
    G(x) = x G_{H^{\pm}}(x) = \frac{x - x^3 + 2x^2 \ \textrm{ln}(x)}{(1-x)^3},
\end{equation}

\begin{equation}
        A(x,y) = \frac{xy}{2} \left[\frac{-3y + x(1+x+y)}{(x-1)^2 (x-y)^2} + \frac{2 (x^3 + x^2 y (x-3) + y^2) \ \textrm{ln}(x)}{(1-x)^3(x-y)^3} - \frac{2 y^2 \ \textrm{ln}(y)}{(x-y)^3 (1-y)} \right],
\end{equation}

\begin{equation}
    B(x,y) = \frac{xy}{2(x-y)} \left[\frac{(1-y)(y-3) - 2 \ \textrm{ln}(y)}{(1-y)^3} - \frac{(1-x)(x-3) - 2 \ \textrm{ln}(x)}{(1-x)^3} \right] - A(y,x),    
\end{equation}

\begin{equation}
    C(x,y) = \frac{xy}{2} \left[ \frac{x + xy + y - 3}{(1-x)^2(1-y)^2} - \frac{2 x \ \textrm{ln}(x)}{(x-y)(1-x)^3} + \frac{2 y \ \textrm{ln}(y)}{(x-y)(1-y)^3} \right], \\
\end{equation}

\begin{equation}
    \begin{split}
        I(x,y) & = \frac{\sqrt{xy}}{12} \left[ \frac{3\left(x^2(3-x) + xy(x-3)(1+x) + y^2(2+x(x-1)) \right)}{(1-x)^2(x-y)^2(1-y)} \right. \\
        & \left. + \frac{2y(4x^2-2xy+y^2) \ \textrm{ln}(x/y)}{(x-y)^4} \right. \\ 
        & \left. +  \frac{2\left(3x^4+x^2y(x^2(x-6) - 4) + xy^2(2+6x + x^3) + y^3(x^2(x-3)-1)\right) \textrm{ln}(x)}{(1-x)^3 (x-y)^4} \right. \\
        & \left. + \frac{2y\left(xy(y(y-5)-2) + x^2(4+y(y-2)) + y^2 (1+y+y^2) \right) \textrm{ln}(y)}{(1-y)^2(x-y)^4}\right],
    \end{split}
\end{equation}

\begin{equation}
    J(x,y) = \frac{\sqrt{xy}}{2} 
        \left[ \frac{1+x+y-3xy}{(1 -x)^2(1-y)^2} + \frac{2x^2 \ \textrm{ln}(x)}{(1-x)^3(x-y)} - \frac{2y^2 \ \textrm{ln}(y)}{(1-y)^3 (x-y)}\right].
\label{eq:J_fun}
\end{equation}

Additionally, we list limits useful to the discussion in the text, namely, when 
both doublets are light compared to $M_{L,E}$ $(x_{L,E}^{(1,2)} \rightarrow \infty)$, or when $M_2 \simeq M_{L,E}$ $(x_{L,E}^{(2)} \rightarrow 1)$:
\begin{equation}
    \lim_{x \rightarrow \infty} F(x) = 1, \ \ \ \ \ \ \lim_{x \rightarrow 1} F(x) = \frac{2}{3},
\label{eq:F_lim}
\end{equation}

\begin{equation}
    \lim_{x \rightarrow \infty} G(x) = 1, \ \ \ \ \ \ \lim_{x \rightarrow 1} G(x) = \frac{1}{3},
\end{equation}

\begin{equation}
    \lim_{x,y \rightarrow \infty} A(x,y) = \frac{1}{6}, \ \ \ \ \ \ \lim_{x,y \rightarrow 1} A(x,y) = \frac{1}{12},
\end{equation}

\begin{equation}
    \lim_{x,y \rightarrow \infty} B(x,y) = \frac{1}{3}, \ \ \ \ \ \ \lim_{x,y \rightarrow 1} B(x,y) = \frac{1}{6},
\end{equation}

\begin{equation}
    \lim_{x,y \rightarrow \infty} C(x,y) = \frac{1}{2}, \ \ \ \ \ \ \lim_{x,y \rightarrow 1} C(x,y) = \frac{1}{12},
\end{equation}

\begin{equation}
    \lim_{x,y \rightarrow \infty} I(x,y) = -\frac{1}{12}, \ \ \ \ \ \ \lim_{x,y \rightarrow 1} I(x,y) = -\frac{1}{12},
\end{equation}

\begin{equation}
    \lim_{x,y \rightarrow \infty} J(x,y) = 0, \ \ \ \ \ \ \lim_{x,y \rightarrow 1} J(x,y) = \frac{1}{12}.
\label{eq:J_lim}
\end{equation}

\end{document}